\documentclass[]{spie}  %>>> use for US letter paper

\usepackage{wrapfig}
\usepackage{sidecap}

\usepackage{graphicx}

\usepackage{amsmath,amsfonts,amssymb}

\title{Daytime Sky Polarization Calibration Limitations}

\author[a,b,c]{David M. Harrington}
\author[d]{Jeffrey R. Kuhn}
\author[e]{Arturo L\'{o}pez Ariste}
\affil[a]{National Solar Observatory, 3665 Discovery Drive, Boulder, CO, 80303, USA}
\affil[b]{Kiepenheuer-Institut f\"{u}r Sonnenphysik, Sch\"{o}neckstr. 6, D-79104 Freiburg, Germany}
\affil[c]{Inst. for Astronomy, University of Hawaii, 2680 Woodlawn Drive, Honolulu, HI, 96822, USA}
\affil[d]{Inst. for Astronomy Maui, University of Hawaii, 34 Ohia Ku St., Pukalani, HI, 96768, USA}
\affil[e]{IRAP - CNRS UMR 5277. 14, Av. E Belin. Toulouse. France}

\authorinfo{Further author information: (Send correspondence to D.M.H)\\D.M.H.: E-mail: harrington@nso.edu, Telephone: 1 808 572 6888}

\begin{document} 
\maketitle

\begin{abstract}
The daytime sky has been recently demonstrated as a useful calibration tool for deriving polarization cross-talk properties of large astronomical telescopes. The Daniel K Inouye Solar Telescope (DKIST) and other large telescopes under construction can benefit from precise polarimetric calibration of large mirrors. Several atmospheric phenomena and instrumental errors potentially limit the techniques accuracy.  At the 3.67m AEOS telescope on Haleakala, we have performed a large observing campaign with the HiVIS spectropolarimeter to identify limitations and develop algorithms for extracting consistent calibrations. Effective sampling of the telescope optical configurations and filtering of data for several derived parameters provide robustness to the derived Mueller matrix calibrations. Second-order scattering models of the sky show that this method is relatively insensitive to multiple-scattering in the sky provided calibration observations are done in regions of high polarization degree. The technique is also insensitive to assumptions about telescope induced polarization provided the mirror coatings are highly reflective. Zemax-derived polarization models show agreement between the functional dependence of polarization predictions and the corresponding on-sky calibrations.
\end{abstract}


\section{Introduction}
\label{sec:intro}  

Polarization calibration of large telescopes and modern instruments is often limited by the availability of suitable sources for calibration.  Several calibration techniques exist using stars or the sun, internal optical systems or a priori knowledge of the expected signals but each technique has limitations.  For altitude-azimuth telescopes, coud\'{e} or Nasmyth instruments, or telescopes with off-axis primaries, the polarization calibration usually requires bright, highly polarized sources available over a wide range of wavelengths, altitude-azimuth pointings.  In night-time astronomy, polarized standard stars are commonly used but provide very limited altitude-azimuth coverage, are faint and have low polarization amplitudes (typically below 5\% \cite{1982ApJ...262..732H, GilHutton:2003di, Fossati:2007ud}). Unpolarized standard stars also exist but are also faint and provide limited altitude-azimuth coverage.  Solar telescopes can use  solardisk-center as a bright, zero-polarization target, provided there is no magnetic field activity. Solar observations often lack bright, significantly polarized targets of known properties. Smaller telescopes can use fixed polarizing filters placed over the telescope aperture to provide known input states that are detected and yield terms of the Mueller matrix as for the Dunn Solar Telescope \cite{SocasNavarro:2011gn, 2006SoPh..235...55S, SocasNavarro:2005bq, SocasNavarro:2005jl, SocasNavarro:2005gv}. Symmetries of spectropolarimetric signatures from the Zeeman effect have been used in solar physics to determine terms in the Mueller matrix \cite{Kuhn:1994jk, 2010SPIE.7735E..4EE}.  Many studies have either measured and calibrated telescopes, measured mirror properties or attempted to design instruments with minimal polarimetric defects (cf. \cite{1991SoPh..134....1A, Giro:2003il, Patat:2006en, Tinbergen:2007fd, Joos:2008dg, vanHarten:2009gi, Roelfsema:2010ca}). Space-based polarimetric instruments such as Hinode also undergo detailed polarization calibration and characterization \cite{2008SoPh..249..233I, Lites:2013hm}.

Many telescopes use calibration optics such as large polarizers, polarizer mosaic masks, polarization state generators, or optical injection systems at locations in the beam after the primary or secondary mirror.  For systems, a major limitation is the ability to calibrate the primary mirror and optics upstream of the calibration system.  These systems include telescopes with large primary mirrors, systems without accessible intermediate foci, or many-mirror systems without convenient locations for calibration optics. System calibrations are subject to model degeneracies, coherent polarization effects in the point spread function and other complex issues such as fringes or seeing-induced artifacts \cite{SocasNavarro:2011gn} \cite{2006SoPh..235...55S}  \cite{Judge:2004jk} \cite{1994A&A...292..713S} \cite{1992A&A...260..543S}. Modern instrumentation is often behind adaptive optics systems requiring detailed consideration of active performance on polarization artifacts in addition to deconvolution techniques and error budgeting \cite{2014SPIE.9148E..6PS, 2014SPIE.9147E..7CH, vanNoortM:2012iz, ovelar:2012elt}. Modeling telescope polarization is typically done either with simple single-ray traces using assumed mirror refractive indices or with ray tracing programs such as Zemax  \cite{1991SoPh..134....1A, Harrington:2006hu, 2014SPIE.9147E..7CH}.

Every major observatory addresses a diversity of science cases.  Often, cross-talk from the optics limit the polarization calibration to levels of 0.1\% to >1\% in polarization orientation  (e.g. ESPaDOnS at CFHT, LRISp at Keck, SPINOR at DST \cite{Barrick:2010jz, Barrick:2010gv, Harrington:2015cq, SocasNavarro:2011gn}). Artifacts from the instruments limit the absolute degree of polarization measurements from backgrounds or zero-point offsets.  Hinode and the DKIST project outline attempts to create error budgets, calling for correction of these artifacts to small fractions of a percent.  The calibration techniques presented here aim to calibrate the cross-talk elements of the Mueller matrix to levels of roughly 1\% of the element amplitudes, consistent with internal instrument errors. We also show that the limitation of the method is not the model for the polarization patterns of the sky, but other instrumental and observational issues. 

The High resolution Visible and Infrared Spectrograph (HiVIS) is a coud\'{e} instrument for the 3.67m AEOS telescope on Haleakala, HI.  The visible arm of HiVIS has a spectropolarimeter which we recently upgrade to include charge shuffling synchronized with polarization modulation using tunable nematic liquid crystals \cite{Harrington:2015dl}. In \cite{Harrington:2015dl}, hereafter called H15, we outline the coud\'{e} path of the AEOS telescope and details of the HiVIS polarimeter.  

The Daniel K Inouye Solar Telescope (DKIST) is a next-generation solar telescope with a 4m diameter off-axis primary mirror and a many-mirror folded coud\'{e} path \cite{2014SPIE.9145E..25M, Keil:2011wj, Rimmele:2004ew} \cite{2014SPIE.9145E..25M,Rimmele:2004ew}. This altitude-azimuth system uses 7 mirrors to feed light to the coud\'{e} lab \cite{2014SPIE.9147E..0FE, 2014SPIE.9147E..07E, 2014SPIE.9145E..25M}. Its stated science goals require very stringent polarization calibration.  Operations involve 4 polarimetric instruments spanning the 380nm to 5000nm wavelength range with changing configuration and simultaneous operation of 3 polarimetric instruments covering 380nm to 1800nm \cite{2014SPIE.9147E..0FE, 2014SPIE.9147E..07E, 2014SPIE.9147E..0ES, SocasNavarro:2005bq}. Complex modulation and calibration strategies are required for such a mulit-instrument system \cite{2014SPIE.9147E..0FE,2014SPIE.9147E..07E, Sueoka:2014cm, 2015SPIE.9369E..0NS, deWijn:2012dd, 2010SPIE.7735E..4AD}. With a large off-axis primary mirror, calibration of DKIST instruments requires external (solar, sky, stellar) sources.  The planned 4m European Solar Telescope (EST), though on-axis, will also require similar calibration considerations \cite{SanchezCapuchino:2010gy, Bettonvil:2011wj,Bettonvil:2010cj,Collados:2010bh}

\subsection{Polarization}

The following discussion of polarization formalism closely follows \cite{1992plfa.book.....C} and \cite{Clarke:2009ty}. In the Stokes formalism, the polarization state of light is denoted as a 4-vector: ${\bf S}_i = [ I,  Q, U, V ] ^T$. In this formalism, $I$ represents the total intensity, $Q$ and $U$ the linearly polarized intensity along polarization position angles $0^\circ$ and $45^\circ$ in the plane perpendicular to the light beam, and $V$ is the right-handed circularly polarized intensity.  The intensity-normalized Stokes parameters are usually denoted as: $[ 1,q, u, v ]^T = [ I,Q, U, V ] ^T / I$. The degree of polarization (DoP) is the fraction of polarized light in the beam: $DoP = \frac {\sqrt{ Q^2 + U^2 + V^2 } } {I} =  \sqrt{q^2 + u^2 + v^2}$. For this work, we adopt a term Angle of Polarization (AoP) from the references on daytime sky polarimetry which defines the angle of linear polarization as $ATAN(Q/U)/2$. The Mueller matrix is a $4 \times 4$ set of transfer coefficients which describes how an optic changes the input Stokes vector (${\bf S}_{i_{input}}$) to the output Stokes vector (${\bf S}_{i_{output}}$): ${\bf S}_{i_{output}} = {\bf M}_{ij} {\bf S}_{i_{input}}$. If the Mueller matrix for a system is known, then one inverts the matrix to recover the input Stokes vector. One can represent the individual Mueller matrix terms as describing how one incident polarization state transfers to another.  In this paper we will use the notation:

\begin{equation}
{\bf M}_{ij} =
 \left ( \begin{array}{rrrr}
 II   	& QI		& UI		& VI		\\
 IQ 	& QQ	& UQ	& VQ	\\
 IU 	& QU	& UU	& VU		\\
 IV 	& QV	& UV		& VV		\\ 
 \end{array} \right ) 
\end{equation}

\subsection{The Daytime Sky as a Calibration Target}

The daytime sky is a bright, highly linearly polarized source that illuminates the telescope optics similar to distant targets (sun, stars, satellites, planets) starting with the primary mirror. A single-scattering Rayleigh calculation is often adequate to describe the sky polarization to varying precision levels and is introduced in great detail in several text books (e.g. \cite{Coulson:1980wq, Coulson:1988us}). 

%\begin{figure}
%\begin{center} 
\begin{wrapfigure}{r}{0.48\textwidth}
\centering
\vspace{-2mm}
\hbox{
\includegraphics[width=0.98\linewidth]{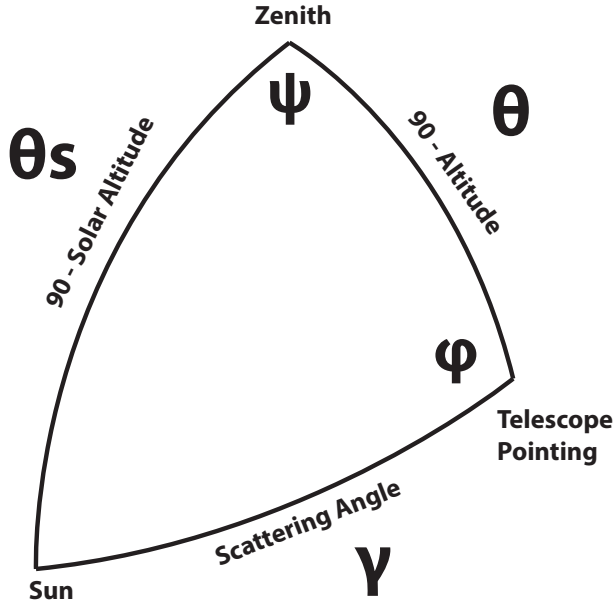} 
}
\caption{\small The celestial triangle representing the geometry for the sky polarization computations at any telescope pointing.  $\gamma$ is the angular distance between the telescope pointing and the sun. $\theta_{s}$ is the solar angular distance from the zenith. $\theta$ is the angular distance between the telescope pointing and the zenith. $\phi$ is the angle between the zenith direction and the solar direction at the telescope pointing. The angle $\psi$ is the difference in azimuth angle between the telescope pointing and the solar direction. The input $qu$ components are derived as SIN and COS of of $\psi$ respectively. The Law of Cosines is used to solve for any angles needed to compute degree of polarization and position angle of polarization.} 
\label{Rayleigh_Sky_Triangle} 
 \end{wrapfigure}
%\end{center}
%\end{figure} 

There are many atmospheric and geometric considerations that change the skylight polarization pattern. The linear polarization amplitude and angle can depend on the solar elevation, atmospheric aerosol content, aerosol vertical distribution, aerosol scattering phase function, wavelength of the observation and secondary sources of illumination such as reflections off oceans, clouds or multiple scattering \cite{Horvath:2002hu, Horvath:2003hn, 2014JQSRT.139....3H, Lee:1998hj, Liu:1997dg, Suhai:2004ed, Gal:2001ed, 2001RSPSA.457.1385G, Vermeulen:2000kq, 2014JQSRT.149..334L, 2009JQSRT.110.1954L,1996JQSRT..55..181L, Pomozi:2001tn, Cronin:2006jy, Cronin:2005kd, Hegedus:2007ce, 2004NJPh....6..162B}.  Anisotropic scattered sunlight from reflections off land or water can be highly polarized and temporally variable \cite{Litvinov:2010td, Peltoniemi:2009tf, He:2010vy, Salinas:2007wh, Ota:2010be, Kisselev:2004if}.  Aerosol particle optical properties and vertical distributions also vary \cite{Wu:1997ki, 2014AMT.....7.4341V, 2006IzAOP..42...68S, Vermeulen:2000kq, Ougolnikov:2005ic, Ougolnikov:2002uh, Ougolnikov:2005vh, 1999KosIs..37..168U, Ugolnikov:2011hn, Ugolnikov:2013dn, Ugolnikov:2009jwa}. The polarization can change across atmospheric absorption bands or can be influenced by other scattering mechanisms \cite{Boesche:2008eb, Boesche:2006iu, Zeng:2008if, Aben:2001ek,Aben:1999el}.  Deviations from a single scattering Rayleigh model grow as the aerosol, cloud, ground or sea-surface scattering sources affect the telescope line-of-sight. Clear, cloudless, low-aerosol conditions should yield high linear polarization amplitudes and small deviations in the polarization direction from a Rayleigh model. Observations generally support this conclusion \cite{Pust:2006fd,Pust:2009fq,Pust:2008bj,Pust:2007fl,Pust:2006gc,Pust:2005hl, 2010SPIE.7672E..0AS, Swindle:2014ue, Swindle:2014wc}.   Conditions at twilight with low solar elevations can present some spectral differences \cite{Ugolnikov:2013dn,Ugolnikov:2011hn,Ugolnikov:2009jwa,2004JQSRT..88..233U,1999KosIs..37..168U}.

An all-sky imaging polarimeter deployed on Haleakala also shows that a single scattering sky model is a reasonable approximation for DKIST and AEOS observatories \cite{Swindle:2014ue, Swindle:2014wc}. The preliminary results from this instrument showed that the angle of linear polarization agreed with single-scattering models to better than 1$^\circ$ in regions of the sky more than 20\% polarized. We show below how to filter data sets based on several measures of the daytime sky properties to ensure second order effects are minimized.  The daytime sky degree of polarization was much more variable, but as shown in later sections, the DoP variability has minimal impact on our calibration method. More detailed models include multiple scattering and aerosol scattering, is also available using industry standard atmospheric radiative transfer software such as MODTRAN \cite{Fetrow:2002uj, Berk:2006ia}.  However, the recent studies on Haleakala applied measurements and modeling techniques to the DKIST site and found that the AoP was very well described by the single-scattering model for regions of the sky with DoP greater than 15\% \cite{Swindle:2014ue, Swindle:2014wc}.  The behavior of the DoP was much more complex and did not consistently match the single-scattering approximation. The technique we develop uses only the AoP.

\subsection{Single Scattering Sky Polarization Model}

	Sky polarization modeling is well represented by simple single-scattering models with a few free parameters.  The most simple Rayleigh sky model includes single scattering with polarization perpendicular to the scattering plane.  A single scale factor for the maximum degree of linear polarization ($\delta_{max}$) scales the polarization pattern across the sky.

\begin{figure} 
\begin{center}
\includegraphics[width=1.0\linewidth, angle=0]{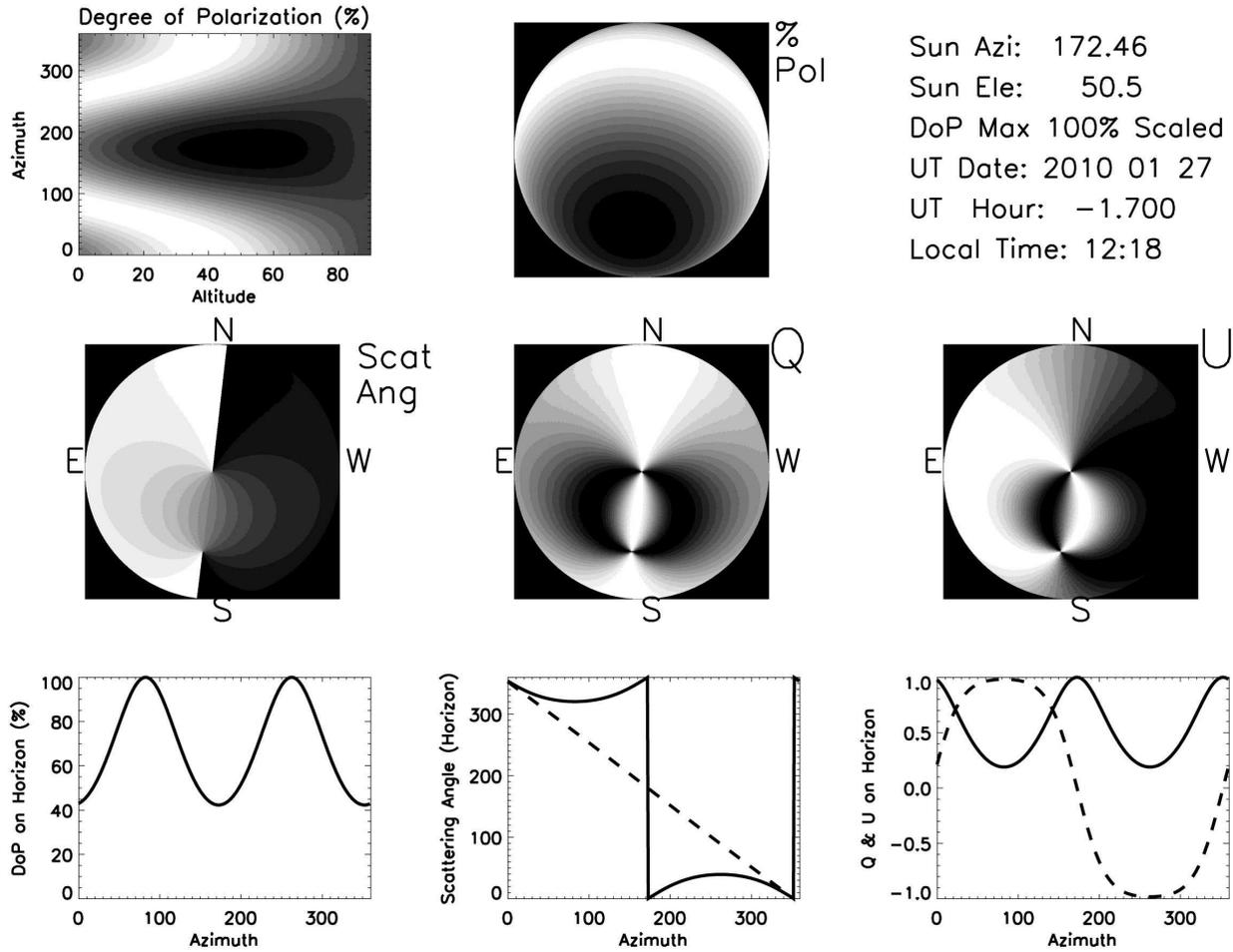}
\caption{ \label{rayleigh_model_haleakala} Various Rayleigh sky model parameters computed in a range of projections for mid morning on January 27th 2010 on Haleakala when the sun was at an elevation of 50.5$^\circ$ and an azimuth of 172$^\circ$. The single scattering model was scaled to a maximum DoP of 100\% ($\delta_{max}$). The top left panel shows the model DoP with white at 100\% and black as 0\% for all altitudes and azimuths plotted on a Cartesian rectangular grid.  The top-middle panel shows the same DoP model data but in an orthographic projection with North up and East left. These two panels show equivalent DoP data just with differing projections. The middle left panel shows the scattering angle in an orthographic projects. This scattering angle shows the input linear polarization angle in the reference frame of the telescope which always has the +elevation axis point .  The middle right and mid panels show $q$ and $u$ in an orthographic projection where white is +1 and black is -1.  The coordinate system for $qu$ was chosen to be +1 in the +altitude direction of an altitude-azimuth system.  This system has a singularity at the zenith where the telescope optics can degenerately point to the zenith with any azimuth.  The +altitude = +$q$ system is referenced to the optical train through the orientation of the primary mirror mount against the sky.  The bottom row of panels show properties of the sky polarization model on the horizon.  The left panel shows the DoP with peaks in the east and west.  The scattering angle in the middle shows sign changes at the solar azimuth of 172$^\circ$.  The right plot shows $q$ as the solid black line and $u$ as the dashed line.  Stokes $u$ changes sign at the solar azimuth of 172$^\circ$.}
\end{center}
\end{figure}

The all-sky model requires knowing the solar location and the scale factor ($\delta_{max}$)  to compute the degree of polarization (DoP) and the angle of linear polarization (AoP) projected on to the sky.  The geometry of the Rayleigh sky model is seen in Figure \ref{Rayleigh_Sky_Triangle}. The geometrical parameters are the observers location (latitude, longitude, elevation) and the time. The solar location and relevant angles from the telescope pointing are computed from the spherical geometry in Figure \ref{Rayleigh_Sky_Triangle}. The maximum degree of polarization ($\delta_{max}$) in this model occurs at a scattering angle ($\gamma$) of 90$^\circ$. The Rayleigh sky model predicts the degree of polarization ($\delta$) at any telescope pointing (azimuth, elevation) as:
	
\begin{equation}
\delta=\frac{\delta_{max}sin^2\gamma}{1+cos^2\gamma}
\label{degpol1}
\end{equation}

\noindent and the spherical geometry is computed as: $cos(\gamma)=sin(\theta_{s})  sin(\theta) cos(\psi) + cos(\theta_{s})cos(\theta)$ where $\gamma$ is the angular distance between the telescope pointing and the sun, $\theta_{s}$ is the solar zenith angle, $\theta$ is the angular distance between the telescope pointing and the zenith, and $\psi$ is the azimuthal angle between the solar direction and the telescope pointing.  The geometry comes from the law of cosines with $\gamma$, $\theta_{s}$, and $\theta$ as the angular distances and $\psi$ representing the interior angle. The spherical triangle formed by the solar location, zenith, and telescope pointing can be seen in Figure \ref{Rayleigh_Sky_Triangle}.   A detailed example of this single scattering model can be seen in Figure \ref{rayleigh_model_haleakala} computed for January 27th 2010. This Figure shows several model parameters either in altitude-azimuth projections or in orthographic projects. The single scattering model has the highest DoP in a band of 90$^\circ$ scattering angle.

\subsection{Solving for Telescope Mueller Matrix Elements}

We model the $3 \times 3$ cross-talk elements ($QUV$ to $QUV$ terms) of the Mueller matrix as a rotation matrix \cite{Harrington:2011fz}. This method makes the assumption that a telescope with weakly polarizing optics can have a Mueller matrix that is well represented by a rotation matrix. We find cross-talk of 100\% but the induced polarization and depolarization terms are less than 5\%. A rotation matrix has been a good fit to our past data, is predicted by our Zemax modeling and is easily described with three Euler angles to produce the 9 terms of the cross-talk matrix \cite{Harrington:2015dl, Harrington:2010km}. We also perform a sensitivity analysis in later sections to show that this approximation is reasonable.  We find in the appendices, that we can neglect the first row and column of the Mueller matrix as the correction to the inner $QUV$ to $QUV$ terms is second order in these neglected terms. 

For our procedure, all Stokes vectors are scaled to unit length (projected on to the Poincar\'{e} sphere) by dividing the Stokes vector by the measured degree of polarization (DoP).  This removes the residual effects from changes in the sky degree of polarization, telescope induced polarization and depolarization. Since we ignore the induced polarization and depolarization, we consider only the 3x3 cross-talk elements as representing the telescope Mueller matrix. We denote the 3 Euler angles as ($\alpha, \beta, \gamma$) and use a short-hand notation where $cos(\gamma)$ is shortened to $c_\gamma$. We specify the rotation matrix (${\bf \mathbb{R}}_{ij}$) using the ZXZ convention for Euler angles as:

\begin{equation}
{\bf \mathbb{R}}_{ij} =
 \left ( \begin{array}{rrr}
 c_\gamma	& s_\gamma	& 0		\\
 -s_\gamma	& c_\gamma	& 0		\\
 0			& 0			&1		\\ 
 \end{array} \right ) 
 \left ( \begin{array}{rrr}
 1			& 0			& 0		\\
 0			& c_\beta		& s_\beta	\\
 0			& -s_\beta		& c_\beta	\\ 
 \end{array} \right ) 
 \left ( \begin{array}{rrr}
 c_\alpha		& s_\alpha	& 0		\\
 -s_\alpha		& c_\alpha	& 0		\\
 0			& 0			&1		\\ 
 \end{array} \right )  =
 \left ( \begin{array}{ccc}
 c_\alpha c_\gamma - s_\alpha c_\beta s_\gamma		& s_\alpha c_\gamma + c_\alpha c_\beta s_\gamma		& s_\beta s_\gamma		\\
-c_\alpha s_\gamma - s_\alpha c_\beta c_\gamma 		&-s_\alpha s_\gamma + c_\alpha c_\beta c_\gamma 	& s_\beta c_\gamma 	\\
 s_\alpha s_\beta								& -c_\alpha s_\beta								& c_\beta				\\ 
 \end{array} \right ) 
\label{eqn_definerot}
\end{equation}

With this definition for the rotation matrix, we solve for the Euler angles assuming a linearly polarized daytime sky scaled to 100\% degree of polarization as calibration input. If we denote the measured Stokes parameters, ${\bf S}_i$, as ($q_m, u_m, v_m$) with i=1,2,3 and the input sky Stokes parameters, ${\bf R}_j$, as ($q_r, u_r, 0$) then the $3 \times 3$ $QUV$ Mueller matrix elements at each wavelength are:

\begin{equation}
{\bf S}_i =
\left ( \begin{array}{r}
q_{m} \\
u_{m} \\
v_{m} \\
\end{array}  \right ) =
{\bf M}_{ij} {\bf R}_j = 
 \left ( \begin{array}{rrr}
 QQ  &  UQ &  VQ     \\
 QU  & UU  & VU         \\
 QV  &  UV  &  VV     \\ 
 \end{array} \right ) 
\left ( \begin{array}{r}
q_{r} \\
u_{r} \\
0 \\
\end{array}  \right )
\label{eqn-rayltransf}
\end{equation}

We have no $V$ input from the daytime sky to constrain the $VQ$, $VU$ and $VV$ terms. Nevertheless, two measurements at different input polarization angles are sufficient to fully specify the rotation matrix. Thus, we use the fact that the sky polarization changes orientation with time and take measurements at identical telescope pointings separated by enough time for the solar sky illumination to change. A set of observations with a changing input angle of polarization yields an over-constrained solvable problem for all six linear polarization terms in the Mueller matrix. 

When using this rotation matrix approximation for the telescope Mueller matrix, the Rayleigh Sky input Stokes parameters multiply each term of the rotation matrix to give a system of equations for the three Euler anlges ($\alpha$, $\beta$, $\gamma$). This system of equations can be solved using a normal non-linear least-squares minimization by searching the ($\alpha$, $\beta$, $\gamma$) space for minima in squared error. This direct solution of this set of equations using standard minimization routines is subject to several ambiguities that affect convergence using standard minimization routines. The details of our methods for deriving Euler angles and an example of how one could plan sky calibration observations are outlined in the Appendix of \cite{Harrington:2011fz}. 

Equating Mueller matrix elements to rotation matrix elements, we can write the system of equations for the three Euler angles. This system of equations can be solved using a normal non-linear least-squares minimization by searching the ($\alpha$, $\beta$, $\gamma$) space for minima in squared error. With the measured Stokes vector (${\bf S}_i$),  i=(1,2,3), the Rayleigh sky input vector (${\bf R}_j$), j=(1,2), and a rotation matrix (${\bf \mathbb{R}}_{ij}$) we define the error ($\epsilon$) as: $\epsilon^2(\alpha,\beta,\gamma) = \sum \limits_{i=1}^{3} \sum \limits_{j=1}^{2} \left [ {\bf S}_i - {\bf R}_i {\bf \mathbb{R}}_{ij}(\alpha,\beta,\gamma) \right ]^2$. For n measurements, this gives us $3 \times n$ terms. This solution is easily solvable in principle but has ambiguities. An alternative method for the direct least-squares solution for Euler angles is done in two steps. First we solve a system of equations for the Mueller matrix elements directly that are not subject to rotational ambiguity. With the estimated Mueller matrix elements in hand, we can then perform a rotation matrix fit to the derived Mueller matrix element estimates. This two-step process allows us to use accurate starting values to speed up the minimization process and to resolve Euler angle ambiguities. When deriving the Mueller matrix elements of the telescope, one must take care that the actual derived matrices are physical. For instance, there are various matrix properties and quantities one can derive to test the physicality of the matrix \cite{Givens:1993cl, Kostinski:1993uh, Takakura:2009bq, 2016OptLE..86..242A}. Noise and systematic errors might give over-polarizing or unphysical Mueller matrices. By fitting a rotation matrix, we avoid unphysical matrices. 

The normal solution for Mueller matrix elements can be computed via the normal least-squares method. We can re-arrange the time-varying Rayleigh sky inputs to (${\bf R}_{ij}$) for $i$ independent observations and $j$ input Stokes parameters. The measured Stokes parameters (${\bf S}_i$) become individual column vectors. The unknown Mueller matrix elements are arranged as a column vector by output Stokes parameter (${\bf M}_{j}$). If we write measured Stokes parameters as ($q_{m_i}, u_{m_i}, v_{m_i}$) and the Rayleigh input Stokes parameters as ($q_{r_i}, u_{r_i}$), we can explicitly write a set of equations for two Mueller matrix elements:

\begin{equation}
{\bf S}_{i} =
\left ( \begin{array}{r}
q_{m_1} \\
q_{m_2} \\
q_{m_3} \\
\end{array}  \right ) = 
{\bf R}_{ij}{\bf M}_j =
\left ( \begin{array}{ll}
q_{r_1}	& u_{r_1}  \\
q_{r_2}	& u_{r_2}  \\
q_{r_3}	& u_{r_3}  \\
\end{array}  \right )
\left ( \begin{array}{r}
QQ	\\
UQ	\\ 
\end{array} \right ) 
\end{equation}

We have three such equations for each set of Mueller matrix elements sampled by sky measurements. We can express the residual error (${\bf \epsilon}_i$) for each incident Stokes parameter (${\bf S}_i$) with an implied sum over $j$ as: ${\bf \epsilon}_i = {\bf S}_i - {\bf R}_{ij} {\bf M}_j$. The normal solution of an over specified system of equations is easily derived in a least-squares sense using matrix notation. The total error $E$ as the sum of all residuals for $m$ independent observations we get: $E = \sum \limits_{i=1}^m {\bf \epsilon}_i^2$. We solve the least-quares system for the unknown Mueller matrix element (${\bf M}_j$) by minimizing the error with respect to each equation. The partial derivative for ${\bf \epsilon}_i$ with respect to ${\bf M}_j$ is just the sky input elements ${\bf R}_{ij}$. Taking the partial with respect to each input Stokes parameter we get:

\begin{small}
\begin{equation}
\frac{\partial E}{\partial {\bf M}_j} = 2 \sum \limits_i {\bf \epsilon}_i \frac{\partial {\bf \epsilon}_i}{\partial {\bf M}_j} =  -2 \sum \limits_i {\bf R}_{ij} \left ( {\bf S}_i - \sum \limits_k {\bf R}_{ik} {\bf M}_k \right ) =  0
\end{equation}
\end{small}

We have inserted a dummy sum over the index $k$. Multiplying out the terms and rearranging gives us the normal equations: $\sum \limits_i \sum \limits_k {\bf R}_{ij} {\bf R}_{ik} {\bf M}_k  = \sum \limits_i {\bf R}_{ij} {\bf S}_i$. This is written in matrix notation is the familiar solution of a system of equations via the normal method: ${\bf M} = \frac{ {\bf R}^T {\bf S} } { {\bf R}^T {\bf R}}$.

This solution is stable provided a diverse range of input states are observed to give a well-conditioned inversion. The noise properties and inversion characteristics of this equation can be calculated in advance of observations and optimized. We can write the matrix ${\bf A}$ with an implied sum over $i$ observations for each term.  As an example for a single element, if we compute the inverse of {\bf A} and multiply out ${\bf A}^{-1}$ for the $QQ$ term we can write:

\begin{equation}
\label{eqn_rayl_modmatref}
{\bf A} =  {\bf R}^T {\bf R} = 
\left ( \begin{array}{rr}
q_{r_i}q_{r_i}	& q_{r_i}u_{r_i}		\\
q_{r_i}u_{r_i}	& u_{r_i}u_{r_i}		\\
\end{array}  \right ),
A_{11} = QQ = \frac { (q_{r_i} q_{m_i})(u_{r_i} u_{r_i}) - (u_{r_i} q_{m_i})(q_{r_i}u_{r_i})}   { (q_{r_i}q_{r_i})(u_{r_i}u_{r_i})-(q_{r_i}u_{r_i})(q_{r_i}u_{r_i}) }
\end{equation}

The solution to the equations for the three sets of Mueller matrix elements is outlined in the Appendix of \cite{Harrington:2011fz}.  In this manner, we can easily implement the usual matrix formalism with a time-series of daytime sky observations to measure six Mueller matrix elements.

\section{Single Scattering Model Limitations}

The assumption of a single-scattering model for computing the daytime sky polarization is incorrect under some circumstances. Multiple scattering, contributions from multiple light sources (upwelling, cloud reflections, ocean reflections) all complicate the computation of the DoP and associated linear polarization angle. In this section we outline a second order scattering model and show how this model can be used to choose calibration observations to avoid such issues.  By planning observations in regions of the sky where multiple scattering issues are minimized, this calibration technique can be efficiently used with a simple single scattering model. 

\subsection{Multiple scattering models}

%\begin{figure}
%\begin{center} 
\begin{wrapfigure}{r}{0.55\textwidth}
\centering
\vspace{-15mm}
\hbox{
\hspace{-2.5em}
\includegraphics[width=1.1\linewidth, angle=0]{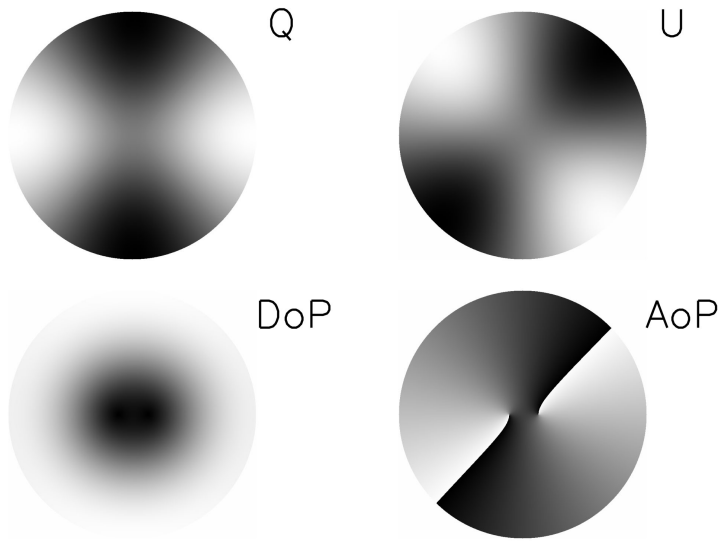}
}
\caption{ \label{second_order_sky} The multiple scattering model with a splitting constant of $\delta$ = 4 ATAN($A$) = 27$^\circ$. All projections are sterographic with North up and East left. The sun is at the Zenith.  The top left panel shows Stokes $q$ and the top right panel shows Stokes $u$ in the altitude-azimuth frame. The grey scale corresponds to white = 1 and black = -1.  The bottom left panel shows the DoP with black as 0 and white as 100\%.  The angle of polarization (AoP) computed as 0.5ATAN(q,u) is shown on the bottom right.  The two polarization zero points are seen as the split singularities near the zenith in the center of the AoP image. The AoP is linearly scaled from black to white from 0 to 180$^\circ$.}
\vspace{-3mm}
 \end{wrapfigure}
%\end{center}
%\end{figure} 

	We show here that the common 2-component multiple scattering model imparts minimal changes the AoP in wide regions of the sky. Several additions to the single-scattering model are possible but behave similarly.  Along any line of sight in the sky, there are contributions from the single scattered sunlight along with multiply scattered light off a range of airborne and ground based sources as well as extinction. Contributions from Mie scattering of water droplets, ice crystals or large aerosols modify the models in complex ways.  As an example of the variations between the single scattered Rayleigh model and a simple multiple scattering model, we follow the mathematical formalism of \cite{2004NJPh....6..162B} to derive general properties of the polarization imprinted from the most common multiple scattering source.

	In their notation, they use $\zeta$ to denote the location of a point on a stereographic projection of the sky. In Cartesian geometry, $\zeta$ = x + iy.  In polar coordinates, $\zeta$ = r $e^{i\phi}$.  In \cite{2004NJPh....6..162B}, they use the term $w$ to represent the polarization pattern across the sky.  By breaking the exponential equation in to an amplitude term $|w|$ and a complex orientation term $\gamma(\zeta)$, they represent the stereographic projection for the sky polarization pattern as: $w(\zeta) = |w| e^{2i\gamma(\zeta)}$. For the single scattering case, this simple relation behaves as $\zeta^2$ can be scaled to an amplitude of 1 and written in polar coordinates ($r,\phi$) as: $w(\zeta) \sim \zeta^2 = r^2 e^{2i(\phi - \frac{\pi}{2})}
$.

In order to add multiple scattering to this equation, we must consider the shift of the zero polarization points away from the solar and anti-solar location. These zero points are Brewster and Babinet points near the sun as well as the Arago and second Brewster point near the anti-solar location.  Several empirical results show that the singularities are found above and below the sun along the solar meridian.  This generally follows from the empirical result that {\it double scattering} is the dominant contribution to multiple scattering in the typical locations surveyed. This double-scattering contribution is generally polarized in the vertical direction as it represents the light scattered in to the line of sight from the integrated skylight incident on all points along the line of sight.  When the sun is low in the horizon, the low DoP regions of the sky are also low on the horizon.  This double scattering contribution is of the same amplitude as the single scattered light when the single scattered light is weak and horizontally polarized, which occurs above and below the sun at low solar elevations during sunrise and sunset. 

The most simple perturbation to the model is to add a constant representing a small additional polarization of assumed constant orientation denoted $A$.   Following \cite{2004NJPh....6..162B}, the zero polarization singularities fall at the locations of $\zeta = \pm iA$ which corresponds to a Cartesian y value of $\pm A$.  To make the singularities at the anti-sun location, the equation was generalized to: $w(\zeta) \sim (\zeta^2 + A^2) (\zeta^2 + \frac{1}{A^2})$.

\begin{wrapfigure}{r}{0.70\textwidth}
\centering
\vspace{-0mm}
\hbox{
\hspace{-0.0em}
\includegraphics[width=1.0\linewidth, angle=0]{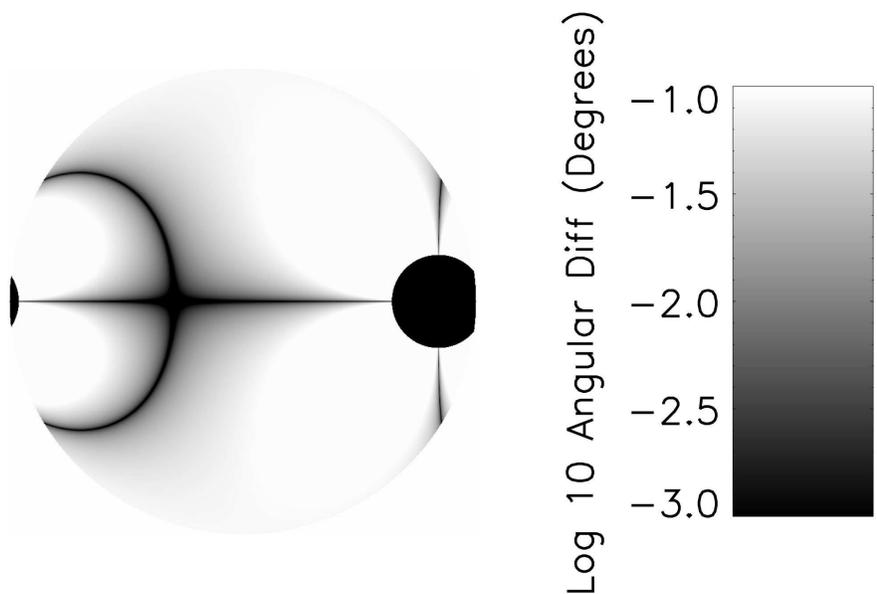}
}
\caption{ \label{sky_angular_residual} The angular differences (in degrees) between the single scattering Rayleigh model and the multiple scattering model outlined above with the {\it double scattering} term.  We use a log color scale and chose a separation of $\delta$ = 4 ATAN($A$) = 12$^\circ$ and the sun is at 10$^\circ$ elevation.  A region within 20$^\circ$ of the sun has been masked and shows up as a black circle.  The scale bar on the right shows the color scheme with white as 0.1 degrees AoP angular difference and black as 0.001 degrees AoP difference. The line of symmetry between the sun and the anti-solar location is a region of minimal difference as is the 90$^\circ$ scattering plane shown as the curved black arc in this stereographic projection.}
 \end{wrapfigure}

A simple example of this 2 term scattering model is shown in Figure \ref{second_order_sky}.  The stereographic projection convention has been used.  In this case, we put the sun on the horizontal axis to match the North = up convention of Figure \ref{rayleigh_model_haleakala}.  However, in this formalism, the Stokes $qu$ parameters are not referenced to the altitude-azimuth frame and there is no singularity at the zenith.  An angular splitting of 27$^\circ$  was chosen and the sun is at an elevation of 89$^\circ$.  This solar elevation puts the sun in the center of the image with the horizon projected on the edge of the circle. 

The calibration method we have pursued is based on the assumption that the angle of linear polarization of the sky polarization pattern is known as a modeled input parameter with a high degree of accuracy.  Variations between the single Rayleigh scattering model and the real input Stokes vector can cause errors in our calibration methodology. 

Figure \ref{sky_angular_residual} shows the AoP variations between the simple single-scattering model and the multiple scattering model considering the {\it double scattering} term in a stereographic projection for a range of multiple-scattering models.  In the regions of highest DoP at scattering angles of 90$^\circ$, the difference between this second order model and the simple Rayleigh model is less than 0.001 degrees.  The band of high DoP following the 90$^\circ$ scattering angle arc shows similar agreement in angle of linear polarization.  Regions near the neutral points show strong angular variation.  This is in agreement with the all-sky imaging polarimetry on Haleakala \cite{Swindle:2014wc, Swindle:2014ue}.

In this section we outlined a second order scattering model which included two components contributing to the polarization pattern of the sky. We showed that by choosing regions of the sky with high DoP, one can avoid several contaminations of the AoP to a small fraction of a degree as shown in Figure 5.  Choose points near the 90$^\circ$ scattering plane and away from the horizon with high airmass to avoid multiple scattering contamination when using this calibration technique.

\subsection{Planning Sky Observations for Diversity \& Efficiency}

\begin{wrapfigure}{r}{0.60\textwidth}
\centering
\vspace{-3mm}
\hbox{
\hspace{-0.5em}
\includegraphics[width=1.0\linewidth, angle=0]{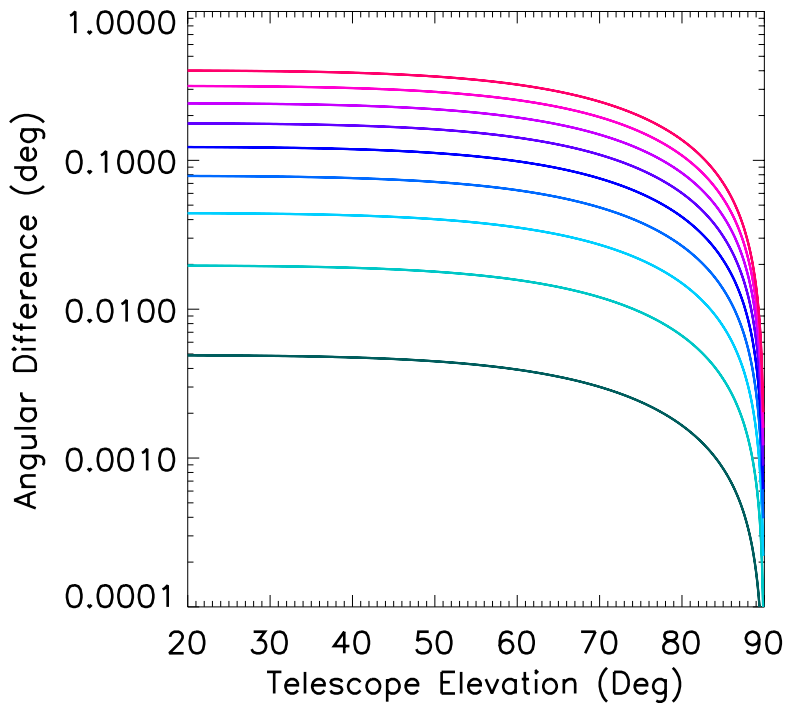}
}
\caption{ \label{sky_angular_residual_one} The AoP variation between the single scattering Rayleigh model and the multiple scattering model here considering {\it double scattering}. For this Figure, the sun was placed on the horizon at an elevation of $\alpha$=0$^\circ$.  Each curve shows a trace from horizon to the zenith (elevation 90$^\circ$ along the 90$^\circ$ scattering plane for maximum DoP. The different colors correspond to different splitting angles $\delta$ = 4 ATAN($A$) of 3$^\circ$ up to 27$^\circ$ in steps of 3$^\circ$.  As the {\it double scattered} term grows stronger and the splitting angle increases, the AoP variations between single Rayleigh scattering and this multiple scattering model increases from 0.005$^\circ$  up to and approaching 0.4$^\circ$.  However, at the 90$^\circ$ scattering location, the angular differences between single and double scattering models drops significantly.}
\vspace{-0mm}
 \end{wrapfigure}

This technique requires a diversity of input polarization angles to minimize noise propagation when deriving telescope Mueller matrices. There is an analogy between the time-dependent Rayleigh sky input polarization and the retardances chosen to create an efficient modulation scheme for polarization measurements. By choosing telescope pointings and observing times such that the solution for Mueller matrix elements is well conditioned (efficient modulation by the daytime sky), a good calibration can be derived.  Polarimeters typically produce intensity modulations by changing the incident polarization state with retardance amplitude and orientation changes. This retardance modulation translated in to varying intensities using an analyzer such as a polarizer, polarizing beam splitter or crystal blocks such as Wollaston prisms or Savart plates. These modulation schemes can vary widely for various optimizations and schemes to maximize or balance polarimetric efficiency over user-chosen Stokes parameters, wavelengths and instrumentation systems \cite{Compain:1999do, 2003isp..book.....D, delToroIniesta:2000cg, deMartino:2003cf, Nagaraju:2007tn, Tomczyk:2010wta} . There have been many implementations of achromatic and polychromatic designs in both stellar and solar communities \cite{Nelson:2010fw,Snik:2012jw, deWijn:2012dd, deWijn:2010fh,Wijn:2011wt, Gisler:2003hy, Hanaoka:2004ku, Xu:2006hk}.  In the notation of these studies, the instrument modulates the incoming polarization information into a series of measured intensities (${\bf I}_{i}$) for $i$ independent observations via the modulation matrix (${\bf O}_{ij}$) for $j$ input Stokes parameters (${\bf S}_j$):  ${\bf I}_{i} = {\bf O}_{ij} {\bf S}_{j}$. This is exactly analogous to our situation where we have changed the matrix indices to be $i$ independent Stokes parameter measurements for $j$ different sky input Stokes parameters: ${\bf S}_{i} = {\bf R}_{ij} {\bf M}_{j}$.

In most night-time polarimeters, instruments choose a modulation matrix that separates and measures individual parameters of the Stokes vector typically called a {\it Stokes Definition} modulation sequence: 

\begin{equation}
\label{normmod}
{\bf O}_{ij} =
 \left ( \begin{array}{rrrr}
 1   	& +1	&  0	&  0	\\
 1 	&  -1	&  0 	&  0	\\
 1 	&  0 	& +1	& 0	\\
 1 	&  0	& -1	&  0	\\ 
 1 	&  0	& 0	&  +1	\\ 
 1 	&  0	& 0	&  -1	\\ 
 \end{array} \right ) 
\end{equation}

%
%\begin{equation}
%{\bf I}_{i} = {\bf O}_{ij} {\bf S}_{j}
%\end{equation}	
%
%\begin{equation}
%{\bf S}_{i} = {\bf R}_{ij} {\bf M}_{j}
%\end{equation}	
%

Other instruments choose a wide range of modulation schemes to balance the efficiencies over a number of exposures. One recovers the input Stokes vector from a series of intensity measurements by inverting the modulation matrix (${\bf O}$) via the normal least squares formalism:  ${\bf S} = \frac{ {\bf O}^T {\bf I} } { {\bf O}^T {\bf O}}$.  The demodulation matrix is typically defined as: ${\bf D}_{ij} = [{\bf O}^T {\bf O}]^{-1} {\bf O}^T$. 

%
%\begin{equation}
%\label{eqn_demod}
%{\bf S} = \frac{ {\bf O}^T {\bf I} } { {\bf O}^T {\bf O}}
%\end{equation}
%
%\begin{equation}
%{\bf D}_{ij} = [{\bf O}^T {\bf O}]^{-1} {\bf O}^T 
%\end{equation}

In our daytime sky technique, the Rayleigh sky input parameters become the modulation matrix (${\bf O}_{ij} = {\bf R}_{ij}$) and the formalism for noise propagation developed in many studies such as \cite{2003isp..book.....D, delToroIniesta:2000cg} apply.  If each measurement has the same statistical noise level $\sigma$ and there are $n$ total measurements then the noise on each demodulated parameter (${\bf \sigma}_i$) becomes: ${\bf \sigma}_{i}^2 = n \sigma^2 \sum\limits_{j=1}^{n} {\bf D}_{ij}^2$. The efficiency of the observation becomes: ${\bf e}_{i} = \left ( n  \sum\limits_{j=1}^{n} {\bf D}_{ij}^2 \right ) ^{-\frac{1}{2}}$.
%
%
%\begin{equation}
%\label{eqn_sigmas}
%{\bf \sigma}_{i}^2 = n \sigma^2 \sum\limits_{j=1}^{n} {\bf D}_{ij}^2
%\end{equation}
%
%\begin{equation}
%\label{eqn_effics}
%{\bf e}_{i} = \left ( n  \sum\limits_{j=1}^{n} {\bf D}_{ij}^2 \right ) ^{-\frac{1}{2}}
%\end{equation}

One must take care with this technique to build up observations over a wide range of solar locations so that the inversion is well conditioned as outlined in the appendices of \cite{Harrington:2011fz}. The path of the sun throughout the day will create regions of little input sky Stokes vector rotation causing a poorly constrained inversion with high condition number. For instance, at our location in the tropics the sun rises and sets without changing azimuth until it rises quite high in the sky. We are constrained to observing in early morning and late evening with the dome walls raised since we may not expose the telescope to the sun. This causes input vectors at east-west pointings to be mostly $q$ oriented with little rotation over many hours. Observations at other times of the year or at higher solar elevations are required to have a well conditioned inversion.  One can easily build up the expected sky input polarizations at a given observing site with the Rayleigh sky polarization equations. Then it is straightforward to determine the modulation matrix and noise propagation for a planned observing sequence to ensure a well-measured telescope matrix with good signal-to-noise.

\section{The HiVIS Daytime Sky Observing Campaign}

From October 2014 to May of 2015 we collected a large data set of daytime sky observations with HiVIS using the new liquid crystal charge-shuffling configuration \cite{Harrington:2015dl}. We obtained over 1700 measurements in our standard setup with 17 spectral orders and 4000 pixels per order. The daytime sky was observed in a grid of telescope pointings (azimuth elevation combinations).  

\begin{wrapfigure}{r}{0.65\textwidth}
\centering
\vspace{-5mm}
\hbox{
\hspace{-0.8em}
\includegraphics[width=1.\linewidth, angle=0]{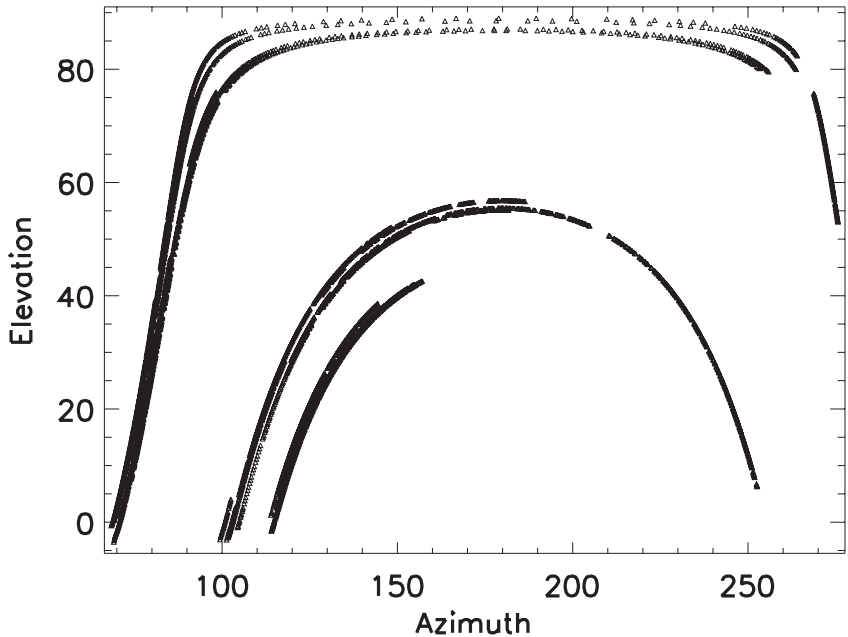}
}
\caption{ \label{daysky_solar_azel} The computed solar azimuth and elevation for all HiVIS daytime sky observations. May data had the sun near the zenith while winter observations (October, December) had the sun low and in the south. The observing allocations were dominated by sunrise to noon times giving far more observations with the sun east and overhead. }
 \end{wrapfigure}

The first subset of telescope pointings were chosen starting North-South with a 60$^\circ$ spacing for azimuths of:  [060, 120, 180, 240, 300, 360] and elevations of: [10, 25, 50, 75].  The second subset of telescope pointings were chosen starting East-West with azimuths of: [030, 090, 150, 210, 270, 330] and elevations of [20, 35, 60, 89].  See \cite{Harrington:2015dl} for a schematic and optical layout of HiVIS. The solar azimuth and elevation is shown in Figure \ref{daysky_solar_azel} for all observations. The sun was low in the south for October to December 2014, while nearly passing through the Zenith in May of 2015. These telescope pointings were used during daytime sky observations over several days:  October 19, 24, 25, 29, 30.  December 01, 11, 14, 15, May 9, 10, 11, 16, 17, 18 for a total of 15 days spread over 7 months.

\subsection{HiVIS Data Extraction}

	As part of routine calibration, modulation matrix elements were derived using our polarization calibration unit \cite{Harrington:2015dl}.  This unit is a wire grid polarizer and a Bolder Vision Optik achromatic quarter wave plate on computer-controlled rotation-translation stages. This polarization state generator unit is mounted immediately in front of the HiVIS slit and dichroics slit window. An alignment procedure was done during initial installation to find the stepper motor rotation positions where the polarizing axis of the polarizer and the fast axis of the quarter-wave retarder are aligned with the Savart plate at nominal wavelengths. By using a standard sequence of polarizer and quarter wave plate retarder orientations, 6 pure Stokes inputs ($\pm q$, $\pm u$, $\pm v$) are used to derive redundant calibration sets.

\vspace{5mm}

\begin{figure}  [ht]
\begin{center}
\includegraphics[width=0.97\linewidth, angle=0]{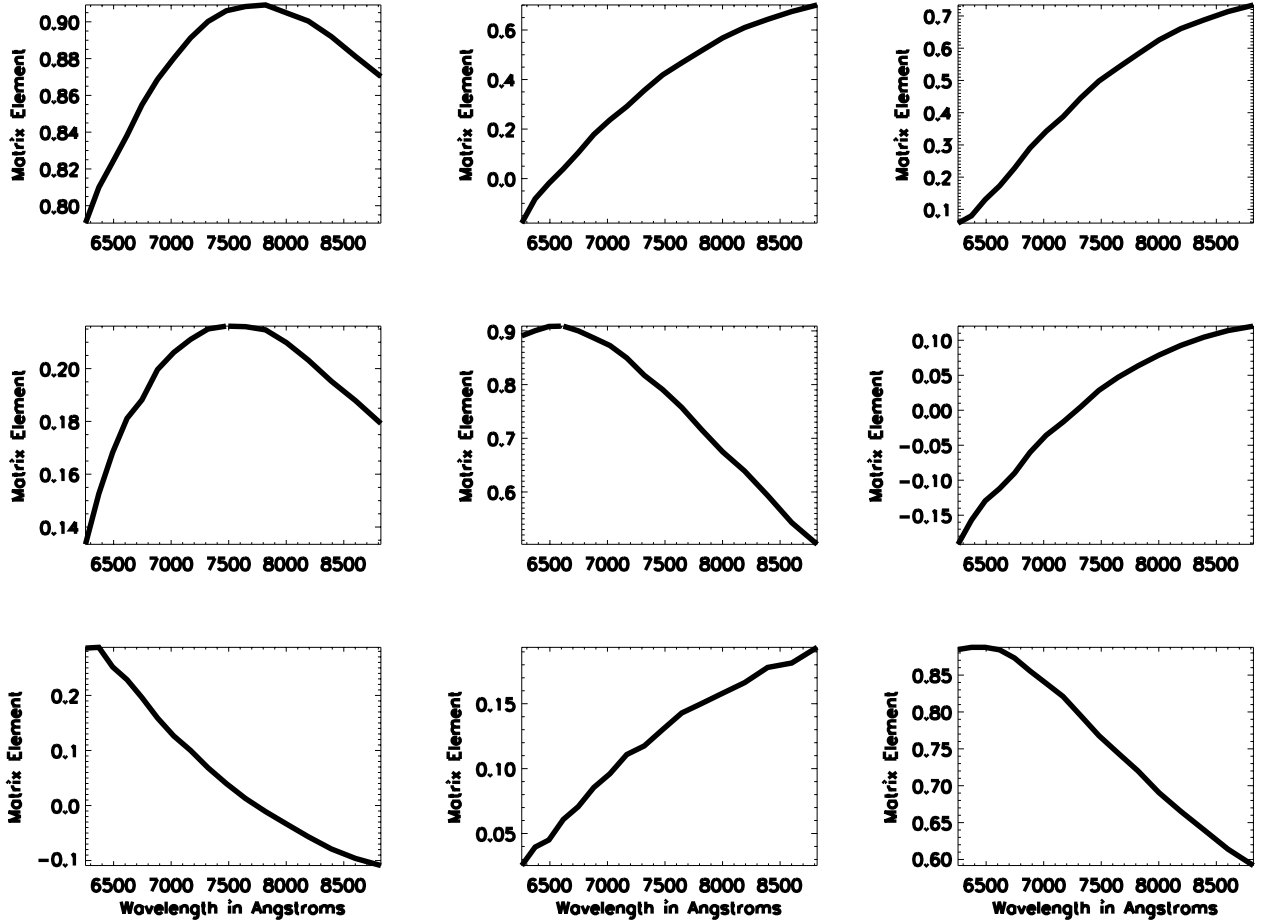}
\caption{ \label{epsaru_modmat_order_median_all} The $quv$ to $quv$ modulation matrix elements for the standard {\it Stokes definition} sequence liquid crystal voltages derived using the Full-Stokes Injection Unit (a polarization state generator) in front of the HiVIS spectrograph slit. The unit is a wire grid polarizer and quarter wave retarder on independently controlled rotation stages for creating known $quv$ inputs.   The matrices shown here use only one ($+$)  of the two polarization calibration unit input Stokes parameter sets. Wavelengths span the $\sim$6300\AA to 8800\AA range. For clarity, only the median value for each spectral order is shown (4000 spectral pixels per order). The liquid crystals were roughly tuned for a standard {\it Stokes definition} modulation set around 7000\AA.  The diagonal elements are roughly 0.9 at these wavelengths. The non-diagonal elements are all non-zero and some have amplitudes above 0.7 within the observed wavelength range. }
\end{center}
\end{figure}

	There is cross-talk in the quarter wave retarder that can be compensated by additional fitting techniques, but this effect is also removed by using the daytime sky calibrations \cite{Harrington:2015dl}.  By demodulating the polarization state generator calibration data at the slit, we decouple the spectrograph polarization response from the telescope.  The average system modulation matrix is shown in Figure \ref{epsaru_modmat_order_median_all} as the average of all October 2013, December 2013 and May 2014 modulation matrices. There is little variation in the derived modulation matrix within the main observing periods of October, December or May. For clarity, only the median modulation matrix value for each spectral order is shown. Calibrations are derived by doing spectral averaging (binning) to 50 spectral pixels per order. The variation in individual modulation matrix elements is small ($<$0.05). To remove any effect by varying system modulation matrix elements, we used calibration observations taken for each major observing season. Typically within each run, full calibration sequences were taken daily with little change shown over timescales of days to 2 weeks.

	These modulation matrices are used to demodulate the dual-beam charge shuffled measurements (3 exposures, 12 intensity spectra) into individual $quv$ measurements. The algorithm for computing the telescope cross-talk elements assumes that the measurements are projected on to the Poincar\'{e} sphere. Each individual demodulated spectrum is divided by the measured degree of polarization in order to create scaled Stokes vectors with 100\% degree of polarization.

\section{Filtering the Data For High Accuracy Calibrations}

	There are several sources of error present when using daytime sky measurements for computing telescope Mueller matrices.  In this section, we outline techniques to reject observations. 	

	One limitation is that single-scattering Rayleigh sky model is only an approximation. In areas of the sky with low degree of polarization (DoP), the computed angle of polarization can vary substantially. This fact immediately suggests removing  data points with low measured DoP as well as to avoid using the low DoP region of the sky for this technique. Cirrus clouds have been shown to rotate the AoP and also cause strong departures from the Rayeigh single scattering model.  Cirrus clouds are know to decrease the measured degree of polarization in addition to rotating the polarization by a large angle.  On Haleakala, occasional small patches of low-laying cumulus can blow over the telescope aperture a few hundred feet above the ground.  If a patch of cloud depolarizes a single exposure of a data set, strong deviations from the Rayleigh sky model can be seen. Figure 8 shows the measured DoP and our scattering angle coverage for this observing campaign.

We experienced and must compensate for several types of errors:
	
\begin{itemize}
\item Rayleigh sky model is inaccurate in low DoP regions
\item Cirrus clouds rotate AoP strongly.
\item Optical window uncertainty (BK7 / Infrasil encoder failure \& motor replacement)
\item Operators manually point telescope to wrong pointing (no computer feedback).
\item Cumulus blow-by in single exposures (often on Haleakala marginal inversion layer)
\end{itemize}

	The measured DoP for each data set is shown in Figure \ref{daysky_solar_angsep_and_degpol_vs_exposure}.  On three separate days there were thick cirrus clouds that impacted the measured DoP seen in measurements 200-300, 400-500 and around 800.  Light cirrus were present on May 17 and 18.  Low-lying cumulus blowing over from the Haleakala crater were possible in certain October and December days. There are several ways to identify and filter this large data set in order to select quality data.

\begin{figure}[htb]
\begin{center}
\hbox{
\hspace{0.0em}
\includegraphics[width=0.47\linewidth, angle=0]{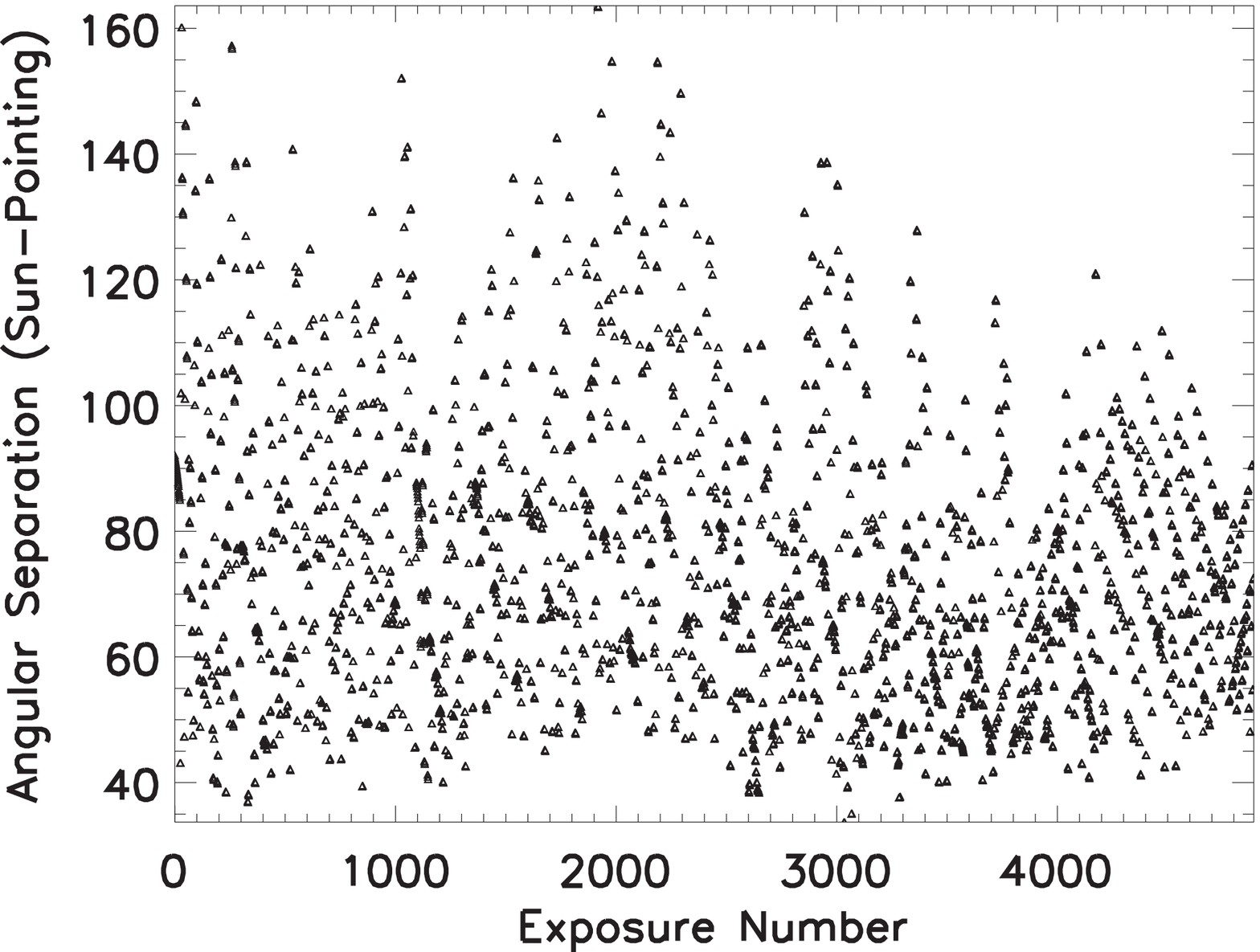}
\hspace{4mm}
\includegraphics[width=0.47\linewidth, angle=0]{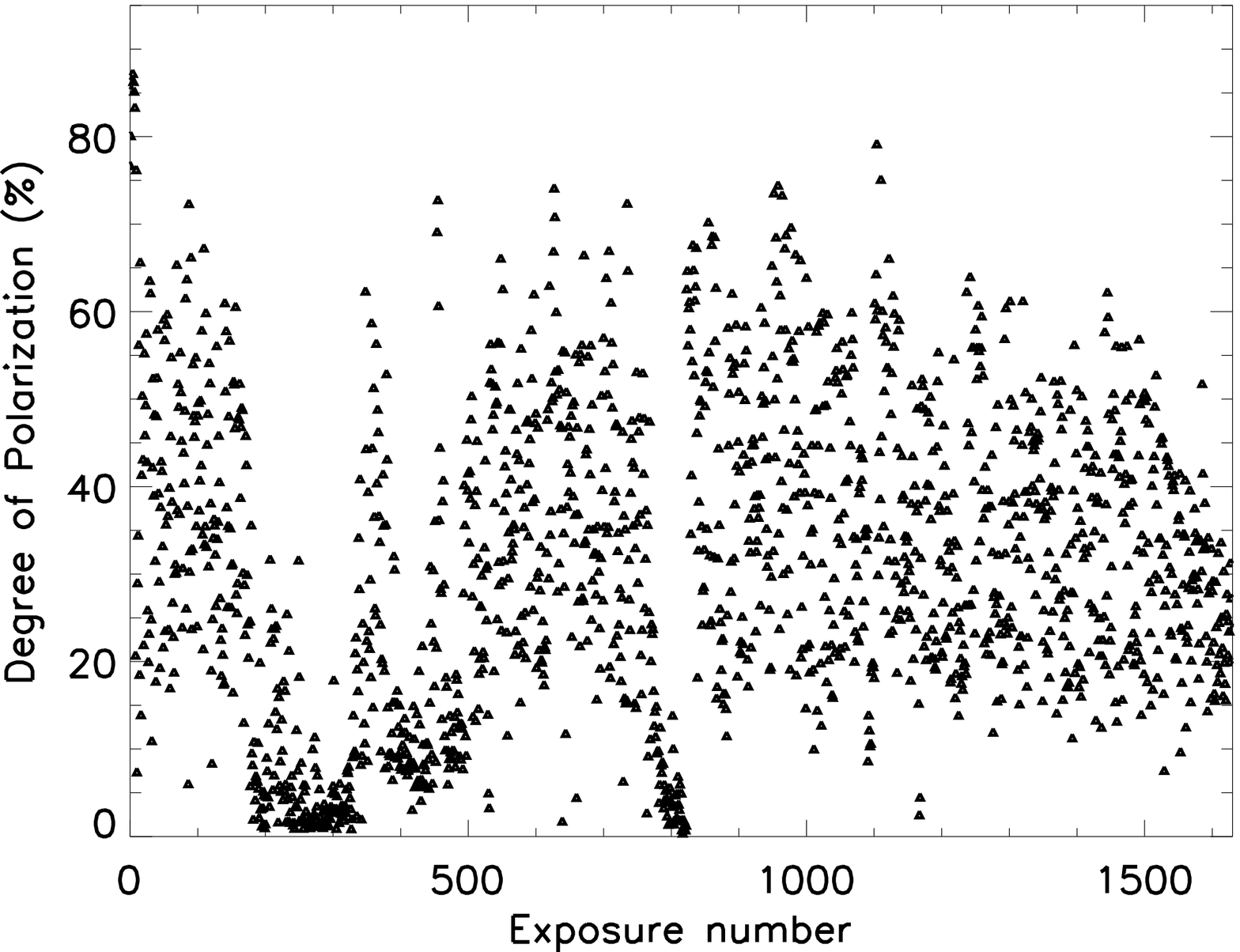}
}
\caption{ \label{daysky_solar_angsep_and_degpol_vs_exposure}  The computed angular separation ($\gamma$) between the telescope pointing and the sun for all 4890 exposures (1630 full-stokes polarization measurements) shown on the left. The measured daytime sky degree of polarization for each exposure set is shown on the right. Clouds, pointing, atmospheric properties and time are all variables.  Note that these measured DoP values are used to scale each measured Stokes vector to 100\% DoP for use in our calibration algorithm.  }
\end{center}
\end{figure}

\subsection{Filter: Measured DoP Threshold}

	A simple data filter which improves calibration quality is to discard observations showing low measured degree of polarization. Low detected polarization is often an indicator of either bad atmospheric conditions or issues with the data. This calibration technique requires knowledge of the angle of polarization (AoP) with reasonable precision to keep noise amplification low.

\begin{wrapfigure}{r}{0.49\textwidth}
\centering
\vspace{-7mm}
\hbox{
\hspace{-0.5em}
\includegraphics[width=0.80\linewidth, angle=90]{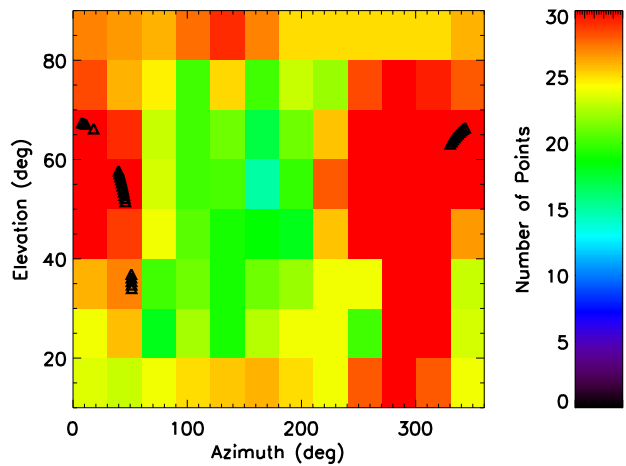}
}
\caption{ \label{number_obs_as_computed} The color coding shows the number of daytime sky polarization observations at each telescope pointing available for the telescope Mueller matrix calculations. The filtering by demodulated degree of polarization (DoP$>$15\%) is shown but no iterative filter (convergence) has been applied. Interpolation (linear) between neighbors on the az-el grid has been applied for clarity. The small black triangles show the altitude-azimuth track for a star, $\epsilon$ Aurigae that we observed in 2015 to illustrate a typical altitude-azimuth combination required for calibration.}
\vspace{-8mm}
 \end{wrapfigure}

	At low DoP values, the AoP uncertainties grow substantially. Figure \ref{number_obs_as_computed} shows the number of daytime sky observations we have in the data set after several filtering processes are applied. The grid of (azimuth,elevation) observation points was linearly interpolated to a continuous map over all observable (azimuth,elevation) optical geometries. The small black triangles show the position of a stellar target ($\epsilon$ Aurigae) we use for calibration purposes. These triangles show a typical azimuth, elevation track for a target marking each individual data set where independent calibrations are required.

\subsection{Filter: Rayleigh DoP Agreement Threshold}

	Several geometrical calculations are required for assessing and filtering data using the Rayleigh sky model. Figure \ref{daysky_solar_angsep_and_degpol_vs_exposure} shows the angular separation between each measurement pointing and the computed solar location. Note, there are 3 exposures per complete full Stokes measurement set so there are only 4890 / 3 = 1630 unique data sets. From the pointing and solar geometry, we derive the input Rayleigh-sky stokes parameters. One way of checking the agreement of the HiVIS data is to compute the Rayleigh sky parameters from the HiVIS measurements at all pointings. We can rearrange the Rayleigh sky polarization equation to give the calculated maximum degree of polarization ($\delta_{max}$) from the HiVIS measured degree of polarization ($\delta$) and the scattering angle ($\gamma$): $\delta_{max}=\delta \frac{1+cos^2\gamma}{sin^2\gamma}$.

\begin{figure} 
\begin{center}
\hbox{
\hspace{-1.5em}
\includegraphics[width=0.39\linewidth, angle=90]{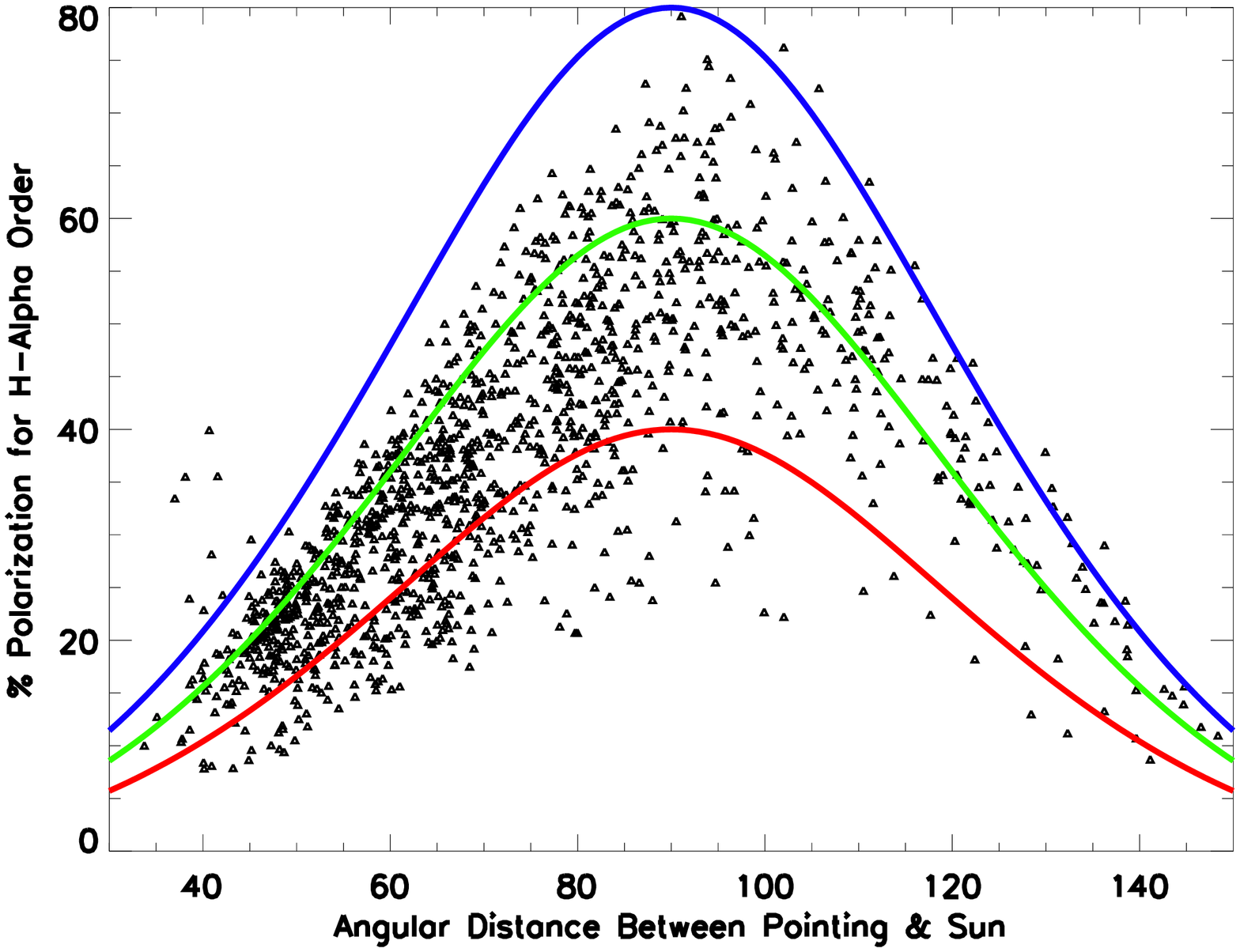}
\hspace{-2.5em}
\includegraphics[width=0.40\linewidth, angle=90]{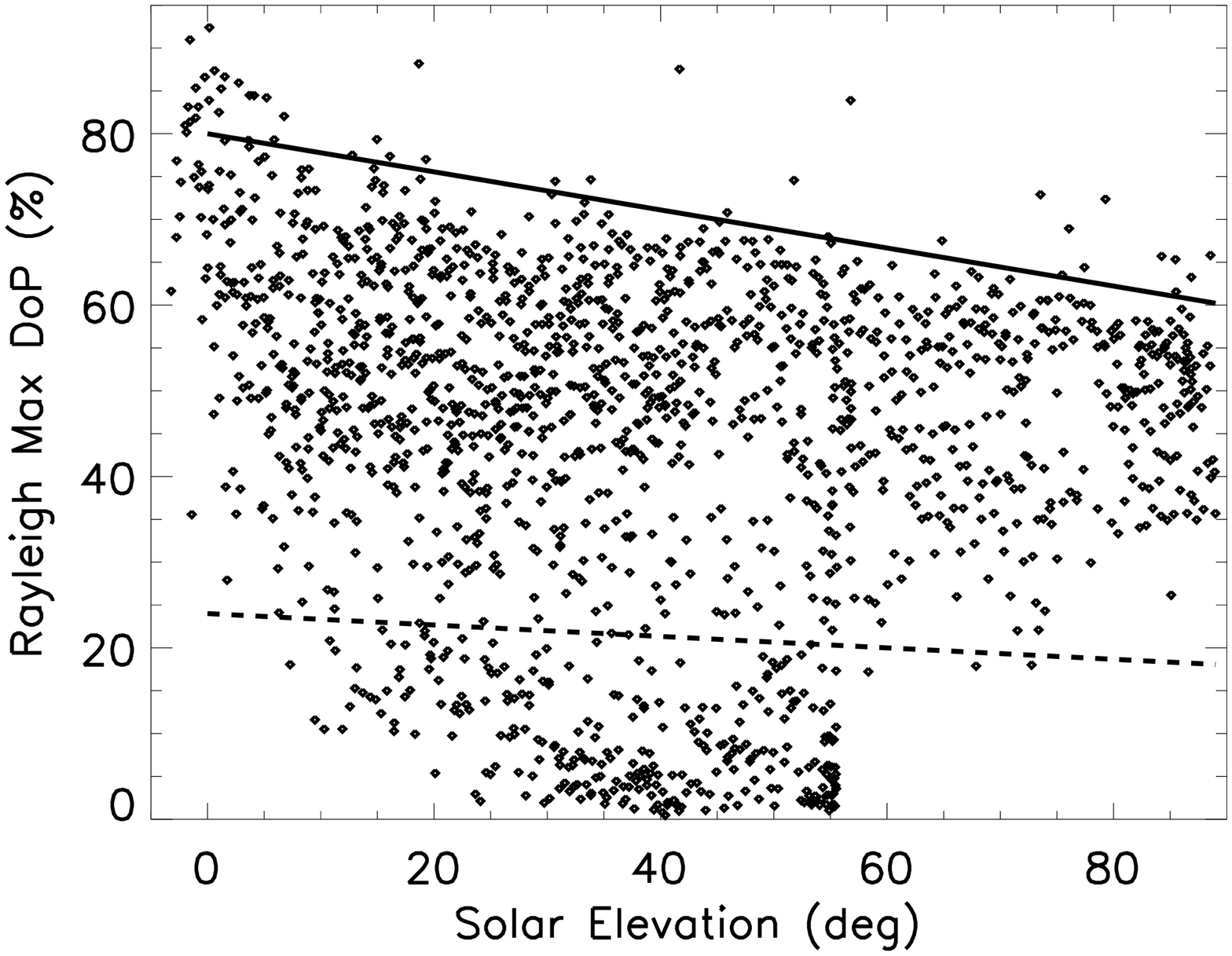}
}
\caption{ \label{degpol_vs_soldis} The left panel shows the measured sky degree of polarization for HiVIS daytime sky observations as a function of angular distance from the sun ($\gamma$). Only points passing a 10\% DoP filter and a 30\% $\delta_{max}$ data filter are included. Colored curves show the Rayleigh sky polarization $\delta$ as function of scattering angle   for $\delta{max}$ amplitudes of 40\%, 60\% and 80\%.  The right hand panel shows the estimated maximum atmospheric degree of polarization ($\delta_{max}$) computed directly from the demodulated HiVIS exposures. The summer observing had a maximum solar elevation of 87$^\circ$ while the winter observing season had a maximum solar elevation of around 55$^\circ$.  As we observed in the afternoons more often (time allocation constraints), and had more time with the sun well above the horizon, there are less data points at angular distances larger than 90$^\circ$.}
\end{center}
\vspace{-7mm}
\end{figure}

	From this equation we can use the data to calculate a measure of atmospheric conditions ($\delta_{max}$), and we can create a data filter to reject HiVIS observations on hazy days with low $\delta_{max}$. Figure \ref{degpol_vs_soldis} shows the computed $\delta_{max}$ as a function of scattering angle derived from the data set. The different color curves show 40\%, 60\% and 80\% $\delta_{max}$ scalings.  The $\delta_{max}$ functions are reasonably constrained by all-sky polarimeter measurements and MODTRAN models \cite{Swindle:2014wc,Swindle:2014ue}. A simple function for the maximum sky degree of polarization $\delta_{max}$ on a clear day is used following Mauna Loa measurements: $\delta_{max} = 80^\circ - 20^\circ \times \frac{90^\circ} {Solar Altitude}$ \cite{Dahlberg:2011wk,Dahlberg:2009jh}.

	By taking this simple relation, a set of data filters can be created. Data points with low $\delta_{max}$ predictions can be rejected as likely influenced by clouds, haze, multiple scattering and other effects. The typical Rayleigh sky dependence on solar elevation is scaled down by 30\% and calculated for every data set to show a minimum acceptable degree of polarization for each measurement.  Figure \ref{degpol_vs_soldis} shows the $\delta_{max}$ values computed from the HiVS measurements. There are a large number of points showing a high maximum degree of sky polarization ($\delta_{max}$) as expected for this high, dry observing site like Haleakala \cite{Dahlberg:2011wk,Dahlberg:2009jh}. There are clusters of points at low DoP values that correspond to days with cirrus clouds. These points are rejected by data filters. 
	
	The measured DoP roughly follows the expected Rayleigh patterns.  The polarization is higher at scattering angles approaching 90$^\circ$ and the predicted maximum sky degree of polarization ($\delta_{max}$) matches Mauna Loa measurements on cloud-free days \cite{Dahlberg:2011wk,Dahlberg:2009jh}.  This rule is only approximate as daytime sky polarization is modified near sunrise and sunset as well as by varying solar elevation.

\subsection{Data Filtering Summary}

\begin{figure} 
\begin{center}
\hbox{
\includegraphics[width=0.44\linewidth, angle=90]{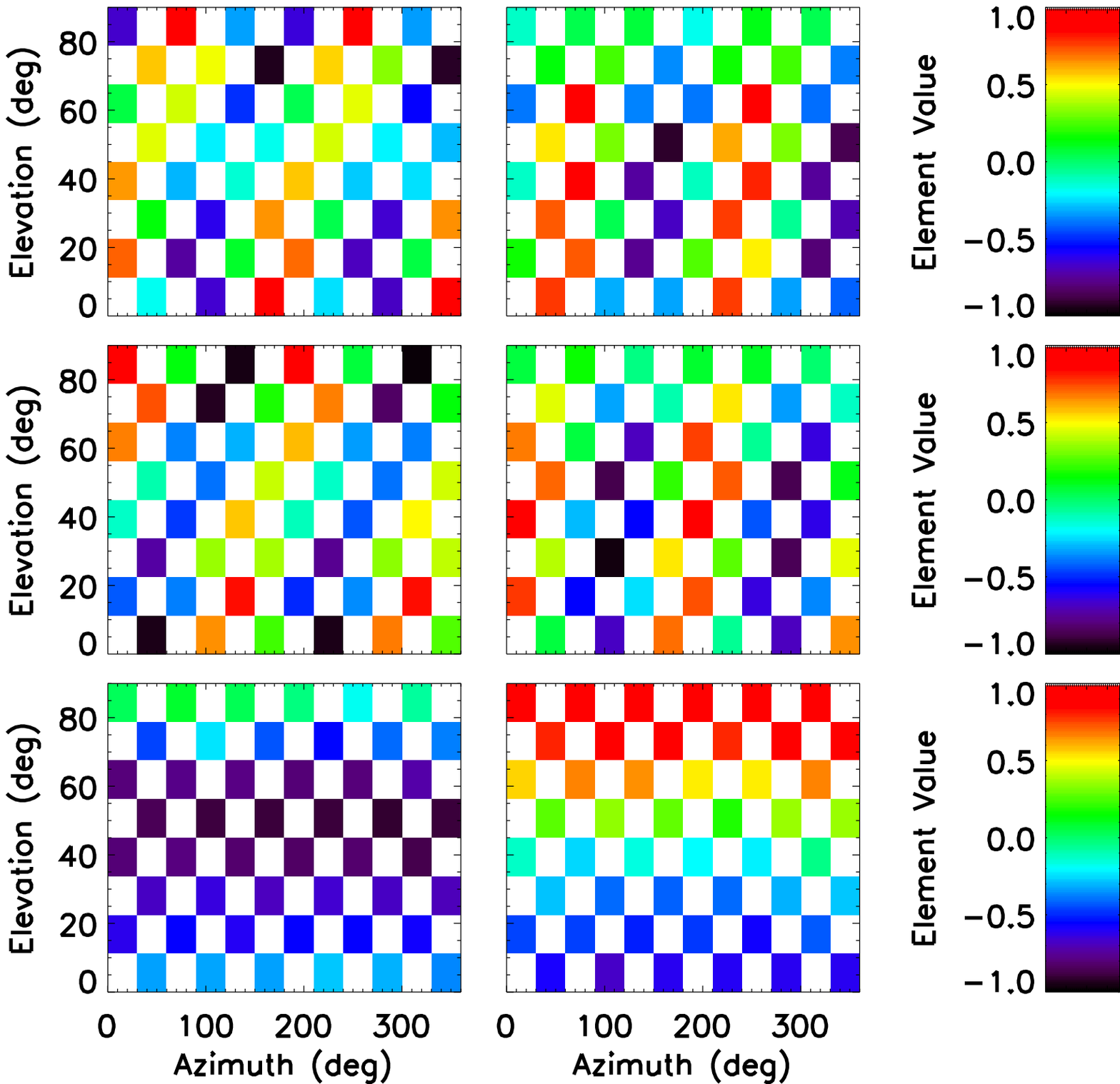}
\hspace{9mm}
\includegraphics[width=0.44\linewidth, angle=90]{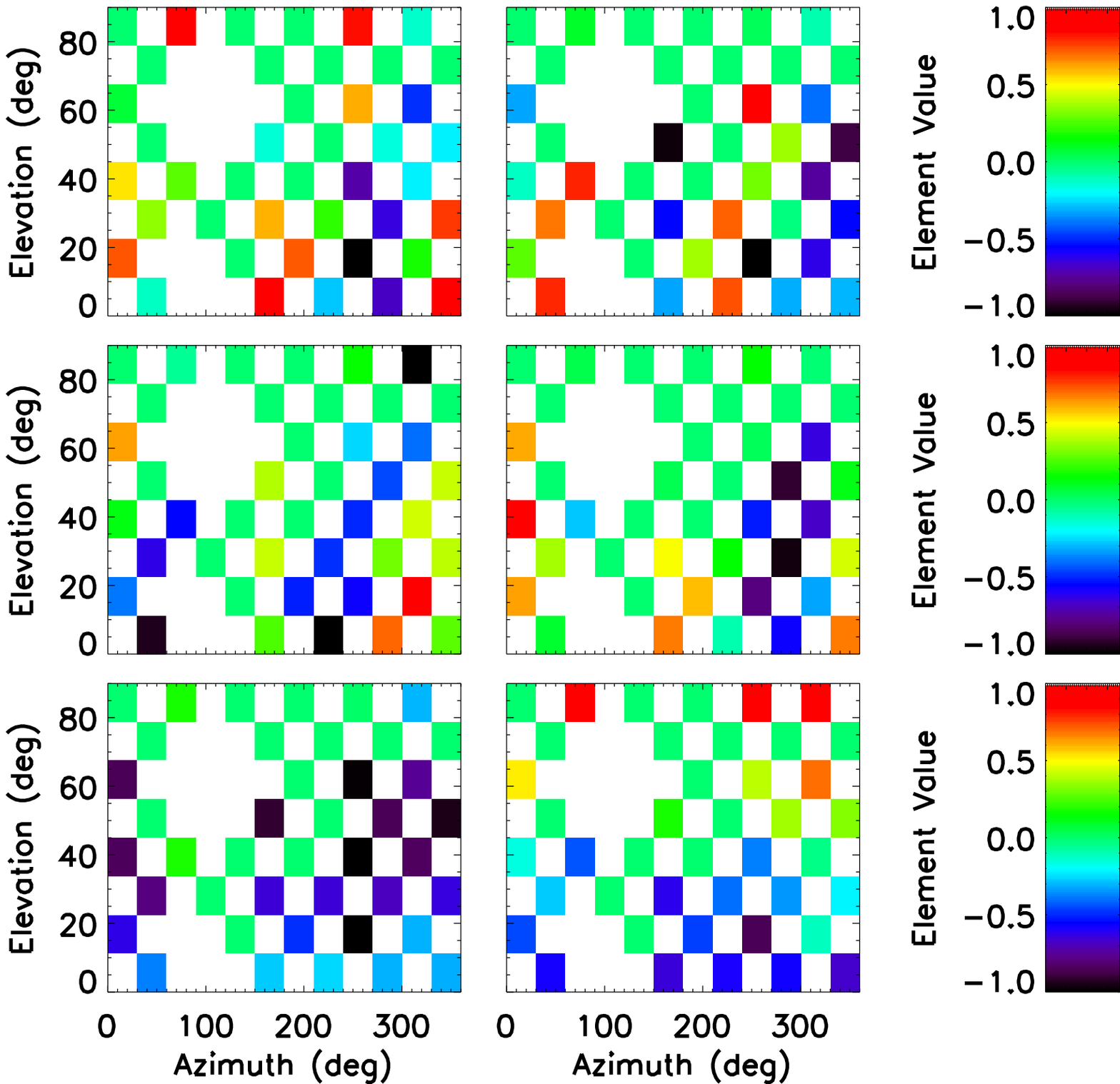}
}
\vspace{10mm}
\hbox{
\includegraphics[width=0.44\linewidth, angle=90]{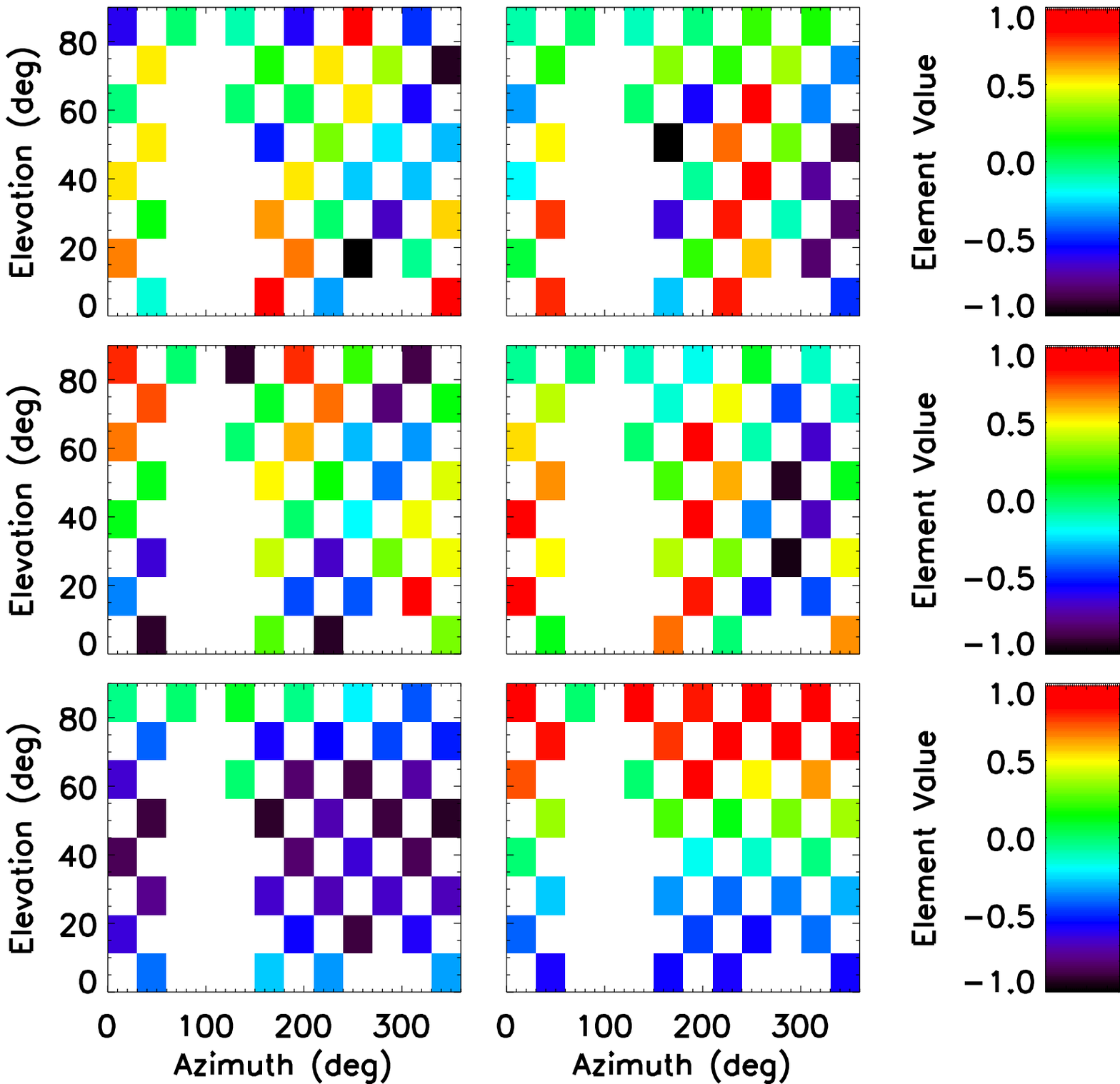}
\hspace{9mm}
\includegraphics[width=0.44\linewidth, angle=90]{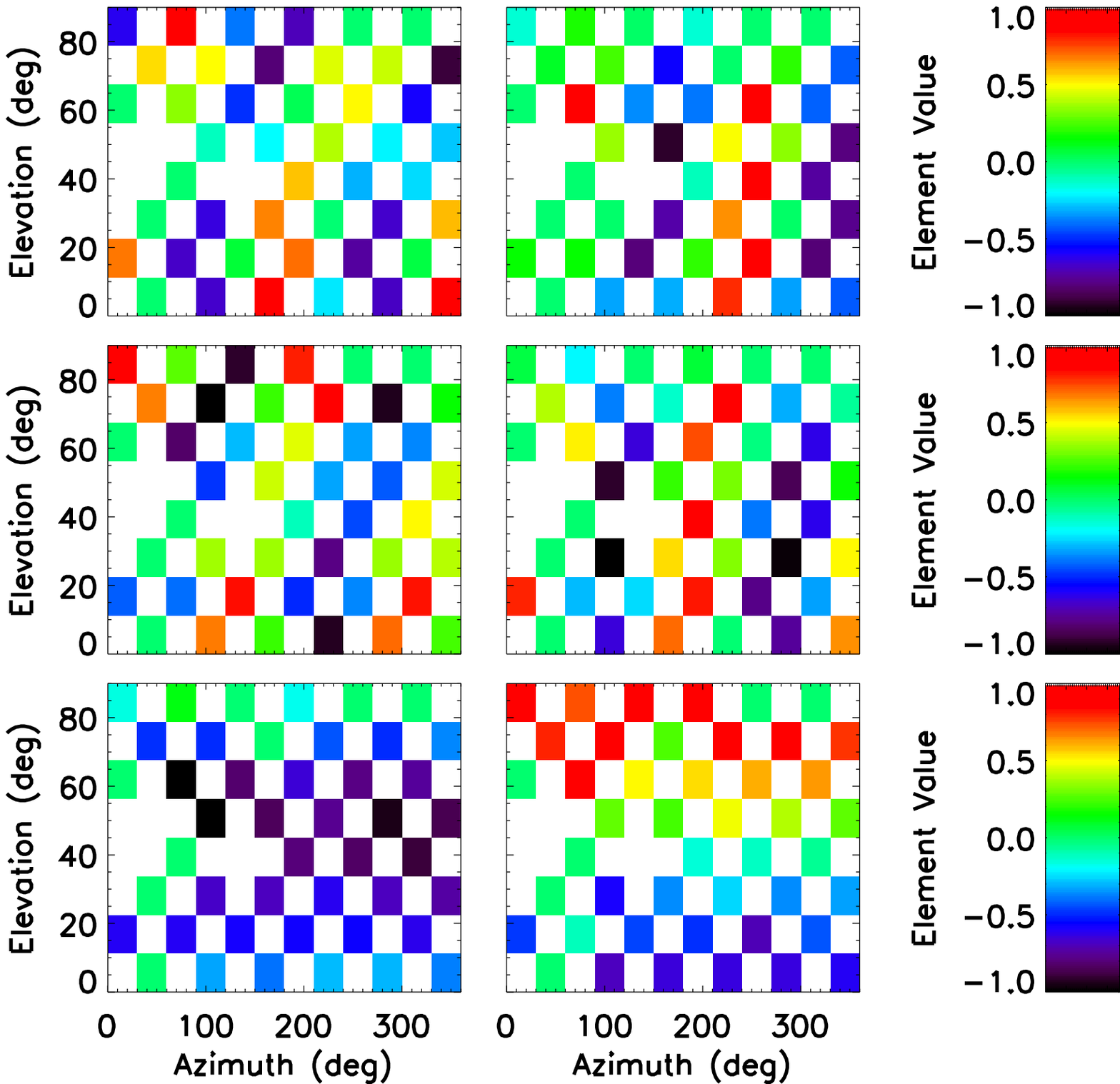}
}
\vspace{5mm}
\caption{ \label{mueller_matrix_estimates_raw} The six Mueller matrix element estimates computed at all azimuths and elevations with sufficient numbers of valid measurements. The grid of points corresponds to the (azimuth, elevation) sampling used for this study.  Only data with valid observations are shown, giving rise to the grid pattern.  Each of the four panels shows estimates for $QQ$ and $UQ$ on top, $QU$ and $UU$ in the middle, $QV$ and $UV$ on the bottom.  The top left group shows all data from October to May with a minimum 15\% DoP measured as well as a minimum of 5 observations per telescope pointing requirement.  The top right group shows the October-only set with a 15\% minimum DoP filter and a minimum of 3 observations per pointing. The bottom left group shows the December data with a 20\% minimum DoP filter and a minimum of 3 observations per pointing.   The bottom right group shows the May-only set with a 30\% minimum DoP filter and a minimum of 5 observations per pointing.  A close inspection of some points near elevations of 90$^\circ$ shows some disagreement between the October and both the December set and May sets. In all the single-month data sets October, December and May data sets, some telescope pointings at elevations of $\sim$40$^\circ$ and azimuths $\sim$90$^\circ$ are not computed due to a lack of sufficient number of data sets.  A calibration campaign must balance the requirements for efficiency, speed, sufficient pointing coverage and having a large enough data set to reliably identify and reject contaminated outliers.}
\end{center}
\end{figure}

We use several methods for ensuring data integrity and solution consistency.  First, we require a signal-to-noise threshold for every spectrum at a nominal wavelength.  Second, we require a minimum number of observations at each (Azimuth, Elevation) combination on the telescope pointing grid.  Third, we reject observations with a low measured DoP after demodulation. Fourth, we reject observations where the computed sky polarization as estimated by the projected maximum degree of polarization ($\delta_{max}$) suggests haze, cloud or other data contamination.  Fifth, we reject (Azimuth, Elevation) grid points with low AoP diversity to ensure a well conditioned solution (fit) to each Mueller matrix element estimate. Sixth, an iteration is done to ensure that the remaining observations give consistent Mueller matrix estimates. The angular distance between observations and Rayleigh model is preserved for a system that is not depolarizing. Thus, data sets showing inconsistent angles between the bulk of the observations are rejected.   Seventh, an iterative process is followed to ensure the Rotation matrix fits give consistently calibrated measurements. Observations with a residual angle between the calibrated observations and the Rayleigh sky model above a threshold are rejected. 

\begin{itemize}
\item Reject observations by measured SNR ($>$500 for all $quv$, spectral order 3 after binned 80x to sampling of 50 spectral pixels per order)
\item Require several observations to compute the six Mueller matrix element estimates ($>$4 points optimal)
\item Reject observations by measured DoP ($>$15\% after demodulation)
\item Reject low $\delta_{max}$ points for agreement with Rayleigh model (40\% typical $\delta_{max}$ for a clear day)
\item Reject pointings with low $qu$ input diversity  ($>$ 20$^\circ$ for each pointing)
\item Reject observations where the calibrated Mueller matrix calibrations show high error outliers (\textbf{$S_{cal}$} $\cdot$ \textbf{M} Angle $<$ 0.1)
\item Reject iteratively by calibrated residual angle btw measurements and theory ( \textbf{$S_{cal}$} $\cdot$ \textbf{R} ) until convergence below a threshold angle (e.g. 25$^\circ$) for consistency
\end{itemize}

\begin{figure}[ht]
\begin{center}
\hbox{
\hspace{0.0em}
\includegraphics[width=0.47\linewidth, angle=0]{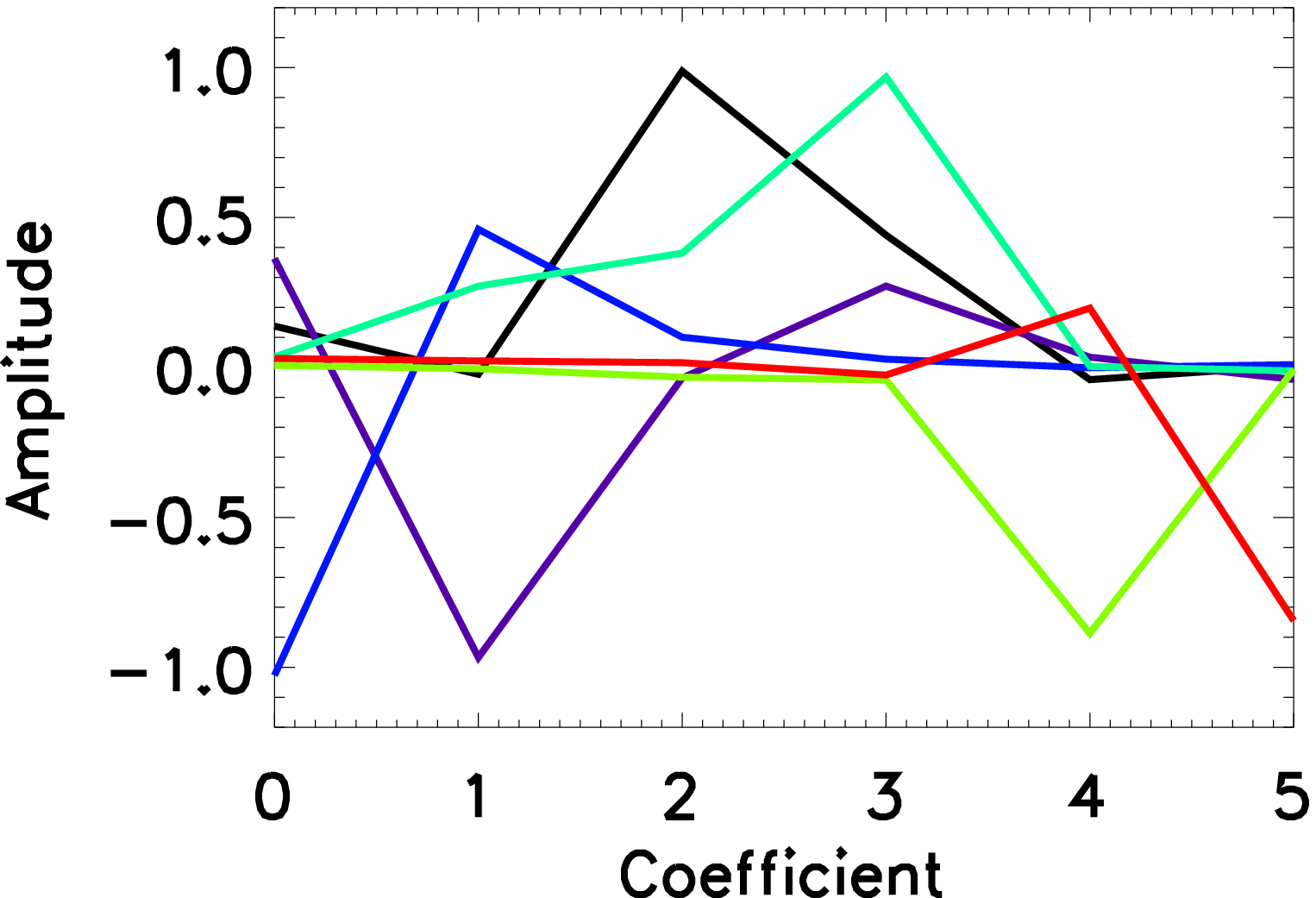}
\hspace{4mm}
\includegraphics[width=0.47\linewidth, angle=0]{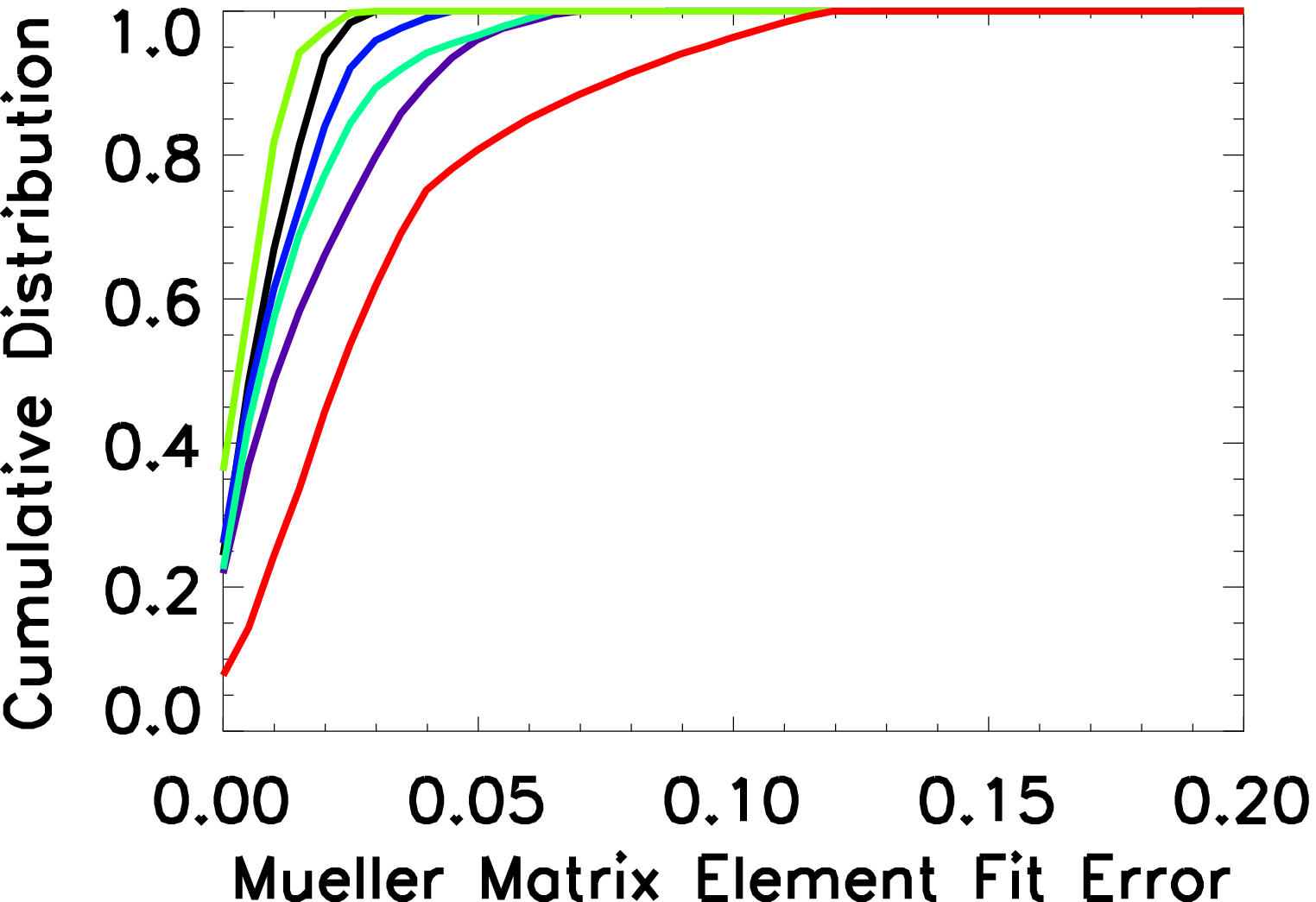}
}
\vspace{2mm}
\caption{ \label{daysky_solar_angsep_and_degpol_vs_exposure}  The left panel shows the 6-term trigonometric function coefficients fit to the rotation matrices derived from all six Mueller matrix estimate terms for spectral orders 5 at a wavelength of 6881{\AA}. Each color corresponds to a different Mueller matrix element.  For instance, black shows the QQ term which is dominated by coefficient 2 and 3 giving a functional form of $C_{2A}C_{2E}+S_{2A}S_{2E}$.  Demodulation was done using the $+$ inputs and a minimum 15\% measured DoP filter have been applied. The iterative consistency and Rayleigh minimum DoP maximum filters have also been applied.   The right hand panel shows the cumulative distribution function (CDF) for the errors between trig-based Mueller matrix estimates and the corresponding rotation matrix fits for spectral order 3.  Each color shows one of the 6 Mueller matrix estimate residual CDFs. The 15\% DoP and consistency-filters have been applied.  }
\vspace{-5mm}
\end{center}
\end{figure}

As examples of some of these filters, Figure \ref{mueller_matrix_estimates_raw} shows three different sets of Mueller matrix element estimates. We show the element estimates on the (azimuth,elevation) grid for 1 - data from all seasons filtered as in the above list, 2 - data from only October but with a minimum of 3 points per (azimuth,elevation) grid point and 3 - data from May with a minimum measured DoP of 30\%.

As the filters reject observations based on season, DoP, diversity or other minimum thresholds, some (azimuth,elevation) grid points become excluded.  Inspection of Figure \ref{mueller_matrix_estimates_raw} also shows that there are some points where there is seasonal disagreement.  The May and October data sets disagree at elevations of 89$^\circ$. 

We find that calibrations are consistent when the data filters are set to the parameters above.  Knowing in advance the impact of several atmospheric factors we outlined here can help when planning a calibration observing campaign.  To know how many telescope pointings must be observed, we now assess the functional dependence and errors when interpolating the telescope model to intermediate pointings.

\section{Mueller Matrices with Azimuth, Elevation and Wavelength}

Once the telescope Mueller matrix elements have been estimated on a grid of azimuth, elevation points, the full telescope Mueller matrix must be interpolated to every possible azimuth, elevation combination for every target that requires calibration.  As seen in \cite{Harrington:2015dl}, the Mueller matrix is smooth trigonometric functions of azimuth and elevation. This is caused by the fold mirror axes crossing in the f/200 coud\'{e} path. 

\begin{wrapfigure}{r}{0.56\textwidth}
\centering
\vspace{-0mm}
\hbox{
\hspace{-0.4em}
\includegraphics[width=0.95\linewidth, angle=0]{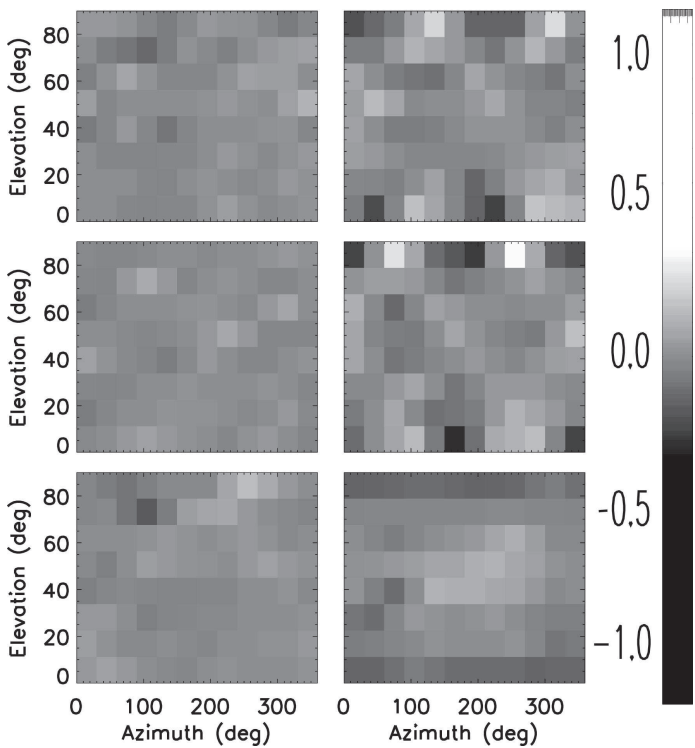}
}
\caption{ \label{sincos_fitting_minus_mm_estimates_reinterpolated} The difference between the trigonometric function fit rotation matrix elements and each of the corresponding Mueller matrix element estimates derived directly from the data.  The grey scale has been highly stretched to highlight differences at $\pm$0.1 amplitudes. The differences were computed only where valid observations are recorded after all the filters are applied. Linear interpolation was performed to all other pointings to make smooth maps. The 15\% minimum measured DoP filter and iterative consistency-filters have been applied.  These represent the disagreement between the empirical Mueller matrix estimates and the trigonometric functions.}
\vspace{-0mm}
 \end{wrapfigure}

The Mueller matrix elements can be modeled as sin and cos functions with a range of possible forms.  As a simple test, we included fits to functions of azimuth and elevation with the forms: $SIN(2Az+2El), SIN(Az+El),  SIN(2El), SIN(Az),  SIN(El), Constant$. 

We will show in later sections that these azimuth-elevation trigonometric functions are a natural consequence of the optical design and are the only terms required to fit the Zemax-predicted Mueller matrix elements from the design. 

These equations can be expanded using trigonometric identities to include 13 possible terms.  Since the domain of the fit is restricted to azimuths of 0$^\circ$ to 360$^\circ$ and elevations of only 0 to 90, care must be taken about the uniqueness of the fit parameters given various combinations of the functions. After testing several functional forms, we found that only the SIN(2Az+2El) and SIN(2Az) terms had significant amplitude. These functional forms are fit to the Mueller matrix element estimates as well as the rotation matrix fits to the estimates. Typically only very small differences between the individual Mueller matrix estimates and their corresponding trigonometric function fits are seen. We use a shorthand notation SIN=S, COS=C and subscripts for A=Azimuth, E=Elevation. The best function we found to fit for each Mueller matrix estimate contains these 6 terms: $S_{2A}C_{2E}, S_{2E}C_{2A}, C_{2A}C_{2E}, S_{2A}S_{2E}, S_{2E}, C_{2E}$.

As an example of the coefficients found when fitting each of the 6 Mueller matrix element estimates, Figure \ref{daysky_solar_angsep_and_degpol_vs_exposure} shows the terms for spectral order 5 at a wavelength of 6881{\AA}.  Each particular Mueller matrix element seems to be dominated by only 1 or 2 terms.

The small differences between the trigonometric function fits to the rotation matrices and the original Mueller matrix element estimates show how well the interpolation method reproduces all Mueller matrix element estimates over the azimuth-elevation grid.  Figure \ref{sincos_fitting_minus_mm_estimates_reinterpolated} shows the difference between the trigonometric fits and the original Mueller matrix elements. The differences were computed only for (azimuth,elevation) points where all filters were applied and a valid result was obtained.  Some slight variation at elevations of 0 and 90 are seen in a few Mueller matrix elements.

\subsection{Interpolation Scheme Errors: Rotation Re-fits to Trig Functions} 

The trigonometric function fits to the rotation matrix element azimuth-elevation dependences cause some interpolation errors. The rotation matrices are re-fit to ensures that the Mueller matrices are strictly rotation matrices. This adds one more step in processing to ensure that any data calibrated at an arbitrary azimuth-elevation is not corrupted by the interpolation process.  In addition, interpolation from our chosen grid to the actual telescope pointing of any desired target adds uncertainty.  This interpolation introduces another source of error. The trigonometric fitting functions creates errors in the Mueller matrix estimates at all intermediate pointings because they are derived from interpolated rotation matrices. We chose a finely sampled azimuth-elevation grid spacing of 1$^\circ$ for this campaign.

To quantify this fitting error, differences between interpolated Mueller matrix elements and the corresponding re-fit rotation matrix elements were derived for telescope pointings in between the nominal observed azimuth-elevation grid at the maximum angular distance.  The cumulative distribution function for these Rotation matrix minus Mueller matrix estimate residuals is shown in Figure \ref{daysky_solar_angsep_and_degpol_vs_exposure}.

\subsection{Wavelength dependence}

\begin{figure} [ht] 
\begin{center}
\includegraphics[width=0.38\linewidth, angle=90]{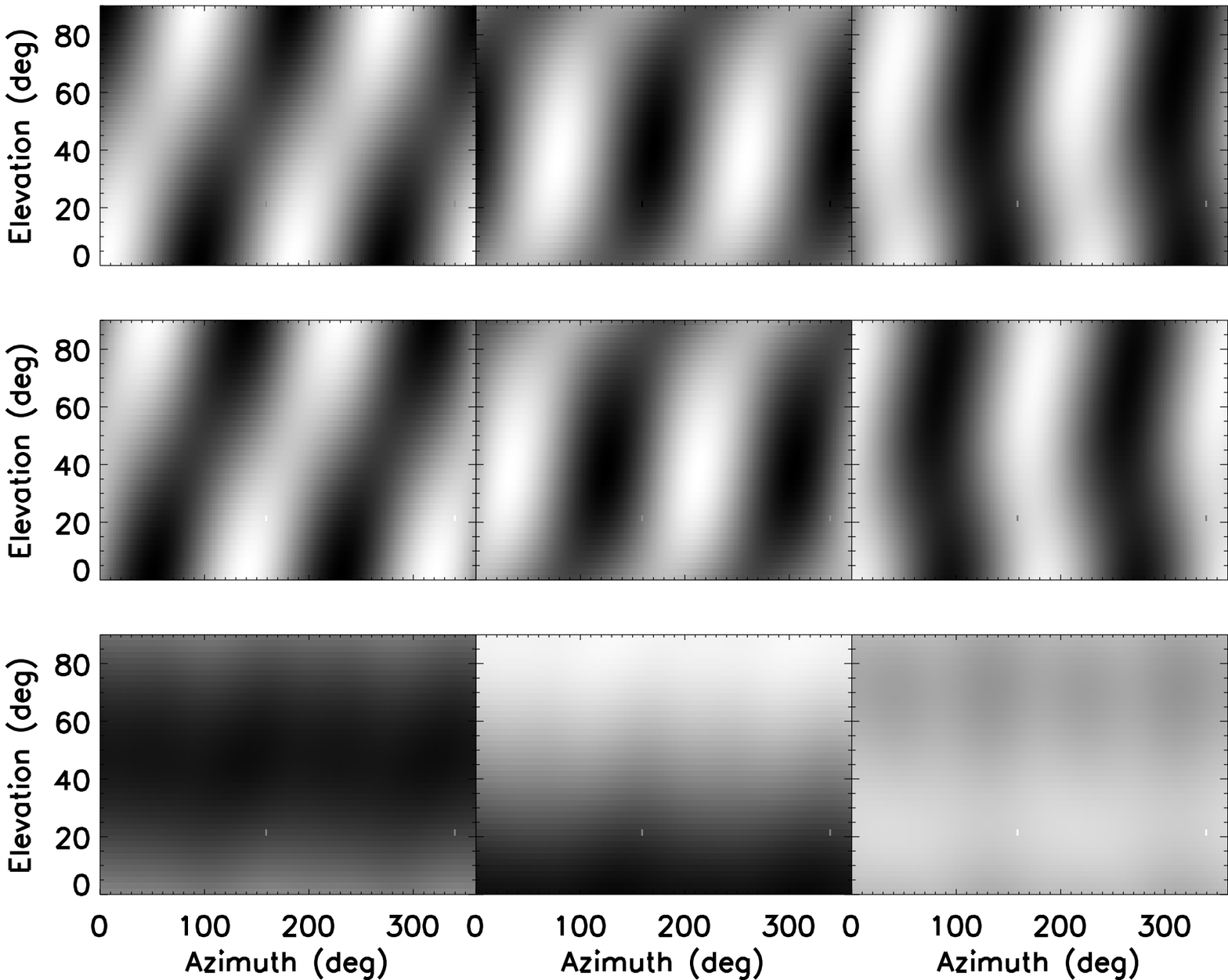}
\includegraphics[width=0.38\linewidth, angle=90]{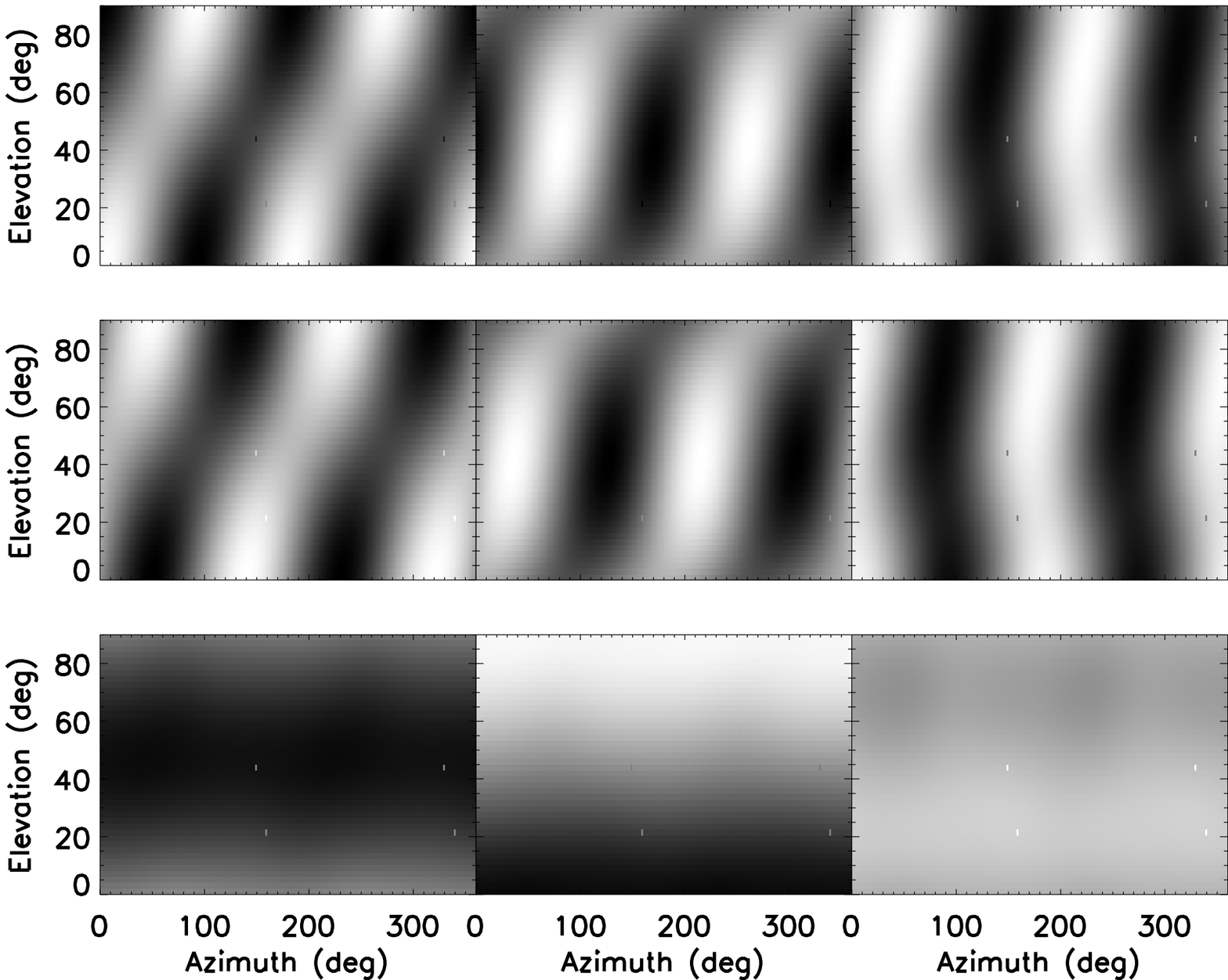} \\
\vspace{5mm}
\includegraphics[width=0.38\linewidth, angle=90]{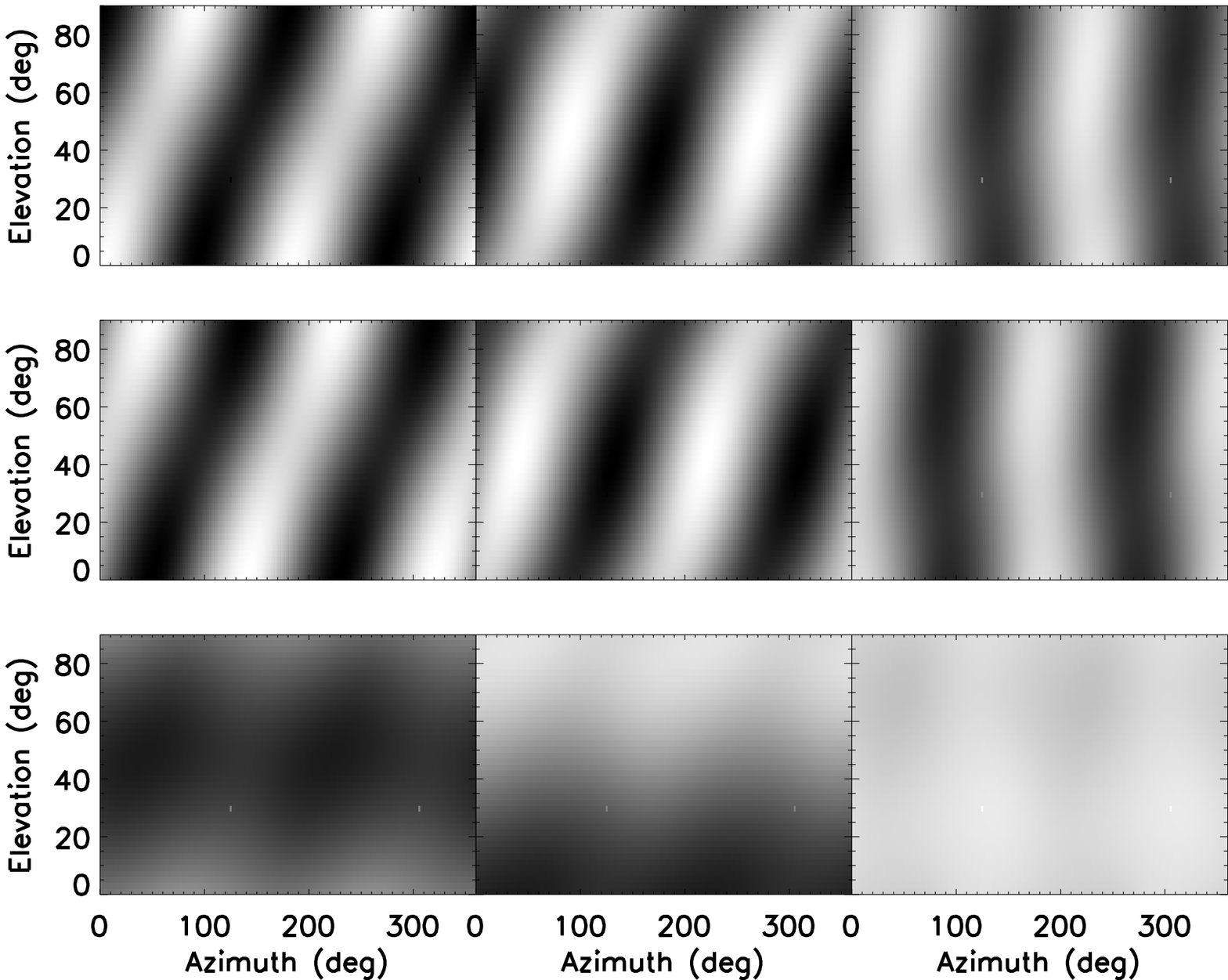}
\includegraphics[width=0.38\linewidth, angle=90]{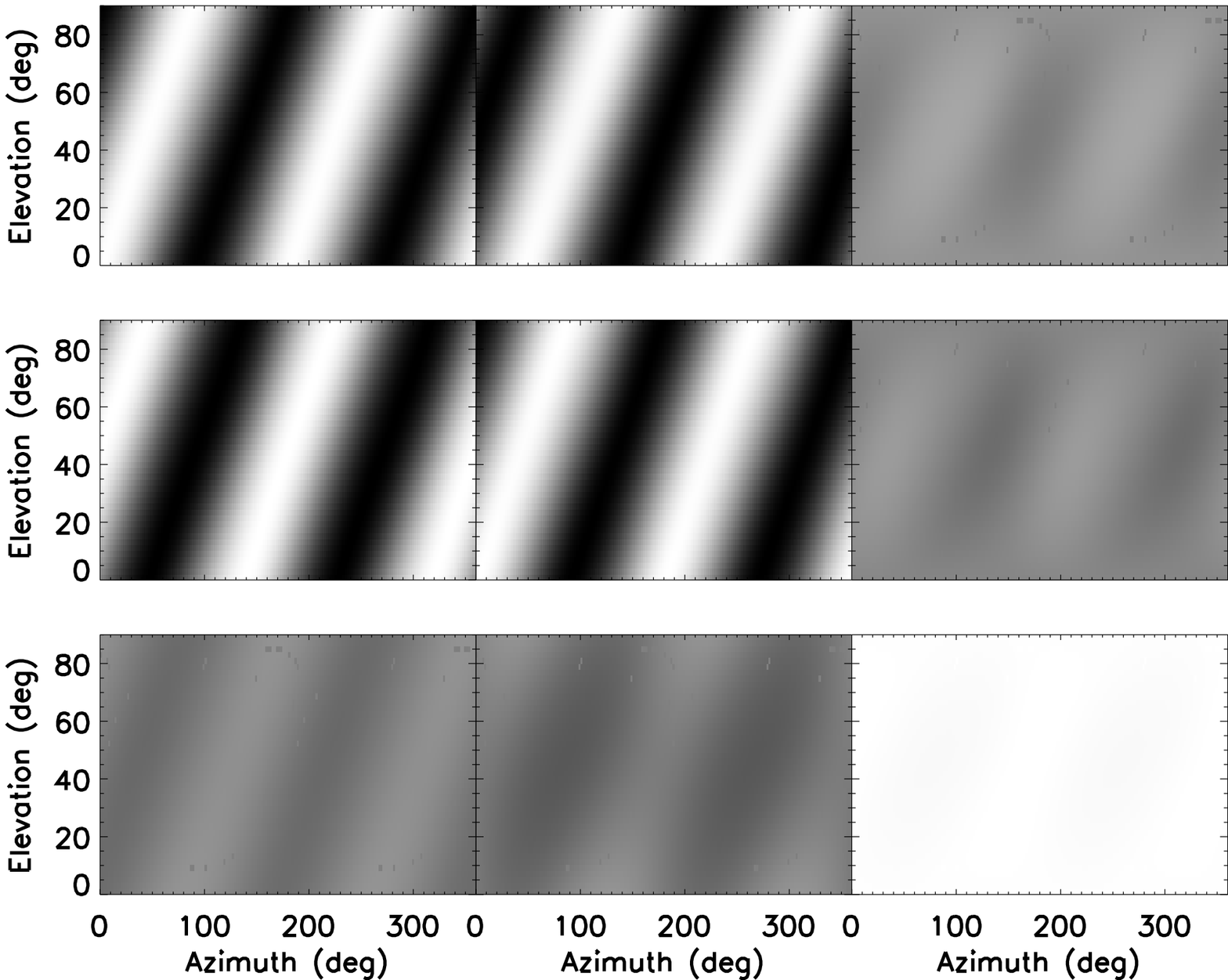}
\caption{ \label{sincos_fitting_high_resolution_map_interpolated_orders} The best fit Mueller matrix derived for 4 different spectral orders as functions of azimuth and elevation. All panels are on a linear grey scale of $\pm$1. The final rotation matrices shown here are fits to the trigonometric function based Mueller matrix maps on a fine altitude-azimuth grid.  The 15\% minimum measured DoP filter and iterative consistency-filters have been applied.  The four spectral orders are [0,5,10,15] corresponding to wavelengths of 6260{\AA}, 6880{\AA} 7650{\AA} and 8600{\AA}. For the longest wavelength in the lower right hand corner, the $VV$ term is essentially $+$1 with minimal dependence on elevation.  The $QU$ terms show the expected geometrical projection from an altitude-azimuth based reference frame to the fixed slit based reference frame. The $VQ$ and $VU$ terms on the right hand side of each panel show nearly $\pm$1 amplitudes at the shorter wavelengths $<$7000{\AA} but are nearly 0 at 8600{\AA}}
\end{center}
\end{figure}

The Mueller matrices for HiVIS are smooth functions of wavelength. Figure \ref{sincos_fitting_high_resolution_map_interpolated_orders} shows the azimuth elevation dependence for HiVIS in four spectral orders numbered [0,5,10,15] corresponding to wavelengths of [6260{\AA}, 6880{\AA} 7650{\AA} and 8600{\AA}].   

For wavelengths short of 7500{\AA}, the linear to circular and circular to linear cross-talk terms are quite large. The $VV$ terms show elevation dependence and are much less than 1. For the last spectral order at 8600{\AA}, the $VV$ term is nearly 1 and shows negligible dependence on elevation. The $VQ$ and $VU$ terms show nearly $\pm$1 amplitudes with strong functional dependence on azimuth and elevation at the shorter wavelengths $<$7000{\AA}, but are nearly 0 at 8600{\AA}. As wavelengths increase, the polarization response of HiVIS goes from severe linear-to-circular cross-talk to very benign cross-talk approaching the nominal geometrical $qu$ variation expected for an altitude-azimuth referenced coordinate frame. 

\subsection{Zemax system modeling of azimuth-elevation dependence}

The optical ray tracing program, Zemax, has the ability to perform fully polarized ray propagation using a Jones formalism.  With this program, we can verify that the trigonometric functions in azimuth and elevation fully capture the behavior of the telescope Mueller matrix. We have used this program in the past to create polarization models of HiVIS \cite{Harrington:2006hu, 2014SPIE.9147E..7CH}. With the polarized ray trace function called POLTRACE, a Zemax user can propagate rays from any pupil coordinate (Px, Py) to any field coordinate (Hx, Hy) in the Zemax file.  By selecting a set of fully polarized inputs in the Jones formalism that also correspond to the Stokes vectors ($quv$), one can determine the polarization response of an optical design. With the Zemax programming language (ZPL), we trace many rays propagated from a grid of coordinates (Px,Py) across the pupil through the optical system to the corresponding focal plane. Typical sampling of 10\% where the pupil coordinates (Px,Py) are scanned in steps of 0.1 achieves 0.0001 level or better agreement to Mueller matrix calculations using more fine pupil sampling. This match between Mueller matrix terms depends on the details of the optical system including optical power, tilted optics and in general the symmetries in the polarization properties of the exit pupil. However, a 0.1 step in Px,Py seems to be a good compromise between model run speed and calculation sensitivity. Tests run at step sizes of 0.01, 0.025, and 0.05 do not vary by more than the 5th decimal place under typical, mostly symmetric, non-vignetted system configurations.

\begin{figure} [ht]
\begin{center}
\hbox{
\hspace{-3.0em}
\includegraphics[width=0.55\linewidth, angle=0]{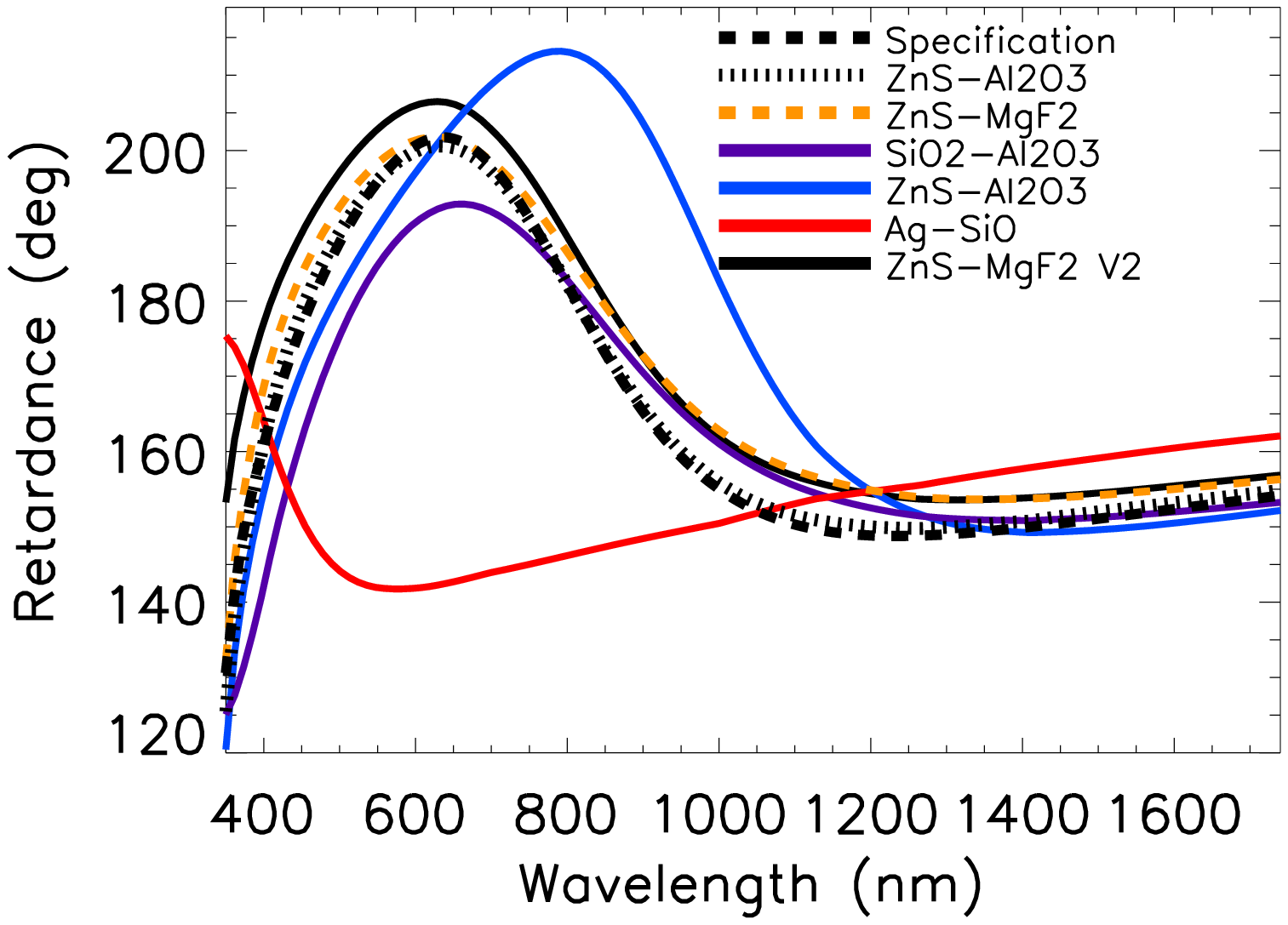}
\hspace{-1.5em}
\includegraphics[width=0.55\linewidth, angle=0]{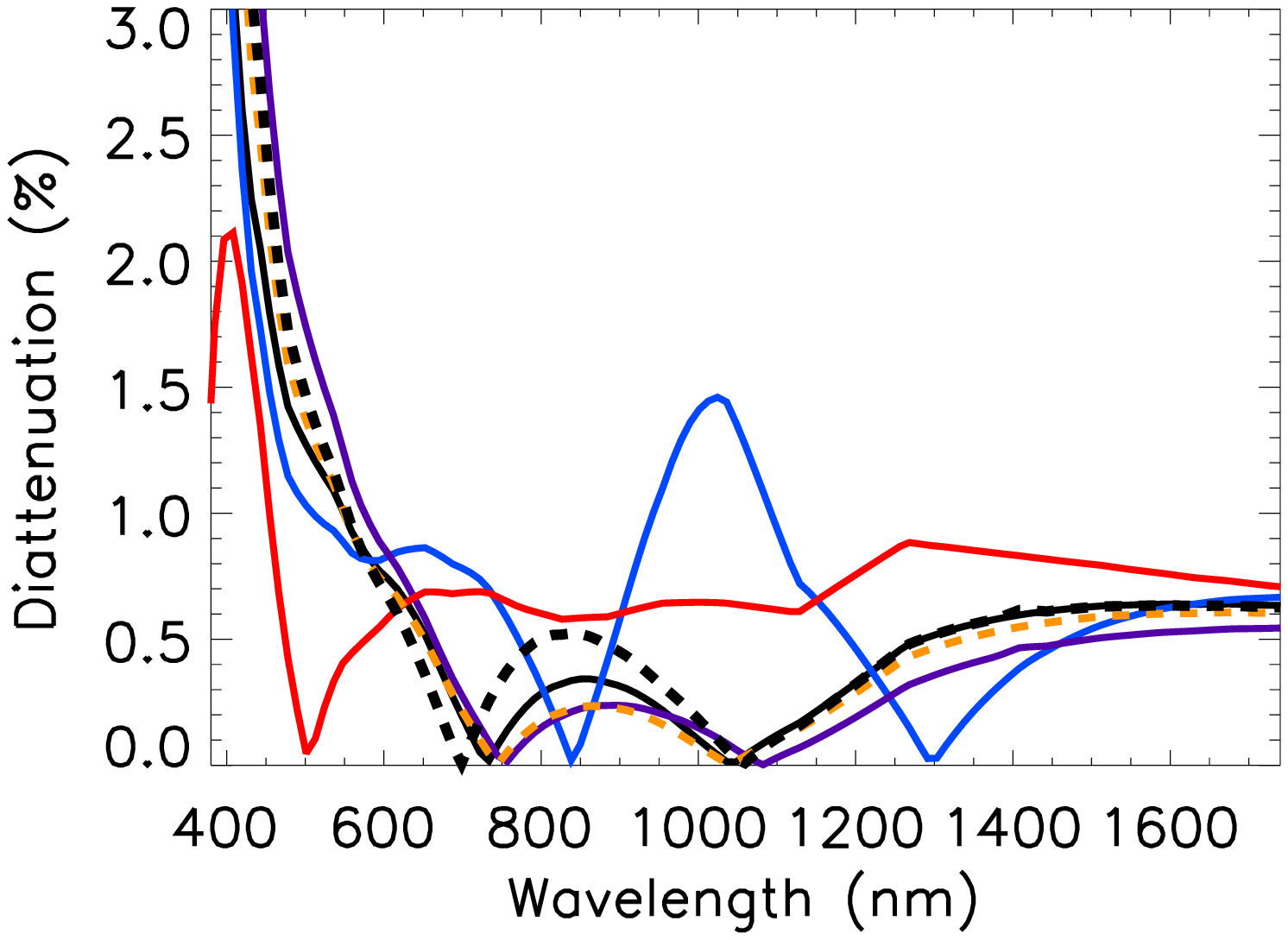}
}
\caption{ \label{coating_phase_comparison} The top panel retardance in degrees phase for a 45$^\circ$ fold mirror coated with various flavors of enhanced protected silver formulas.  The solid black line shows a typical enhanced protected silver specification.  The other lines show the retardation caused by various coating formulations.  Common materials and coatings include  Zinc Sulfide (ZnS), Sapphire (Al2O3), Fused Silica (SiO2) over silver (Ag).  The retardation of multi-layer coatings often crosses the nominal 180$^\circ$ phase at two wavelengths in the visible.  The red curve shows a single fused silica coating and this curve never reaches 180$^\circ$.   }
\end{center}
\end{figure}

The Zemax electric field calculations in the Jones formalism are turned in to Stokes vector formalism for each of the pure $quv$ input states. The POLTRACE function outputs electric field vector amplitudes (Ex, Ey, Ez) and phases following the Jones formalism for every ray traced.  For simplicity in large f/ number beams, we project this 3D field on to a 2 dimensional surface ignoring the z components along the direction of propagation.  As the HiVIS polarimeter operates at f/40, this is a reasonable assumption for this analysis. 

The computed Stokes intensity is the square of the electric field amplitudes (Ex*Ex + Ey*Ey). This incoherent average is also a reasonable approximation for seeing-limited systems or systems not fully sampling the polarized diffraction limited point spread function. Stokes Q goes as the X and Y intensity difference:  (Ex*Ex - Ey*Ey).  Stokes U is computed from X and Y electric field amplitudes as well as phase variations:  2.*Ex*Ey*COS($\delta$). Stokes V is similarly computed with both XY field amplitudes and phases: 2.*Ex*Ey*SIN($\delta$).  The term $\delta$ represents the phase difference.

A key parameter for determining the polarization response of an optical system is the coating formulation. The output polarization models are very sensitive to the coating model thicknesses.  We performed some experiments with predicting AEOS and HiVIS polarization response using various coating formulations. As we do not have access to the coating or coating formulas for the AEOS mirrors, we show a range of representative functions derived from common coatings.  Figure \ref{coating_phase_comparison} shows some of the formulas commonly used in enhanced protected silver mirrors. The phase retardation and diattenuation for these coatings is shown in Figure \ref{coating_phase_comparison} is for a 45$^\circ$ reflection using enhanced protected silver coated mirrors. 

The retardance of these one- and two-layer coatings matches vendor-provided curves.  For a typical two-layer coating, the retardance has two wavelengths where the nominal 180$^\circ$ phase crossing occurs.  The exact thickness and materials in the coating determine which wavelengths, but overall the crossings typically occur in the blue-green and red-near-infrared regions of the spectrum.  Usually the retardance is 10$^\circ$ to 30$^\circ$ above 180$^\circ$ for the intermediate bandpass and below 180$^\circ$ outside these wavelengths.   Single layer protective coatings such as the fused silica over silver have a retardation always below 180$^\circ$. This is seen as the red curve in Figure \ref{coating_phase_comparison}. 

As expected, all the enhanced protected silver formulations with ZnS as an over-coating show diattenuation values below 1\% for wavelengths longer than 550nm.  This includes all of the double-layer coatings. The formulas with a fused silica over-coating show higher induced polarization levels.

Most multi-layer coatings have polarization responses that are strong functions of the angle of incidence. At near-normal angles of incidence, the coatings shown in Figure \ref{coating_phase_comparison} will all display the nominal 180$^\circ$ phase and minimal diattenuation.  As the angle of incidence is increased past 45$^\circ$, the retardation will be over 40$^\circ$ above nominal.   As Zemax propagates light ray-by-ray through an optical system across both pupil and field, the angle of incidence is accounted for in the ray-trace.  Any variations in angle of incidence across a beam footprint from asymmetries, off-axis rays, decentered optics and / or vignetting will be propagated to the focal plane.

\begin{wrapfigure}{r}{0.60\textwidth}
\centering
\vspace{-4mm}
\hbox{
\hspace{-0.3em}
\includegraphics[width=0.98\linewidth, angle=90]{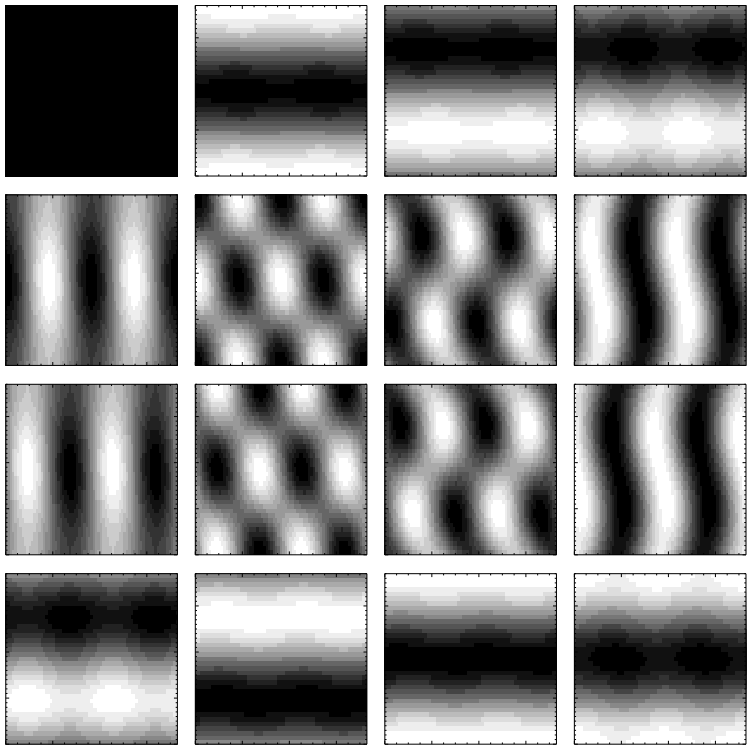}
}
\caption{ \label{zemax_model_626nm_eagd} The Zemax simulated Mueller matrix at 626nm for 360$^\circ$ azimuth variation and a fully articulated 0 to 180$^\circ$ elevation range using a nominal enhanced protected silver coating on all surfaces except the primary mirror (aluminum + aluminum oxide). The full elevation range was modeled to demonstrate the elevation dependence of some Mueller matrix elements. The mirror geometry provides two degenerate optical configurations when pointing to a particular point on the sky with different Mueller matrix calculations  (under the substitution Azimuth --  Azimuth $\pm$ 180,  Elevation --  180 - Elevation.  The intensity to $QUV$ terms have been linearly scaled to $\pm$1.5\%.  The $QUV$ to $QUV$ terms have been scaled to $\pm$1. The II term has not been displayed because it is set to 1 in this model calculation. }
\vspace{29mm}
 \end{wrapfigure}

To demonstrate the typical functional dependence of polarization, we computed the system Mueller matrix as functions of azimuth and elevation for all spectral orders HiVIS samples.  Figure \ref{zemax_model_626nm_eagd} shows a typical output Mueller matrix.  The intensity to polarization and the polarization to intensity terms were linearly scaled to $\pm$1.5\%.  Note that this predicted induced polarization and depolarization is quite small, further supporting the assumption that the telescope is only weakly diattenuating. The polarization to polarization terms ($quv$ to $quv$) have been scaled to $\pm$1.  The functional dependence is exactly as found using our trigonometric functions above.  The $qu$ to $quv$ terms have a strong azimuthal dependence.  The $qu$ to $v$ terms are dominated by elevation dependence.

\begin{figure}  [ht]
\begin{center}
\includegraphics[width=0.24\linewidth, angle=0]{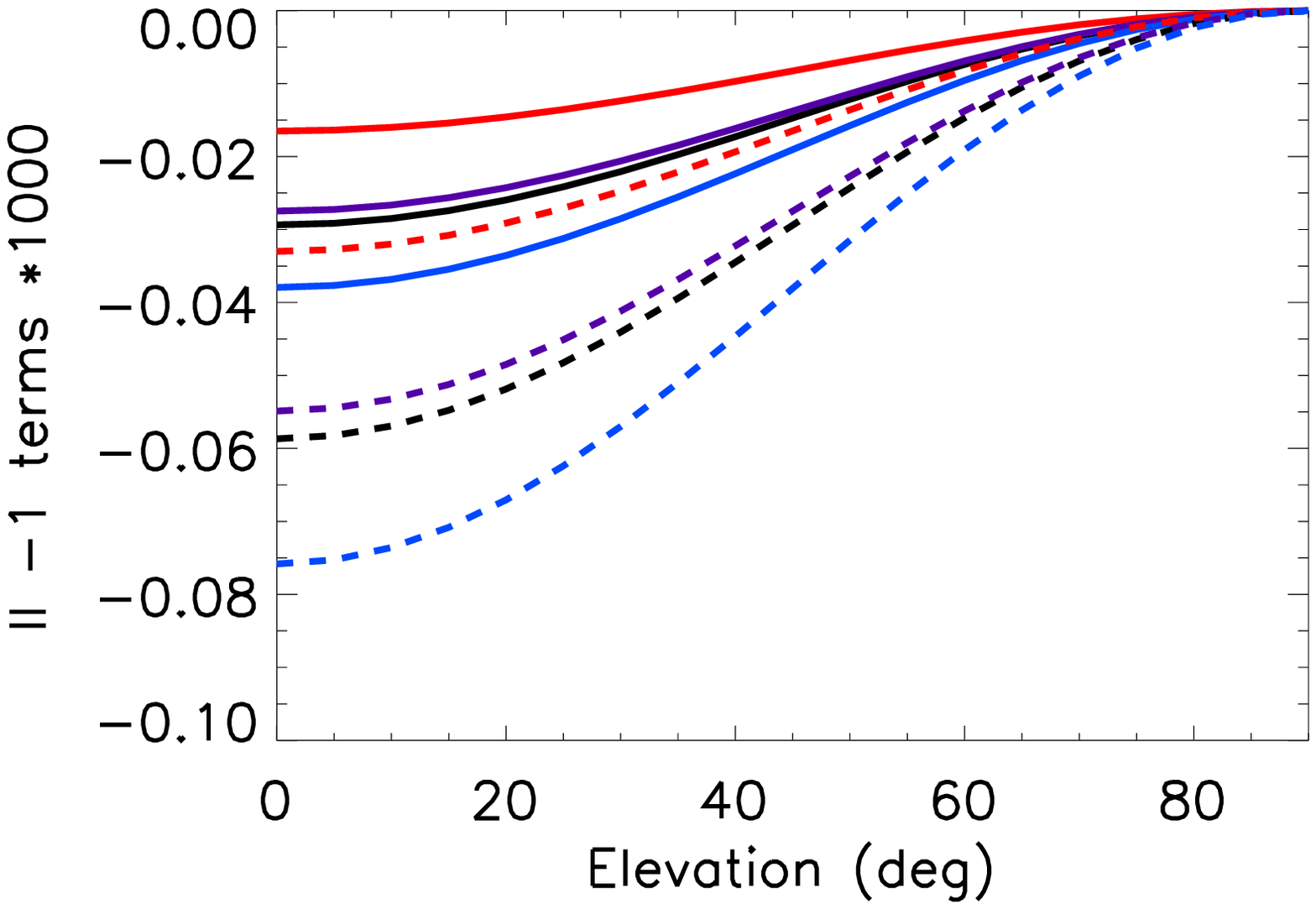}
\includegraphics[width=0.24\linewidth, angle=0]{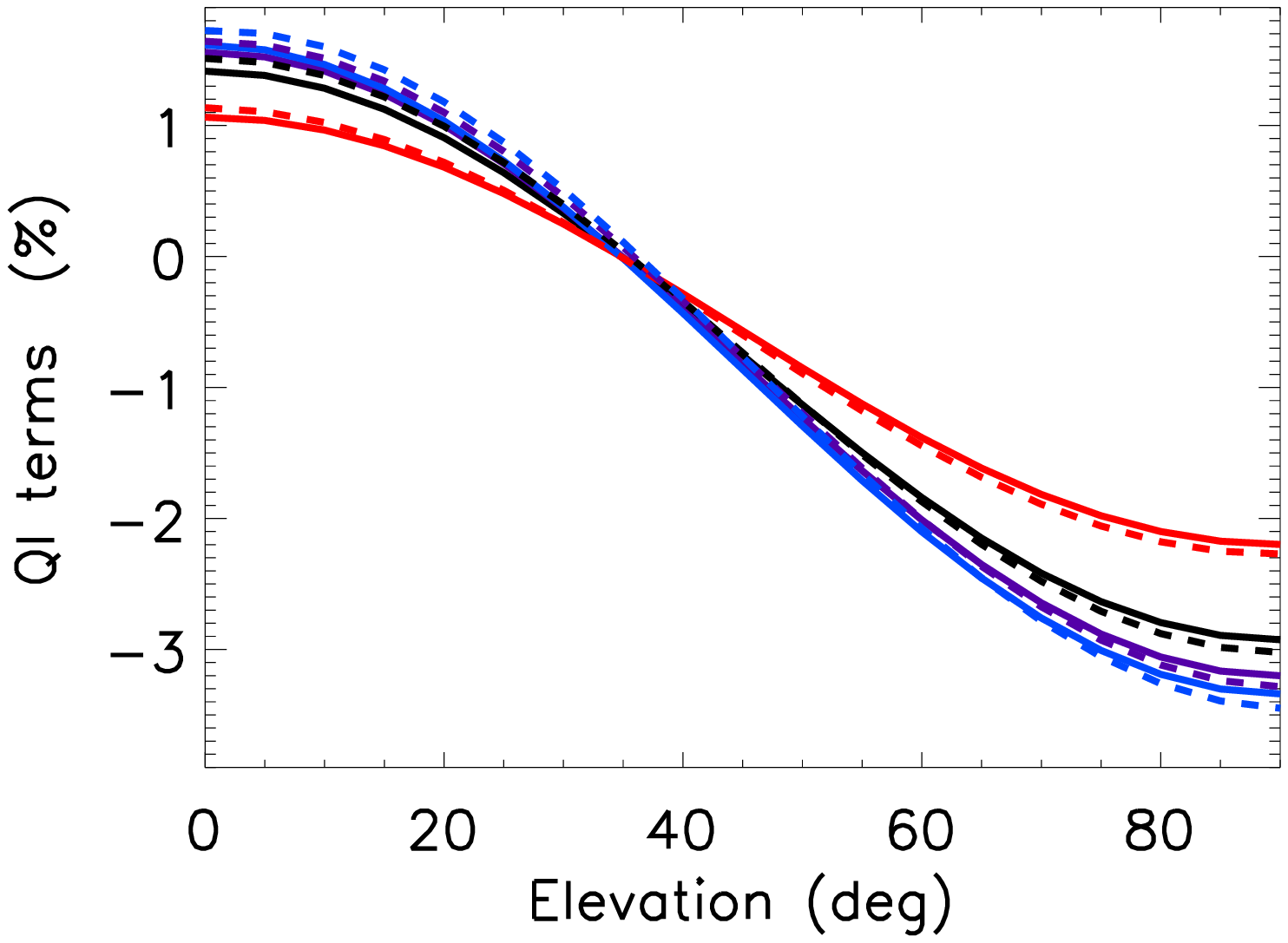}
\includegraphics[width=0.24\linewidth, angle=0]{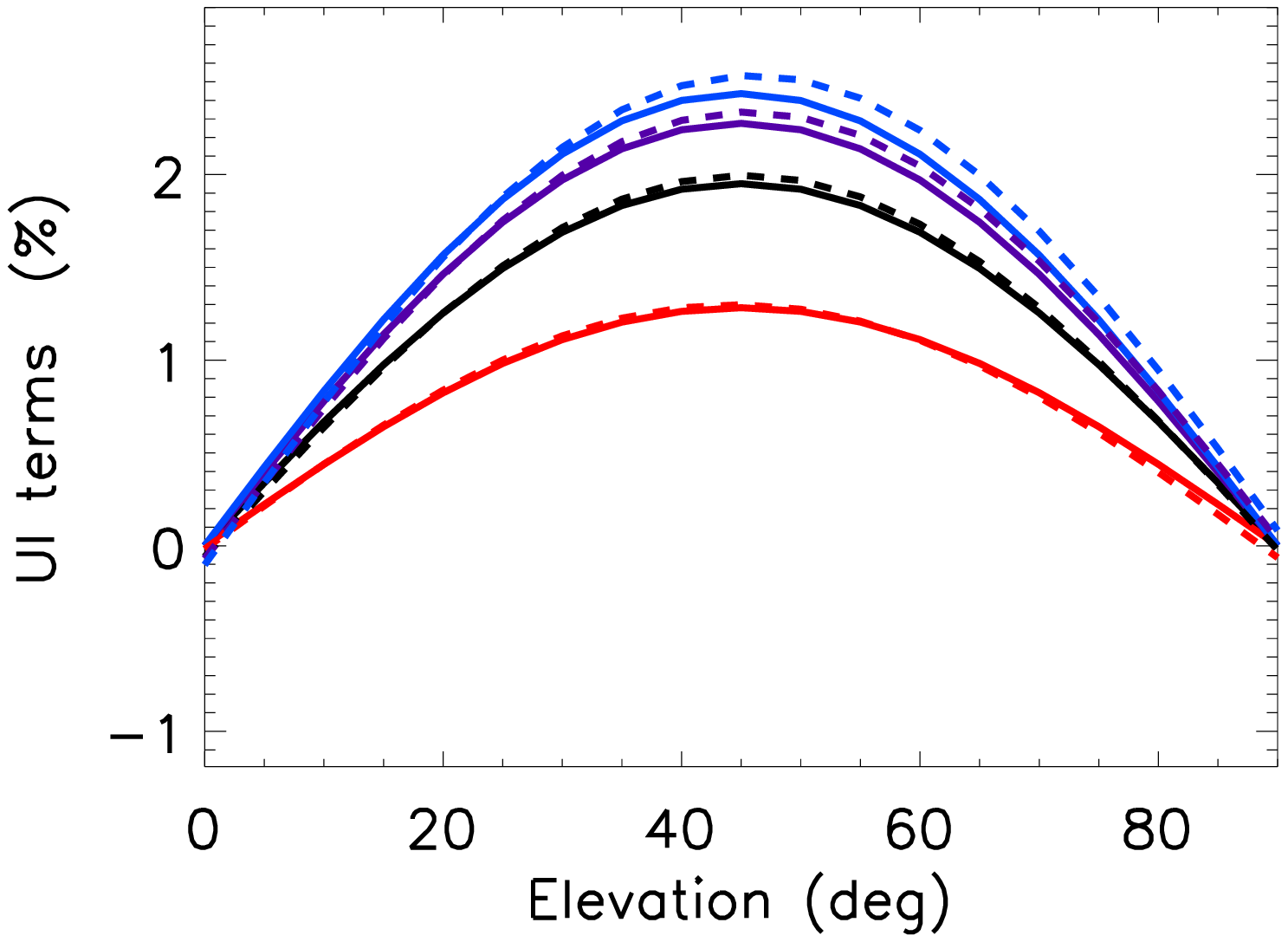}
\includegraphics[width=0.24\linewidth, angle=0]{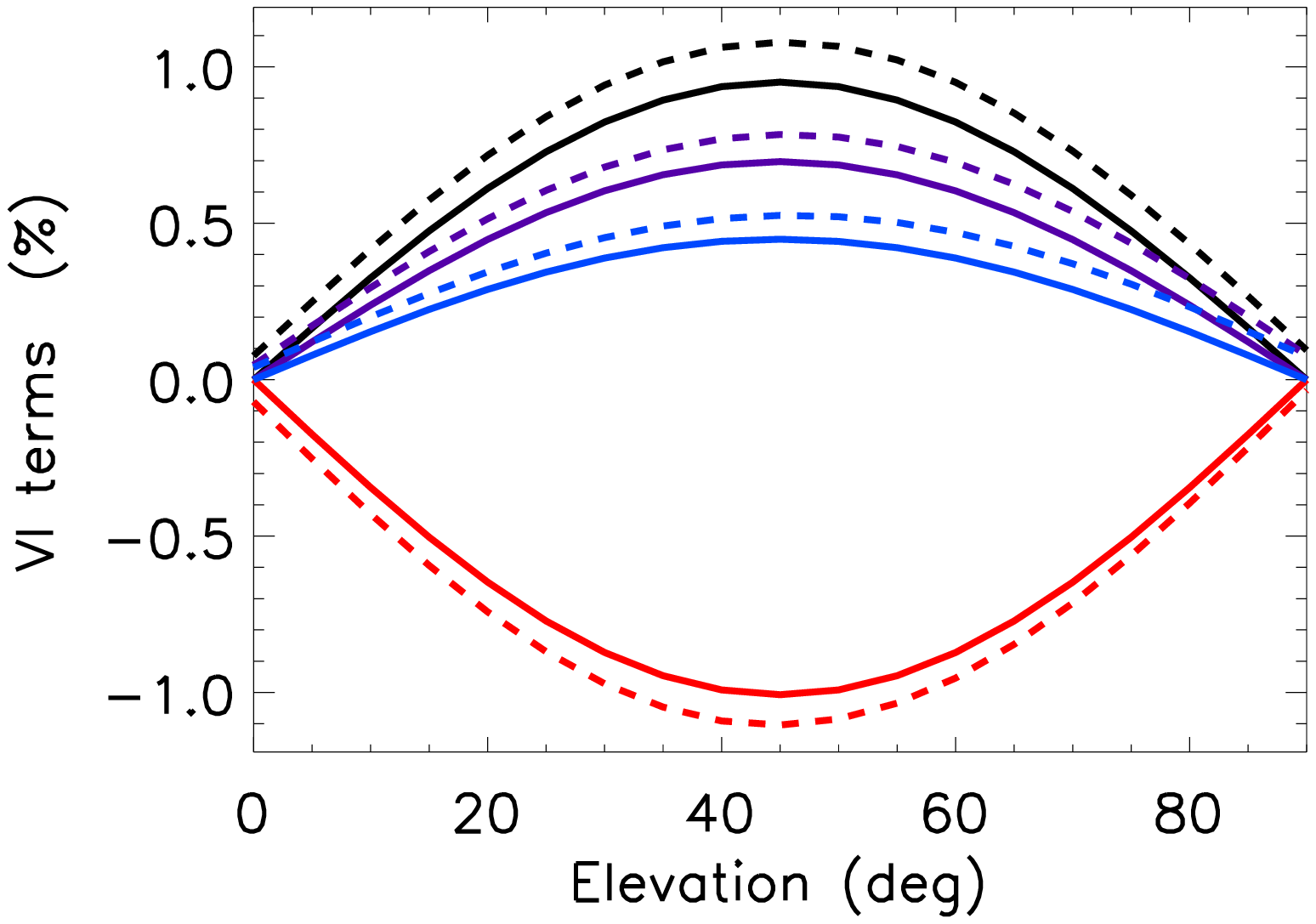}
\includegraphics[width=0.24\linewidth, angle=0]{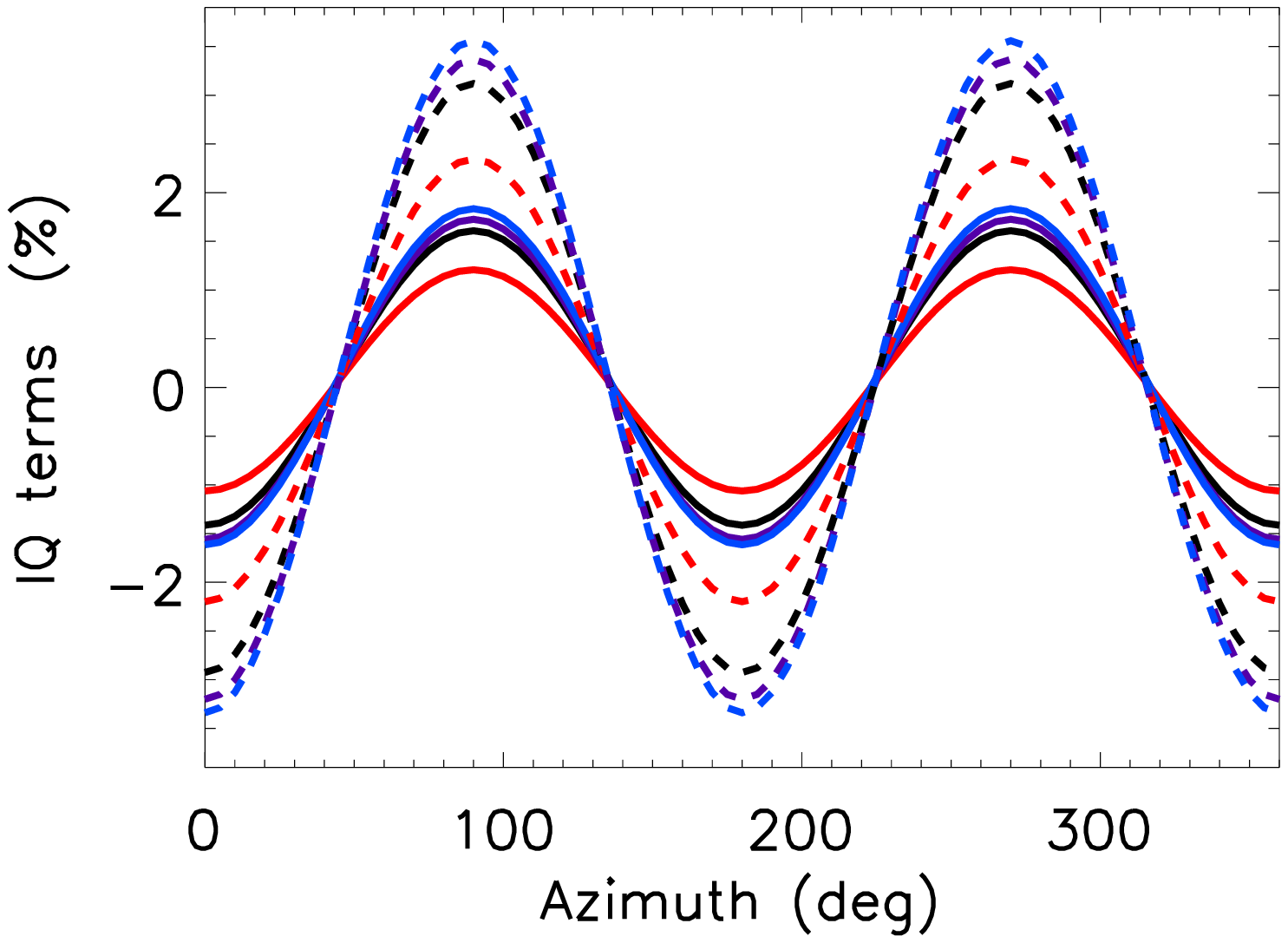}
\includegraphics[width=0.24\linewidth, angle=0]{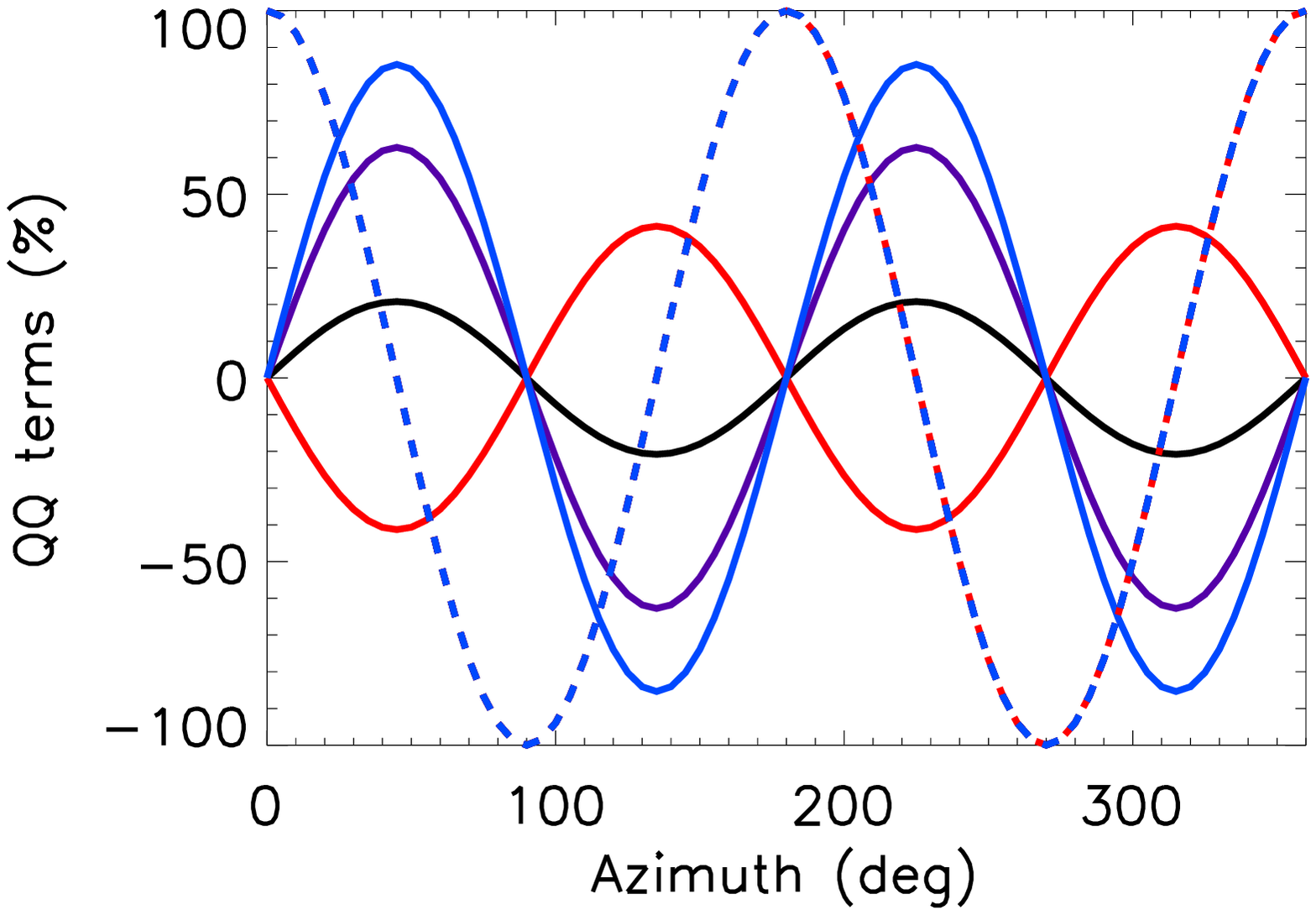}
\includegraphics[width=0.24\linewidth, angle=0]{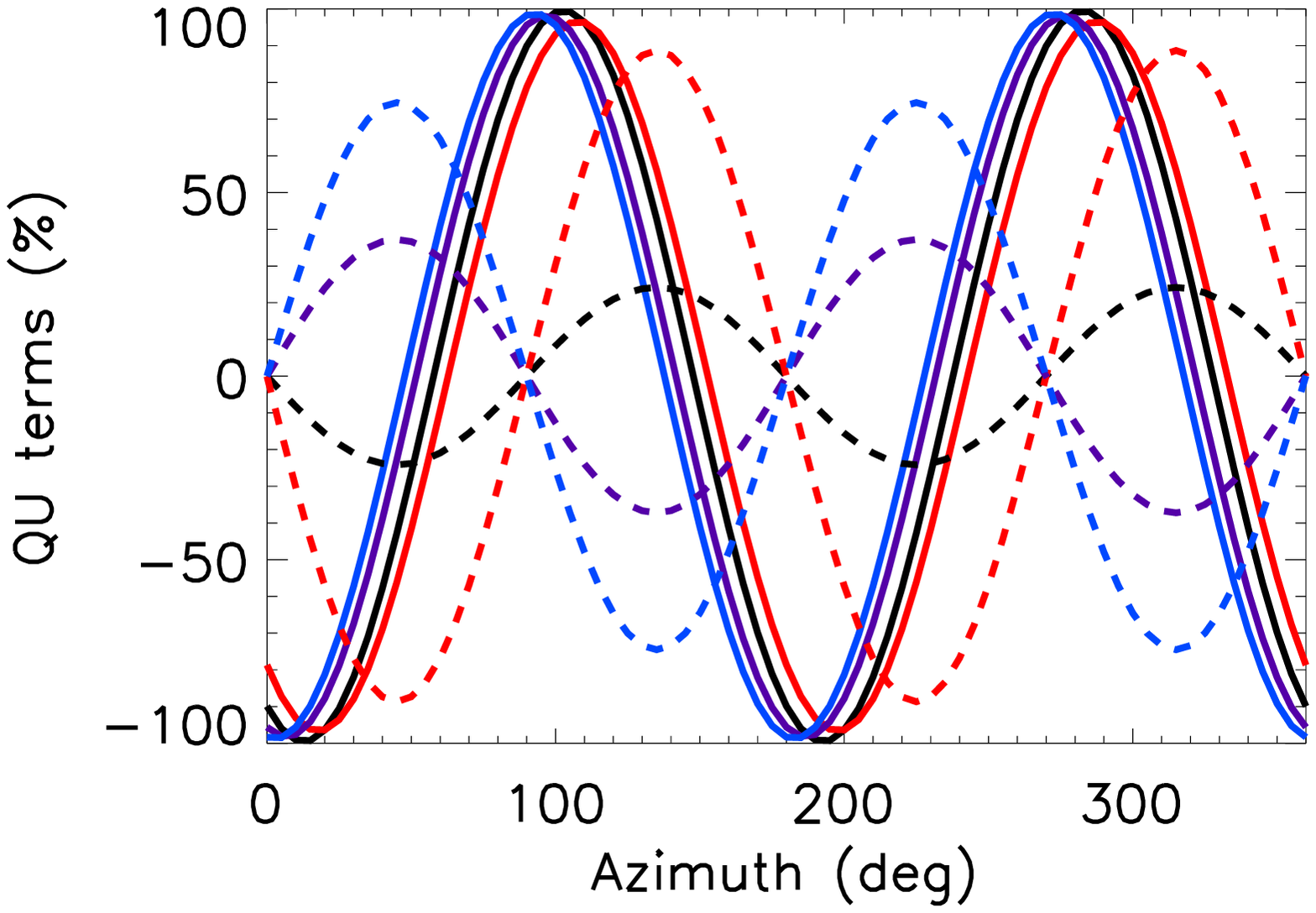}
\includegraphics[width=0.24\linewidth, angle=0]{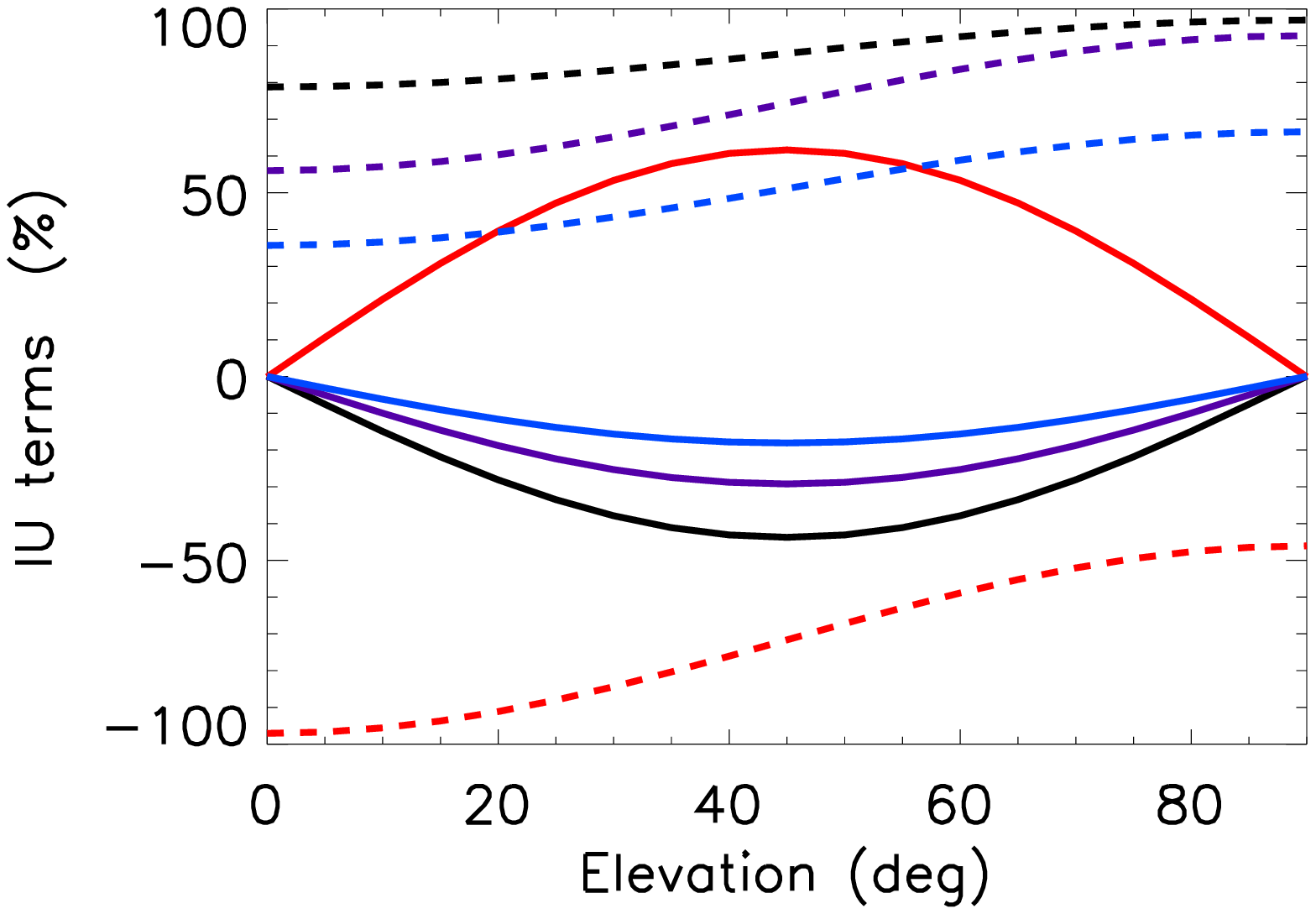}
\includegraphics[width=0.24\linewidth, angle=0]{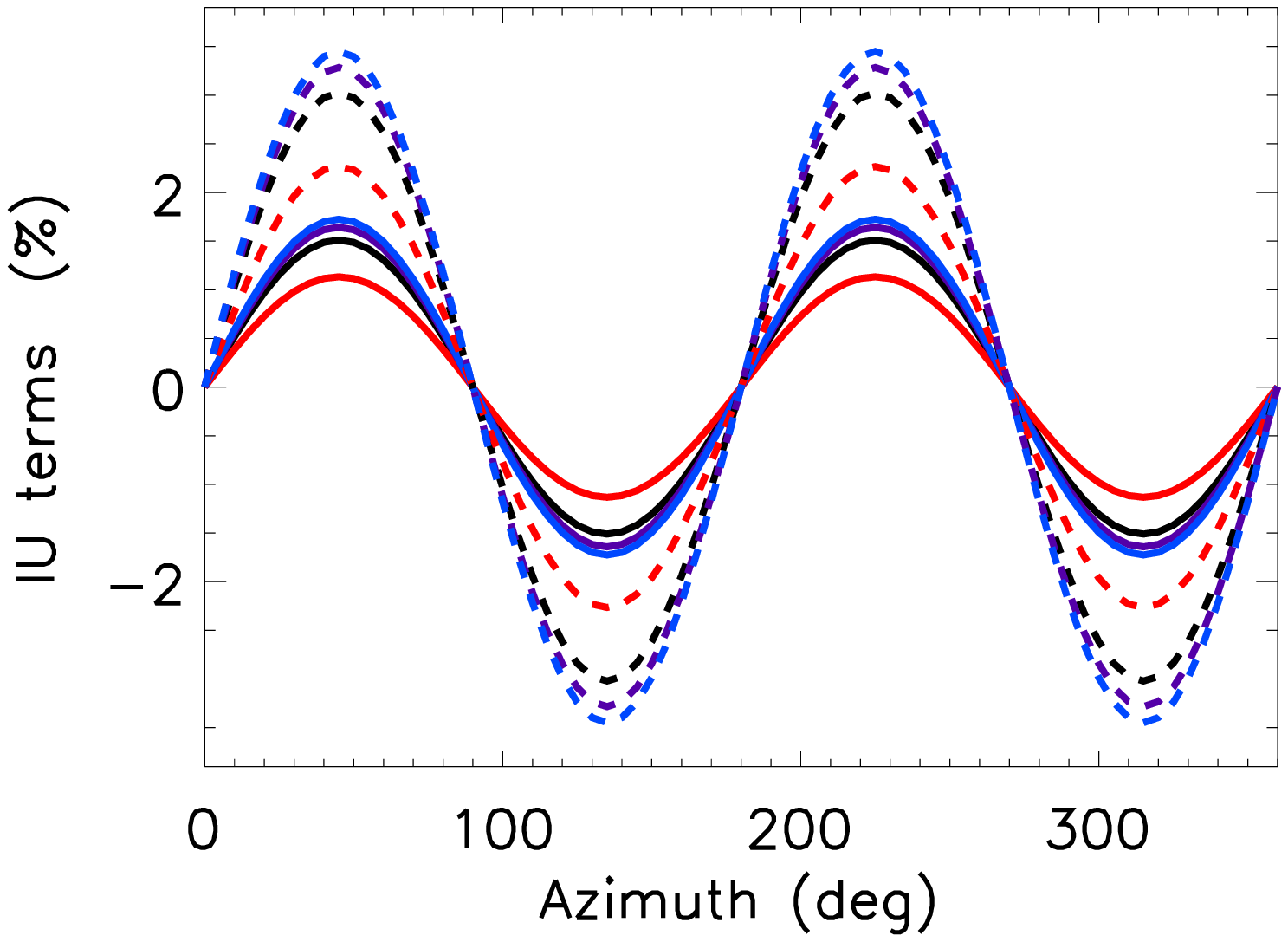}
\includegraphics[width=0.24\linewidth, angle=0]{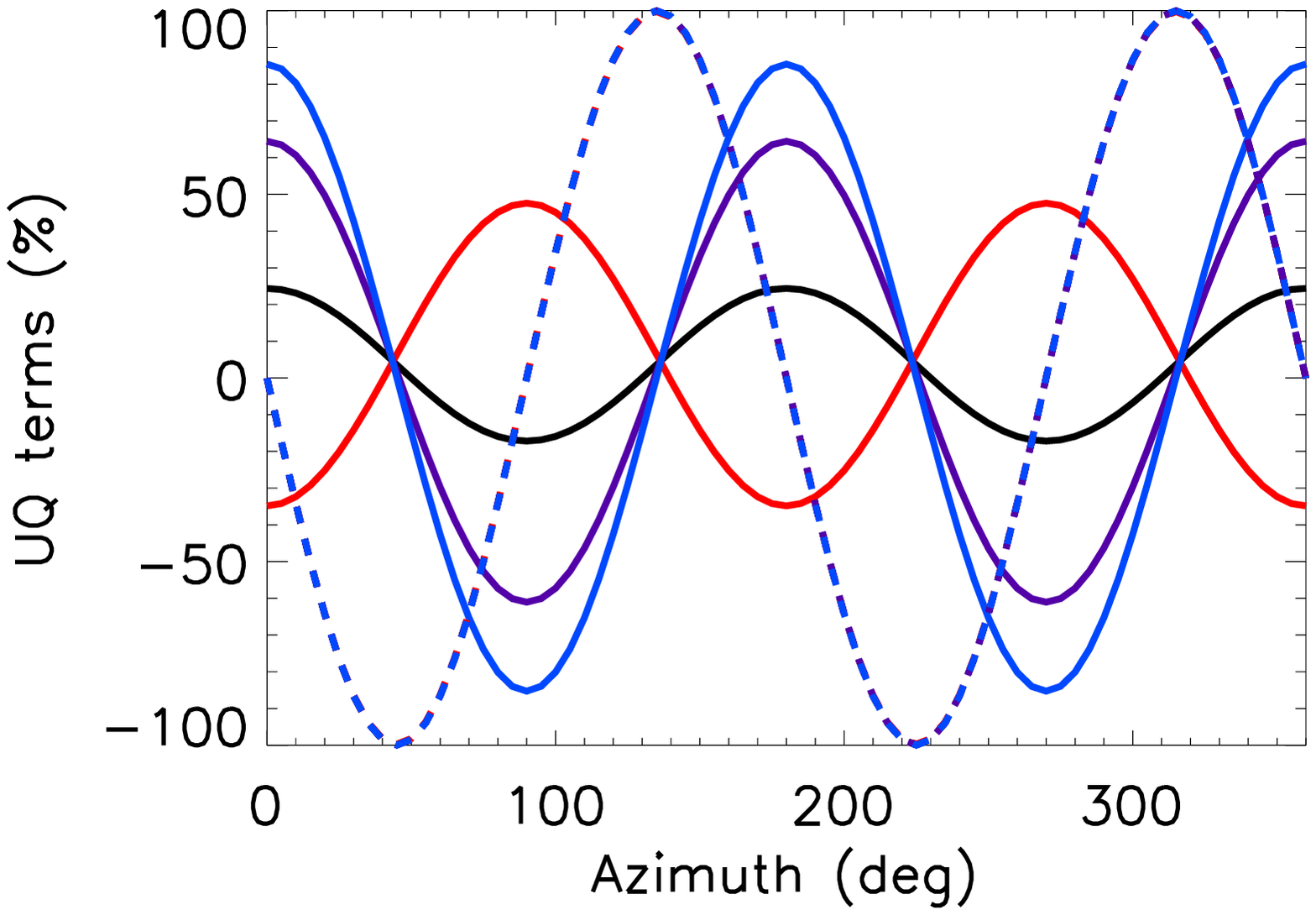}
\includegraphics[width=0.24\linewidth, angle=0]{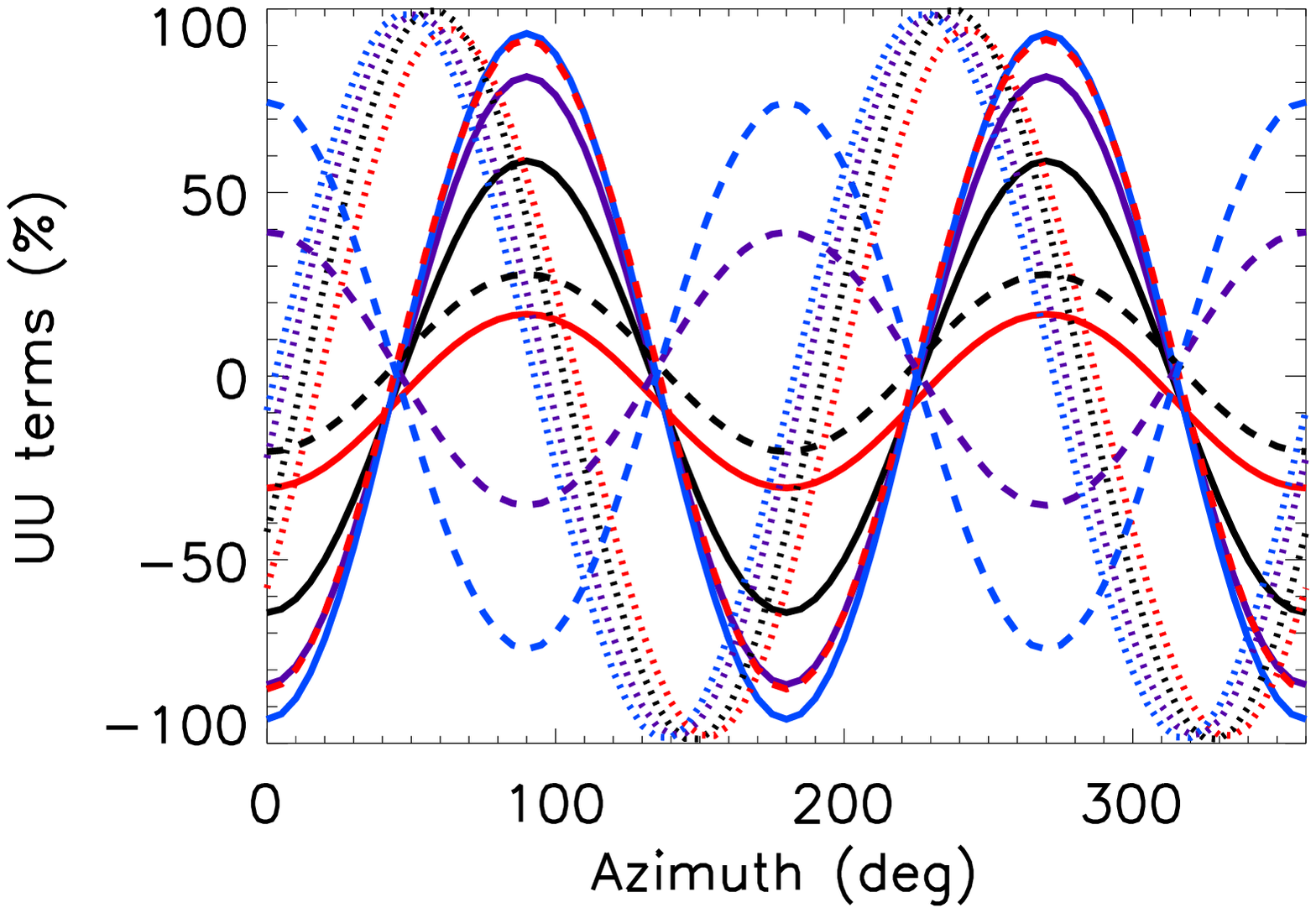}
\includegraphics[width=0.24\linewidth, angle=0]{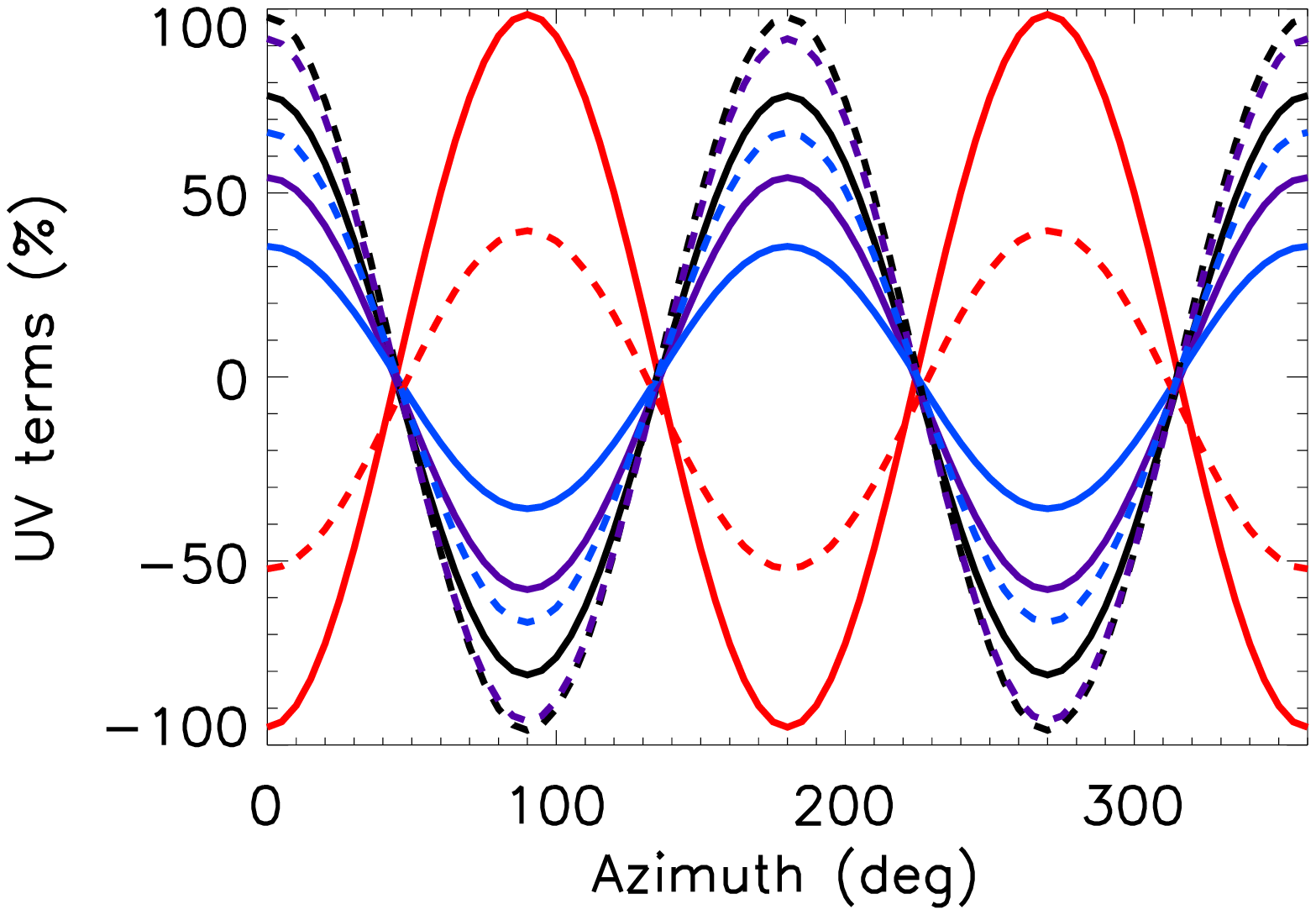}
\includegraphics[width=0.24\linewidth, angle=0]{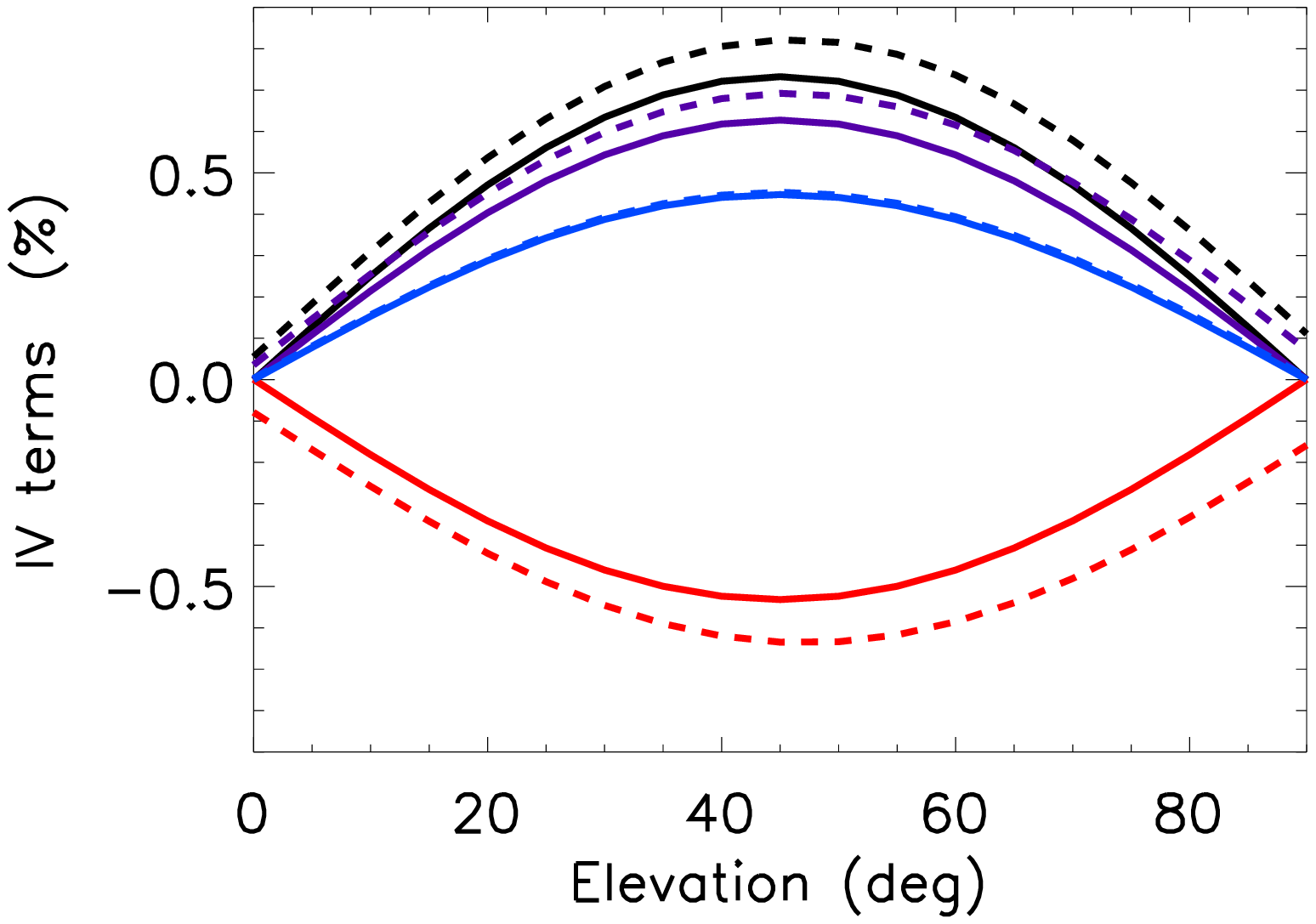}
\includegraphics[width=0.24\linewidth, angle=0]{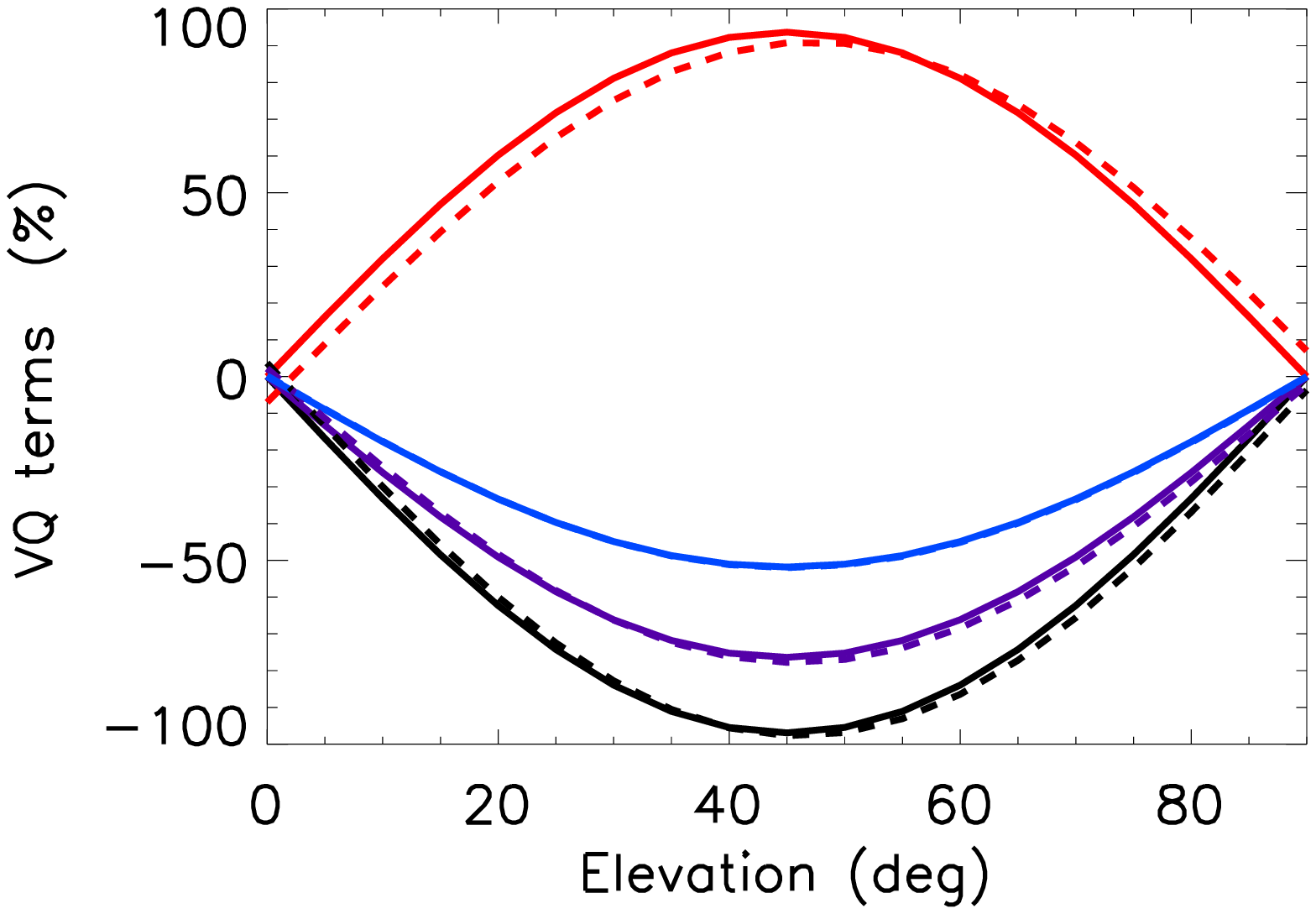}
\includegraphics[width=0.24\linewidth, angle=0]{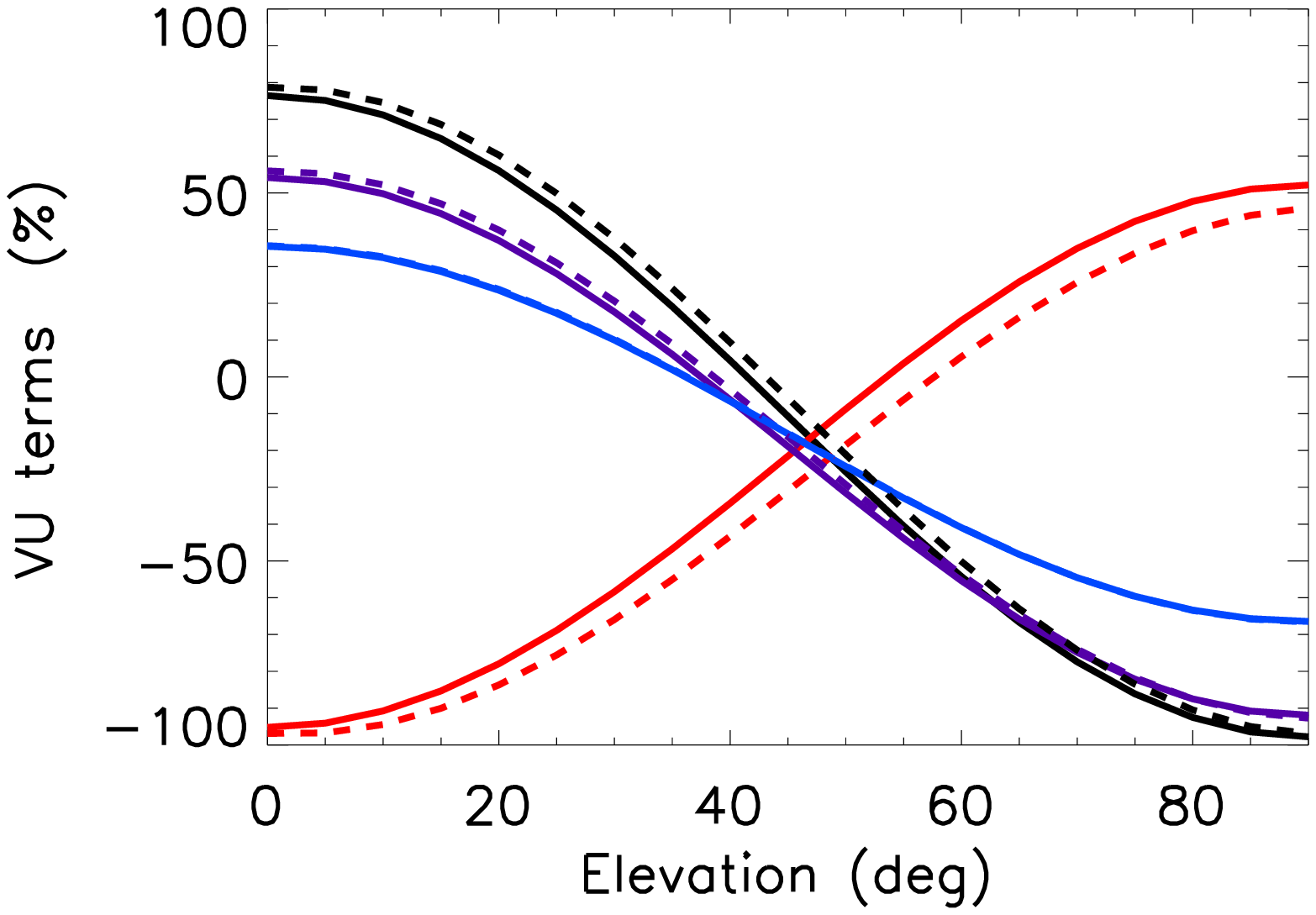}
\includegraphics[width=0.24\linewidth, angle=0]{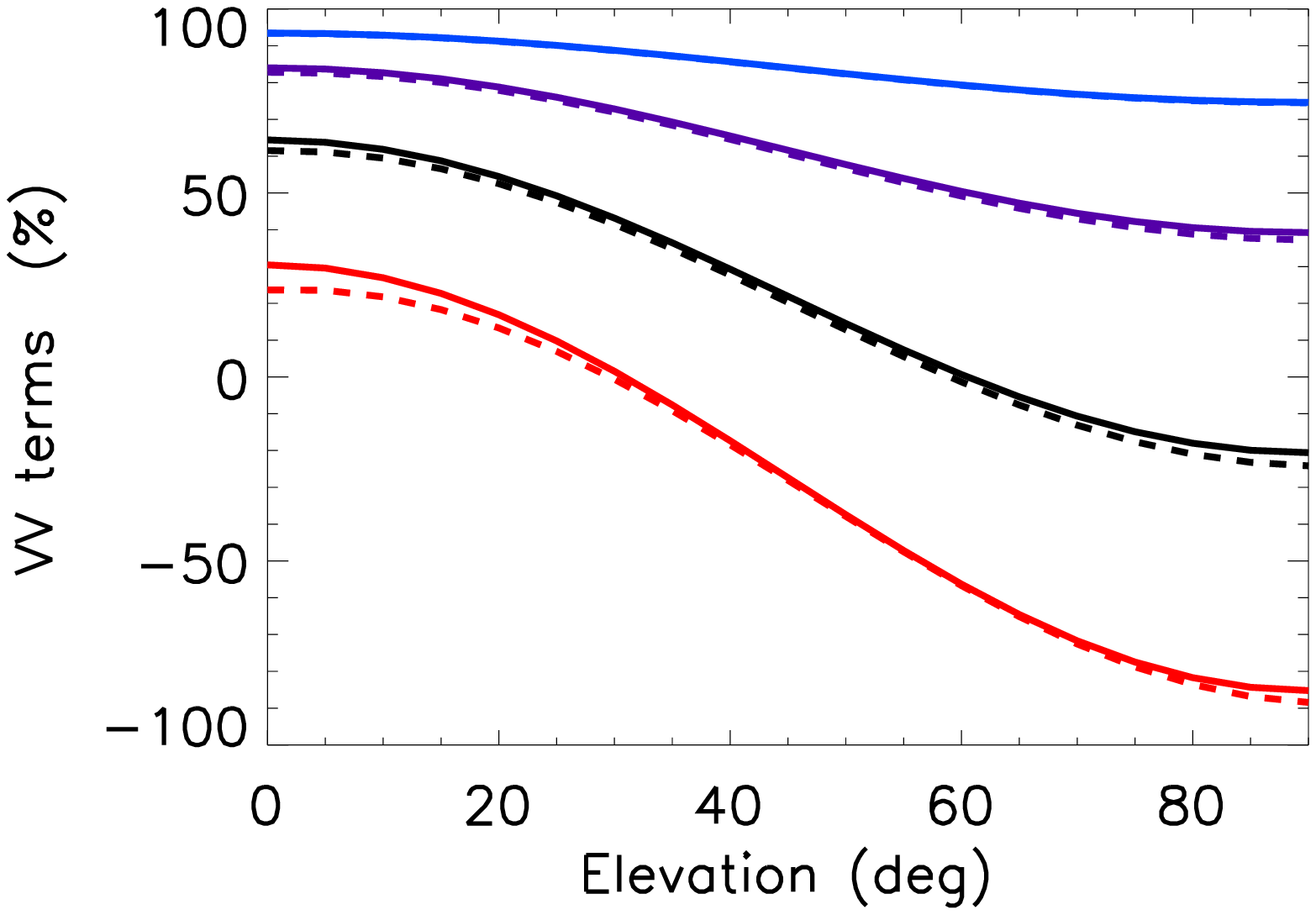}
\caption{ \label{hivis_functional} Comparison of coatings for all Mueller matrix elements at 600nm.  Each panel shows either azimuth or elevation dependence at particular pointings as appropriate for the element.  Telescope pointing locations were chosen to demonstrate how each coating formula changes the amplitude and relative zero-crossing for the different Mueller matrix elements. Solid lines is one limiting case in either Azimuth or Elevation with dashed being the opposite extreme to demonstrate maximal polarization effects (crossed mirrors, maximal differences).  The polarization to intensity terms are dominated by elevation dependence.  The $QU$ to $QU$ terms are strongly azimuth dependent.   Red shows the fused-silica over silver single layer formula Ag-SiO2.  Black is MgF2 over silver.  Dark blue is fused silica over sapphire over silver.  Light blue is zinc sulfide over sapphire over silver. As we do not know the actual coating formula for the various AEOS mirrors, these functions represent the range of possibilities for an optical system using mirrors with common formulas.}
\end{center}
\end{figure}

To demonstrate how the coating formula changes polarization predictions, we computed Mueller matrix predictions for all coating formulas in Figure \ref{coating_phase_comparison}. The full range of optical motion possible in the HiVIS system was used to show the dependence of coating retardation on the Mueller matrix element zero-crossings as well as the relationship between coating formula predictions.  Figure \ref{hivis_functional} shows the functional dependence of the predicted Mueller matrices at chosen azimuth-elevation locations.  Some terms are dominated by azimuthal dependence and are plotted versus azimuth while other terms are strongly dependent on elevation only and are plotted with elevation.   Different coating formulas have very different telescope pointings where minimal or maximal values occur.  The sign and even functional form of some Mueller matrix elements can change with coating formula.

\subsection{Mueller Matrix functional dependence summary}

The system Mueller matrices are smooth functions of telescope pointing and wavelength.   Using relatively simple trigonometric functional dependencies, calibration of any data set at an arbitrary telescope pointing is possible.  The errors inherent in projecting from a coarsely sampled azimuth-elevation grid to any arbitrary location are less than or comparable to other errors presented in this paper.  More accurate calibration can be obtained by calibrating the telescope at the actual pointings for a priority target. 

Zemax modeling has been performed to verify the amplitude and basic functional dependence of the system Mueller matrix with azimuth and elevation given common coating formulas.  Good agreement is seen between Zemax predictions and measured amplitudes.  The induced polarization and depolarization terms are consistent with results previously presented for HiVIS \cite{Harrington:2015dl,Harrington:2010km,2008PASP..120...89H,Harrington:2006hu,2014SPIE.9147E..7CH,Harrington:2011fz}.

\section{Other Instrument Limitations}

The HiVIS system also has other limitations to calibration precision arising from instrumental issues.  As an example, during this campaign we discovered an optical misalignment that leads to unstable continuum polarization. As part of the InnoPOL campaign, we designed, built and installed a diffraction limited f/200 laser simulator system \cite{2014SPIE.9147E..7CH}. It was discovered with this laser simulator that the HiVIS fore-optics were delivering the beam to the spectrograph such that the pupil image was being vignetted by the echelle grating. Small changes in the illumination caused by guiding errors and atmospheric seeing caused a variable vignetting that influenced the continuum polarization. This vignetting is likely the source of many continuum polarization instabilities reported in \cite{Harrington:2015dl}.  If the continuum polarization is unstable for point sources, stellar continuum estimates are subject to an additional source of error.  This error is reduced for continuum sources. Vignetting and illumination of the edges of the optics caused some of the scattered light issues reported in \cite{Harrington:2015dl}, degrading the polarization calibration of highly polarized sources.  This was shown as the asymmetry of continuum polarization between charge shuffled beams shown in \cite{Harrington:2015dl}.

\subsection{Intensity to $quv$ cross-talk.}

With such a large data set, we were able to investigate the median intensity to $quv$ cross-talk following a simple cross correlation \cite{Harrington:2015cq}. The $quv$ spectra are known to contain very large continuum variations with measured DoP above 85\%. However, with so many spectra, we can compute the average continuum-subtracted daytime sky polarization spectrum to very high shot noise statistical limits. This allows us as a very sensitive test of the intensity to polarization cross-talk from artifacts in the data reduction pipeline.

\begin{figure} 
\begin{center}
\hbox{
\hspace{-1.5em}
\includegraphics[width=0.49\linewidth, angle=0]{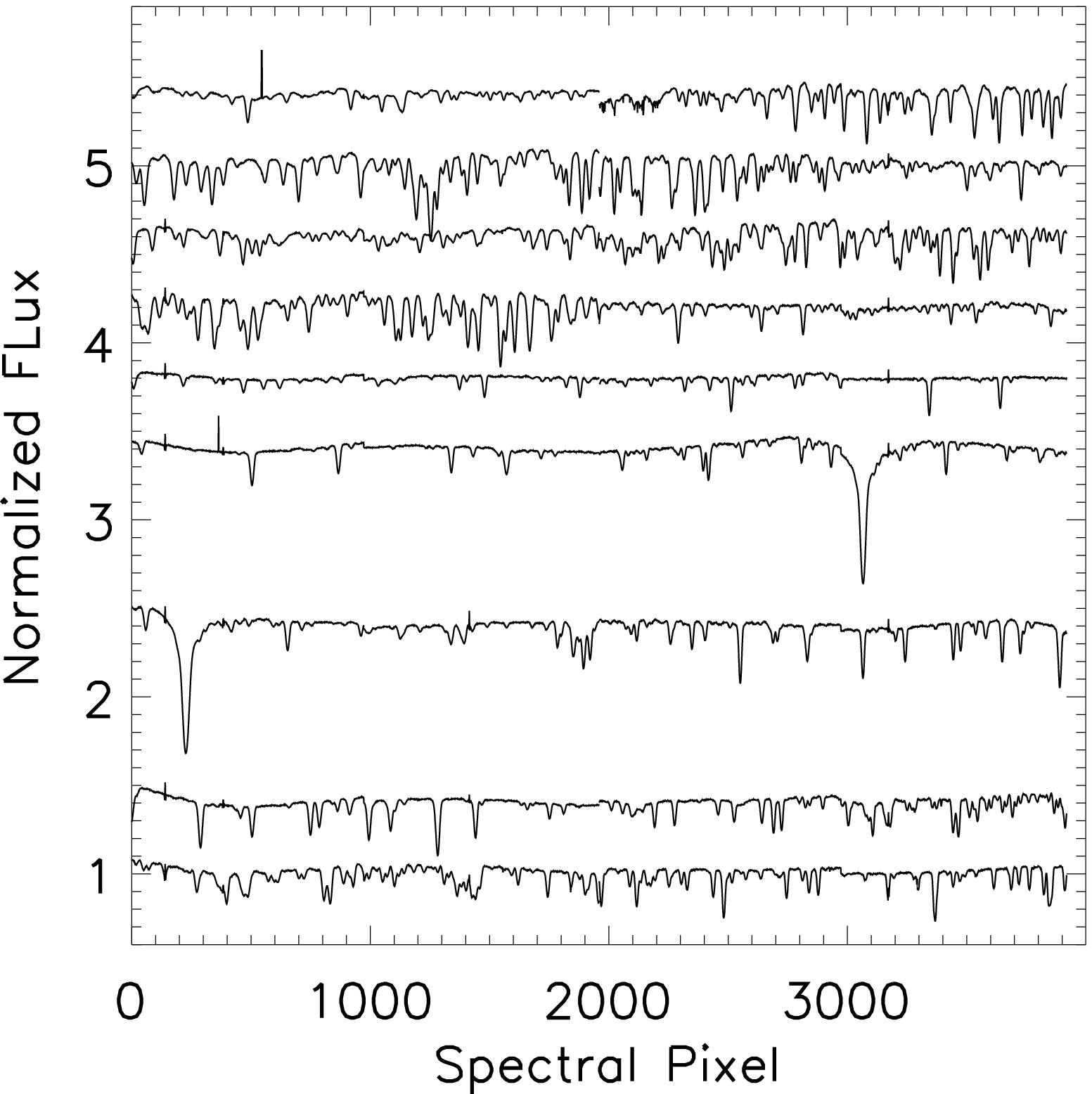}
\hspace{0.5em}
\includegraphics[width=0.49\linewidth, height=0.49\linewidth, angle=0]{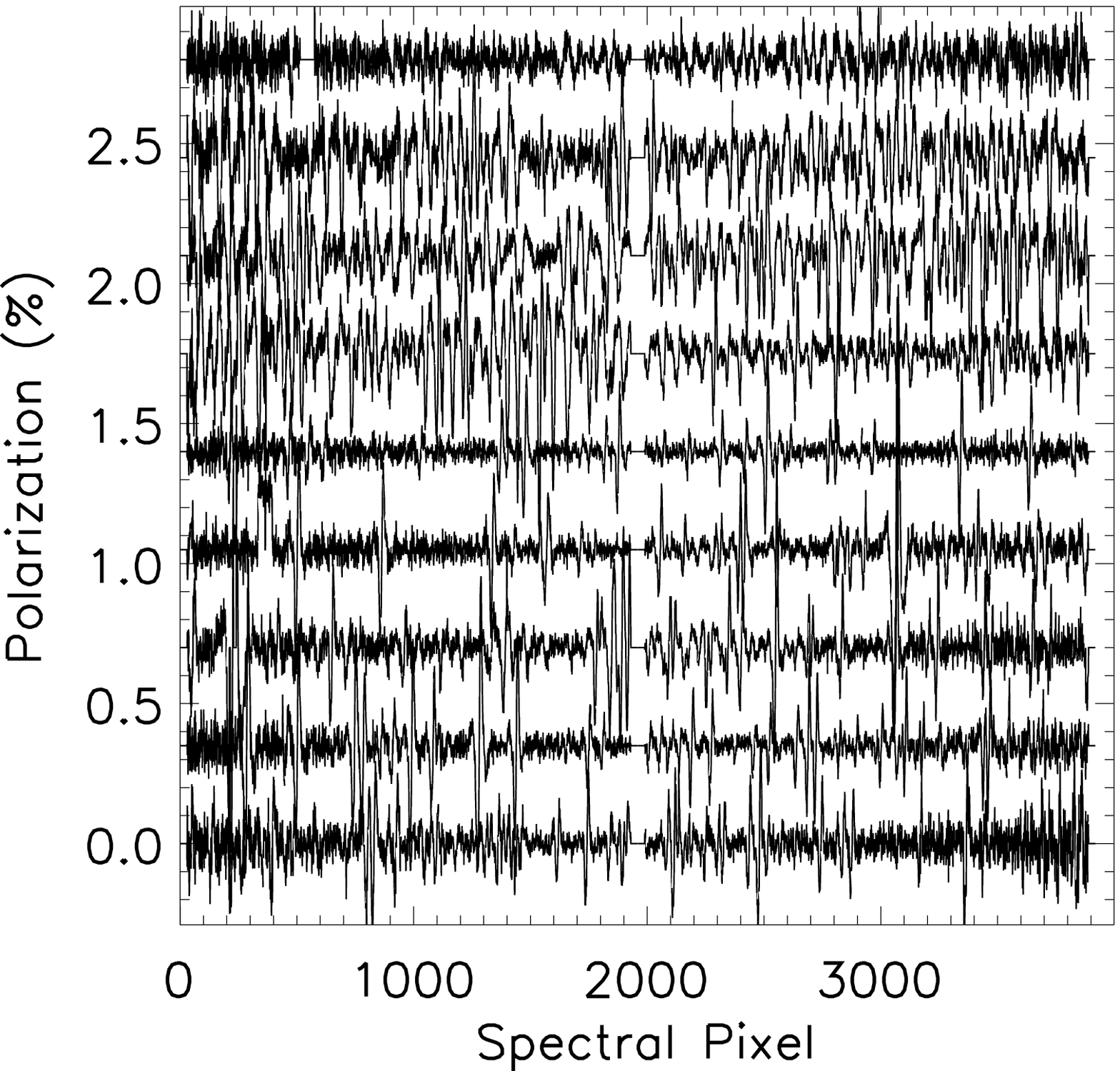}
}
\caption{ \label{hivis_average_iv_normalized} The average intensity and $v$ polarization spectra for May 16th and 17th.  The top panel shows the continuum normalized intensity spectrum filtered by a 60 pixel boxcar smooth to remove low amplitude ripples.  The bottom is Stokes $v$. Select spectral orders are shown with a wide range in number of spectral lines and associated line depths. CCD edges were trimmed (from 2048 pixels down to 2000 pixels per CCD).  The gap in the middle of each spectrum represents the CCD mosaic gap in addition to the 24 pixels trimmed from the edge of each device.  Clear $v$ line polarization spectra are seen in all lines for all displayed wavelengths with amplitudes of 0.1\% to 0.5\%. }
\end{center}
\vspace{-5mm}
\end{figure}

These intensity spectra were normalized by a continuum fitting process. The spectra were median-filtered in wavelength. Note that HiVIS has 2 amplifiers used in reading out each CCD and there are 2 CCDs in the mosaic focal plane.  The median smoothed intensity spectra for each amplifier was fit with an individual polynomial.  The intensity spectra were then divided by these polynomial fits to create continuum-normalized spectra.  There was some additional small level fluctuation with wavelength that was subsequently removed with a 60-pixel boxcar smooth fit.

\begin{wrapfigure}{r}{0.49\textwidth}
\centering
\vspace{-7mm}
\hbox{
\hspace{-0.5em}
\includegraphics[width=0.70\linewidth, angle=90]{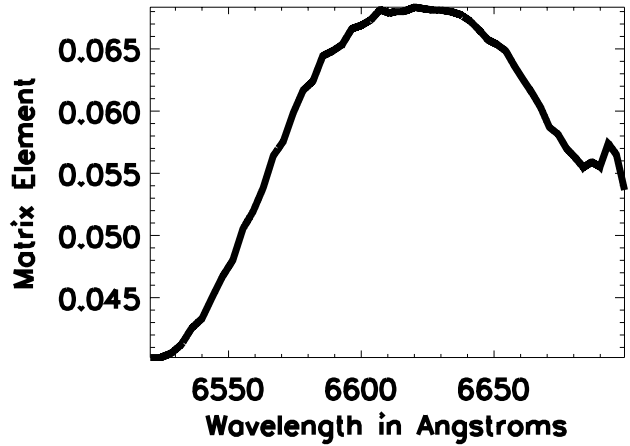}
}
\caption{ \label{epsaru_modmat_order3} The modulation matrix element for the $UU$ term for spectral order 3 at 6620{\AA}. The chromatic variation across the 4000 spectral pixels accounts for variation of 0.2 which is comparable to other error sources presented in this paper.}
\vspace{-10mm}
 \end{wrapfigure}

The corresponding intensity and average continuum subtracted median $v = V/I$ spectra for selected spectral orders are shown in Figure \ref{hivis_average_iv_normalized}. The intensity to polarization cross talk is immediately apparent at levels of 0.1\% to 0.3\% for these continuum measurements. As shown in \cite{Harrington:2015dl}, there is substantial blending between the charge shuffled and modulated exposures when looking at continuum sources with a wider slit length.  We recently installed several new dichroic masks in addition to slits of different lengths and widths to test and overcome these limitations.  The high-sensitivity spatial profiles presented in the appendix of \cite{Harrington:2015dl} show that roughly 15 spatial pixels of separation was required in the old optical configuration to obtain 0.1\% intensity contrast.  An investigation after the new optical alignment may change these numbers substantially.

\subsection{Calibration at reduced spectral sampling}

For this paper, we performed the demodulation and calibration analysis at very low spectral sampling. The HiVIS data were spectrally averaged by 4000x to a single spectral measurement per order. This achieved high SNR for all spectral orders but it does neglect real spectral variation with wavelength.  There are known variations with wavelengths for the demodulation matrices as well as the measured sky polarization across spectral orders that have been neglected for this study.  

Figure \ref{epsaru_modmat_order3} shows the median modulation matrix derived for the $UU$ term for spectral order 3 at 6620{\AA}. The variation across the spectral order is roughly 0.02 in amplitude. Since the calibration process uses only a single point per spectral order, there are potential errors at all wavelengths arising from using the average value when the modulation varies from 0.045 to 0.065 across the order.  Similar amplitude and slowly varying spectral dependence is seen in all terms of the modulation matrix.  As the other errors presented in this paper are of order 0.01 to 0.05, this spectral binning is a contributor to the calibration uncertainties.  Performing calibration at increased spectral sampling should remove this error source.

\subsection{Modulator Stability}

The temporal drifts or other instabilities of an instrument can be a major limitation to any calibration effort. During this campaign, we did not change any HiVIS configurations in order to minimize system drifts. 

\begin{wrapfigure}{r}{0.55\textwidth}
\centering
\vspace{-0mm}
\hbox{
\hspace{-0.5em}
\includegraphics[width=1.\linewidth, angle=0]{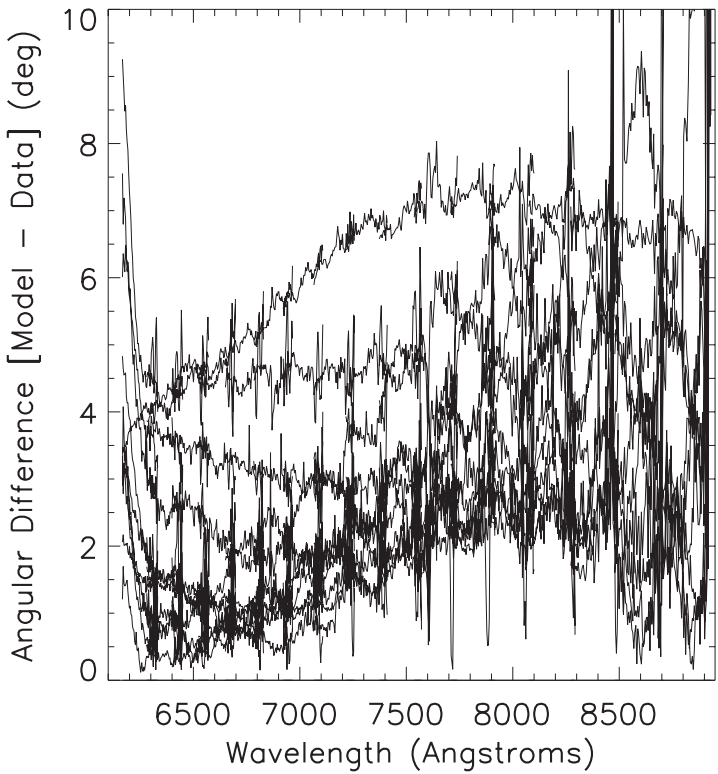}
}
\caption{ \label{angular_residuals_hivis} The angular residual variation between calibrated HiVIS measurements and the computed Rayleigh sky model on the Poincar\'{e} sphere. Calibrations and observations were taken at a telescope azimuth of 180 and elevation of 50. The minimum measured DoP filter of 30\% was applied.  }
\vspace{0mm}
 \end{wrapfigure}

The polarization calibration procedures were followed and standard modulation matrices were recorded. Longer term variations in the system modulation matrix recorded over the campaign showed variations in terms of up to 0.05, but this is compensated by calibrations specific to each night of data collection.  This Figure shows the difference between the average modulation matrix and the modulation matrix derived for the three main observing sessions (October, December, May).  The variations are roughly 0.06 peak to peak for modulation matrices with amplitudes ranging from -1 to +1.

Though we did perform calibrations for every observing run, these liquid crystals are temperature dependent and are known to drift.  Our internal calibrations show the temperature sensitivity \cite{Harrington:2010km}.  Unfortunately, the temperature control of the AEOS coud\'{e} room resulted in temperature changes of up to 5$^\circ$ in some cases.  These drifts certainly contributed to calibration errors similar to those presented in \cite{Harrington:2015dl}.

\subsection{Limitation Summary}

In this section we presented several other instrument calibration limitations specific to HiVIS. The angular error histograms presented \cite{Harrington:2015dl} showed the angle between theoretical sky polarization and the calibrated observations. We can compute the angular error between calibrated daytime sky spectra and the single scattering model for our data set in a similar manner. Figure \ref{angular_residuals_hivis} shows such angular differences with wavelength for a data set at azimuth 180$^\circ$ and elevation 50$^\circ$.

Angular variations of $\sim$6$^\circ$ on the Poincar\'{e} sphere correspond to Mueller matrix element and Stokes vector values of 0.1. There are several errors presented in this section that can create calibration uncertainties of $\sim$0.1 in a Stokes parameter. Liquid crystal temperature instability, reduced spectral sampling and optical misalignment could all contribute to errors in calibrated data.

\section{Summary}

We have presented a 6 month long campaign of 1600 polarized daytime sky spectra to test the algorithms to compute telescope Mueller matrix calibrations.  There are several considerations for planning a calibration observing campaign including the AoP diversity and the functional dependence expected for the telescope Mueller matrix with azimuth and elevation. Plan observations to have a large input polarization angular diversity to provide a well conditioned matrix inversion (least squares solution) and to ensure good estimates for the Mueller matrix elements. We presented several data-based filters necessary to avoid potential problems with data collection, sky model uncertainties, sky polarization variations and instrumental limitations.  

A single scattering model for the linear polarization angle of the daytime sky can be used if calibration data is taken away from regions of naturally low sky polarization. This avoids contamination from multiple scattering as shown in our analysis of a second-order scattering model.  We showed a 2-term scattering model for the sky polarization that consists of a single scattering model plus a constant polarized background from typical multiple scattering sources. The regions of sky 90$^\circ$ from the sun are highly polarized and the AoP is much less sensitive to multiple scattering.  Predictions with large multiple scattering contributions showed that the angle of polarization agreed to better than 0.01 degrees linear polarization rotation in the highly polarized region of the sky near the 90$^\circ$ scattering plane. 

We showed in the appendices that treating a weakly polarizing telescope Mueller matrix as a rotation matrix is insensitive to induced polarization and depolarization terms. By neglecting induced polarization and depolarization, we fit the $quv$ cross-talk elements as a rotation matrix. We showed that this assumption is second-order in small terms from the first row and column of the Mueller matrix.  For our system with many enhanced protected silver coated mirrors, the induced polarization is predicted by Zemax to less than 2\% while cross-talk is 100\%. Our observations of unpolarized standard stars supports this low number, within the limitations of the optical misalignments reported here. 

Several data rejection filters can be applied to ensure quality calibrations.  We reject observations with low measured signal to noise ratio. For this work we used  SNR$>$500 for all $quv$ in spectral order 3 at 6620{\AA} after spectral binning by a factor of 80x to sampling of 50 spectral pixels per order.  We require at least n$>$3 observations to compute the six Mueller matrix element estimates.  We reject observations if the measured DoP is less than 15\%.  For every observation, a calculation of the Rayleigh sky maximum DoP ($\delta_{max}$ ) using the data is required to be above 40\% to guarantee a clear sky.  Telescope pointings where the data set provides a limited range in the input angle of polarization are also rejected ($<$ 20$^\circ$) because the low angular diversity will amplify errors. Iterative filters based on convergence criteria were shown to reject statistical outliers. After initial calibrations are done to the data set, we reject individual outliers where there is a large residual angle on the Poincar\'{e} sphere between the calibrated measurements and the single scattering theory (\textbf{$S_{cal}$} $\cdot$ \textbf{R}) until the group of points converge below a threshold angle (e.g. 25$^\circ$) or the data becomes too sparse to compute a Mueller matrix.

We showed how observations taken on a sparse grid azimuth and elevation telescope pointing combinations can be interpolated on to a continuous function set. By using Zemax optical models of AEOS and HiVIS with representative enhanced protected silver mirror coatings, we can show the expected functional dependence of the Mueller matrix with azimuth and elevation.  The Zemax models show that simple functional dependence is expected and the assumption of weakly polarizing optics is both predicted in Zemax and observed by HiVIS using daytime sky calibrations.  The interpolation from a sparse grid of (azimuth,elevation) measurements adds some errors but with careful planning of calibration observations, the errors from interpolation can be minimized or removed by calibrating along the same pointings as the observations.

\section{Acknowledgements}
Dr. Harrington acknowledge support from the InnoPol grant: SAW-2011-KIS-7 from Leibniz Association, Germany, and by the European Research Council Advanced Grant HotMol (ERC-2011-AdG 291659) during the period 2013-2015 for this work.  Dr. Kuhn acknowledges the NSF-AST DKIST/CryoNIRSP program. This program was partially supported by the Air Force Research Laboratory (AFRL) through partial salary support for Dr. Harrington until October 2015. This work made use of the Dave Fanning and Markwardt IDL libraries.

\appendix    %>>>> this command starts appendixes

\section{Second Order Scattering Model for the Sky}

	We summarize here the mathematics used to expand a scattering model with an additional constant polarization term in addition to the single-scattering Rayleigh term. In the notation of \cite{2004NJPh....6..162B}, they use $\zeta$ to denote the complex location of a point on a stereographic projection of the sky.  In Cartesian geometry, $\zeta$ = x + iy.  In polar coordinates, $\zeta$ = r $e^{i\phi}$.  In \cite{2004NJPh....6..162B}, they use the term $w$ to represent the polarization pattern across the sky.  By breaking the exponential equation in to an amplitude term $|w|$ and a complex orientation term $\gamma(\zeta)$, they represent the stereographic projection for the sky polarization pattern as: $w(\zeta) = |w| e^{2i\gamma(\zeta)}$. For the single scattering case, this simple relation behaves as $\zeta^2$ can be scaled to an amplitude of 1 and written in polar coordinates ($r,\phi$) as: $w(\zeta) \sim \zeta^2 = r^2 e^{2i(\phi - \frac{\pi}{2})}$. 

%
%\begin{equation}
%w(\zeta) = |w| e^{2i\gamma(\zeta)}
%\label{multiscat_general}
%\end{equation}
%

%\begin{equation}
%w(\zeta) \sim \zeta^2 = r^2 e^{2i(\phi - \frac{\pi}{2})}
%\label{multiscat_simple}
%\end{equation}

In order to add multiple scattering to this equation, we must consider the shift of the zero polarization points away from the solar and anti-solar location. These zero points are Brewster and Babinet points near the sun as well as the Arago and second Brewster point near the anti-solar location.  Several empirical results show that the singularities are found above and below the sun along the solar meridian.  This generally follows from the empirical result that {\it double scattering} is the dominant contribution to multiple scattering in the typical locations surveyed. This double-scattering contribution is generally polarized in the vertical direction as it represents the light scattered in to the line of sight from the integrated skylight incident on all points along the line of sight.  When the sun is low in the horizon, the low DoP regions of the sky are also low on the horizon.  This double scattering contribution is of the same amplitude as the single scattered light when the single scattered light is weak and horizontally polarized, which occurs above and below the sun at low solar elevations during sunrise and sunset. 

The most simple perturbation to the model is to add a constant representing a small additional polarization of assumed constant orientation denoted $A$.   Following \cite{2004NJPh....6..162B}, the zero polarization singularities fall at the locations of $\zeta = \pm iA$ which corresponds to a Cartesian y value of $\pm A$.  To make the singularities at the anti-sun location, the equation was generalized to:
	
\begin{equation}
w(\zeta) \sim (\zeta^2 + A^2) (\zeta^2 + \frac{1}{A^2})
\label{multiscat_add_constant}
\end{equation}

The \cite{2004NJPh....6..162B} notation showed the four zero polarization singularities as simple functions of the solar position and the constant $A$.  If you denote the solar elevation as $\alpha$ and use the stereographic projection where the sun is on the y axis at the location $\zeta$ = $iy_s$ = $i(1-tan\frac{\alpha}{2})(1+tan\frac{\alpha}{2})$, the four zero polarization singularities are located:
	
\begin{equation}
(1)\ \zeta_+ = i\frac{y_s+A}{1-Ay_s} \quad   (2)\   \zeta_-= i\frac{y_s - A}{1+Ay_s} \quad  (3)\   \frac{-1}{\zeta_+^\ast} \ \   (4) \ \  \frac{-1}{\zeta_-^\ast}
\label{multiscat_singularities}
\end{equation}

To make the polarization function symmetric across the sky (antipodally invariant) and to scale the DoP to 100\%, the polarization equation can be normalized and written in terms of these singularity locations as:

\begin{equation}
w(\zeta) = -4 \frac{ (\zeta - \zeta_+)(\zeta-\zeta_-)(\zeta+\frac{1}{\zeta_+^\ast})(\zeta+\frac{1}{\zeta_-^\ast})} 
                               {(1+r^2)^2 |\zeta_+ + \frac{1}{\zeta_+^\ast}| |\zeta_- + \frac{1}{\zeta_-^\ast}| }
\label{multiscat_add_constant}
\end{equation}

The denominator was chosen with the complex modulus terms to ensure the amplitude $|w|$ is always 1.  With this simple equation, you can relate the constant $A$ to the angular separation of the polarization zero points as $\delta$ = 4 ATAN($A$).

\section{Details of Data Set Filters}

We outline in this section details of some of the filters applied to the HiVIS data set as we enforce consistency and quality across the daytime sky observing campaign.

\subsection{Filter: Input Angular Diversity Threshold}

	This computational algorithm relies on the assumption that we can solve for a set of rotation matrix angles to match the input data with the Rayleigh sky model through a least squares process.  If the input data lacks enough angular diversity, the least squares solution becomes badly conditioned leading to very large noise propagation errors. We require that all (Azimuth, Elevation) grid points in the computation have sufficient range of input polarization angles (large angular diversity). We find that a minimum of $\sim$20$^\circ$ angular diversity is required to give a well conditioned solution.  
	
	The high elevation telescope pointings near the zenith only have 30$^\circ$ of input diversity in our survey largely because the telescope cannot observe the Zenith while the sun is high above the horizon. Given the nearly East-West oriented rise and set azimuths for the sun during the summer, there is not much diversity in the input polarization angles along East-West telescope azimuths. For telescope pointings at low elevations in the North, the telescope sees a large input AoP diversity for all seasons. Given the winter data set with the sun rotating through all azimuths lower in the South, the input diversity is at least 3x higher at these telescope pointings.

%As an example, Figure \ref{angular_diversity_solang_dop} shows the angular diversity for the data set after several of the previous filters have been applied.  
%
%\begin{wrapfigure}{r}{0.55\textwidth}
%\centering
%\vspace{-5mm}
%\hbox{
%\hspace{-0.5em}
%\includegraphics[width=0.80\linewidth, angle=90]{HIVIS_DAYSKY_2013_AZEL_COVERAGE_Filtered_MinMax_SolDis_Demod0_DegPol20.eps}
%}
%\caption{ \label{angular_diversity_solang_dop} The sky linear polarization angle diversity at each telescope pointing. The grid pattern comes from the chosen sampling of (azimuth,elevation) combinations.  Only points observed are shown.  Angular diversity is computed as the range of polarization angles across all observations at a single telescope pointing. The angular diversity goes from zero to over 80$^\circ$ rotation of the polarization angle for some telescope pointings. The input AoP angular diversity color scale is shown in degrees AoP at right. Greater angular diversity improves the fitting processes used to compute each Mueller matrix element estimate. The {\it measured DoP} filter has been applied at a threshold of 20\% which reduces the angular diversity of the data set but rejects most bad data sets and reduces errors from the single-scattering model imperfections.  }
%\vspace{-0mm}
% \end{wrapfigure}

\subsection{Filter: Consistency and convergence of Matrix Element Estimates}

\begin{figure} [ht]
\begin{center}
\includegraphics[width=0.36\linewidth, angle=90]{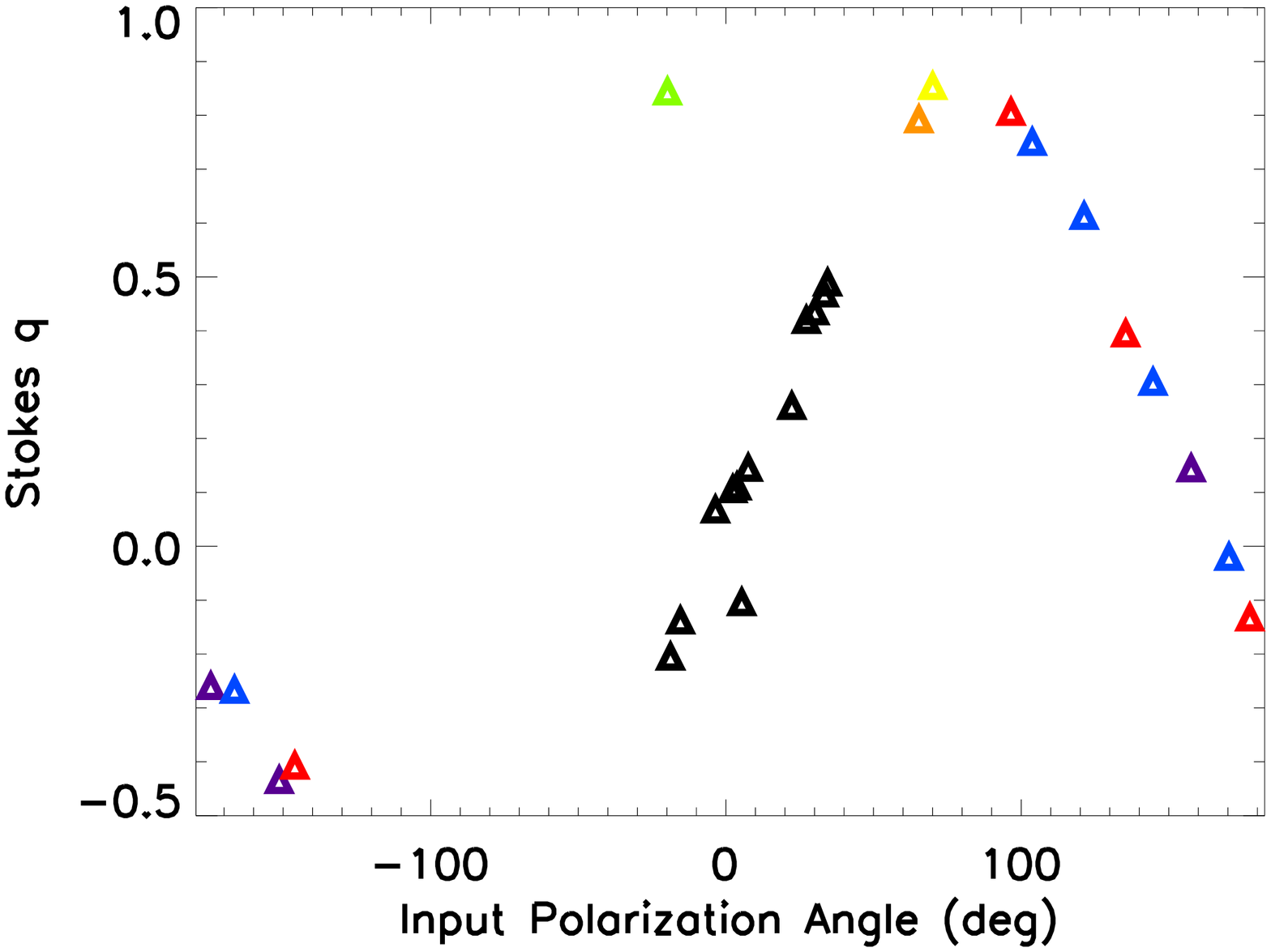}
\includegraphics[width=0.36\linewidth, angle=90]{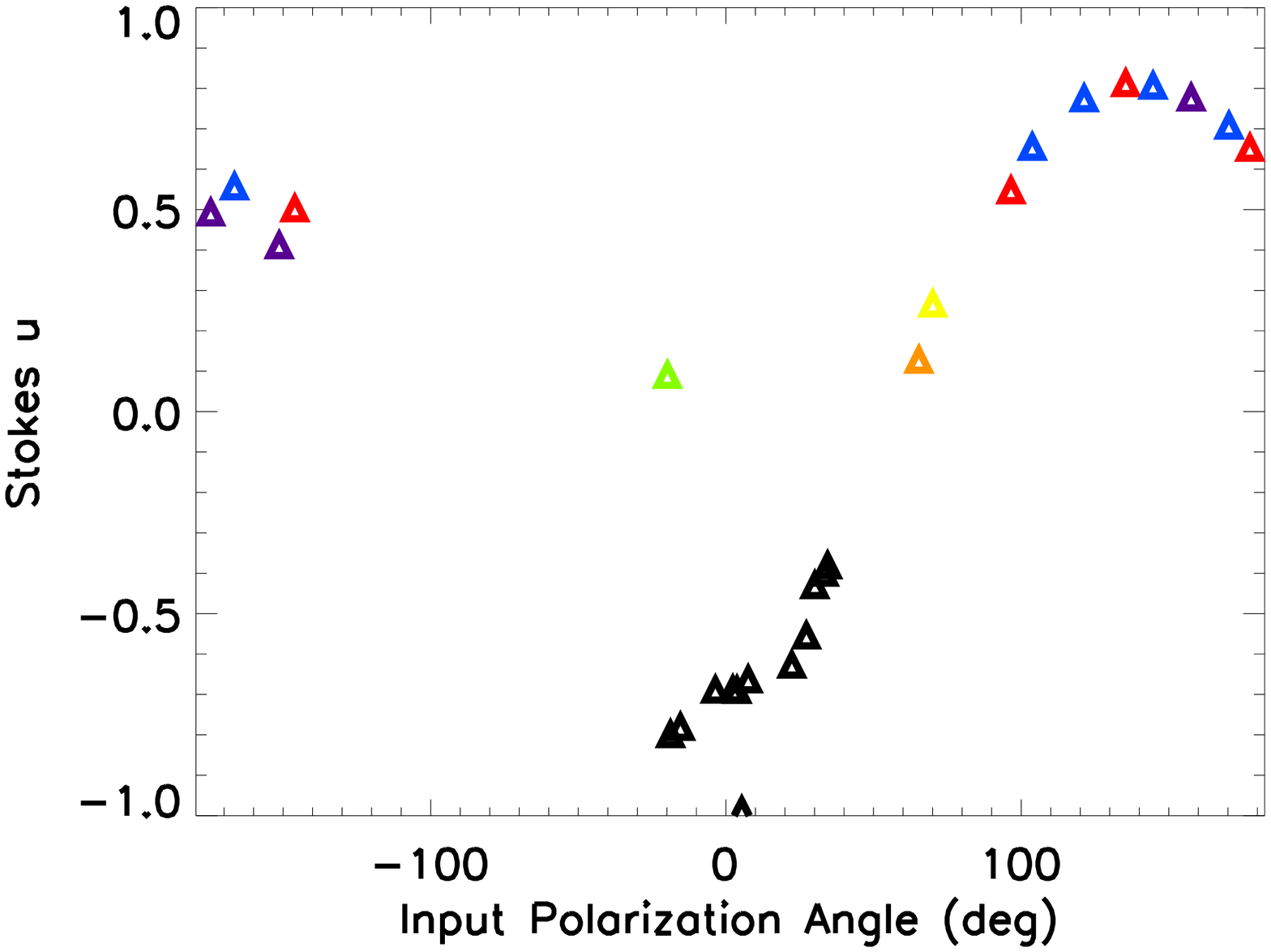}
\includegraphics[width=0.36\linewidth, angle=90]{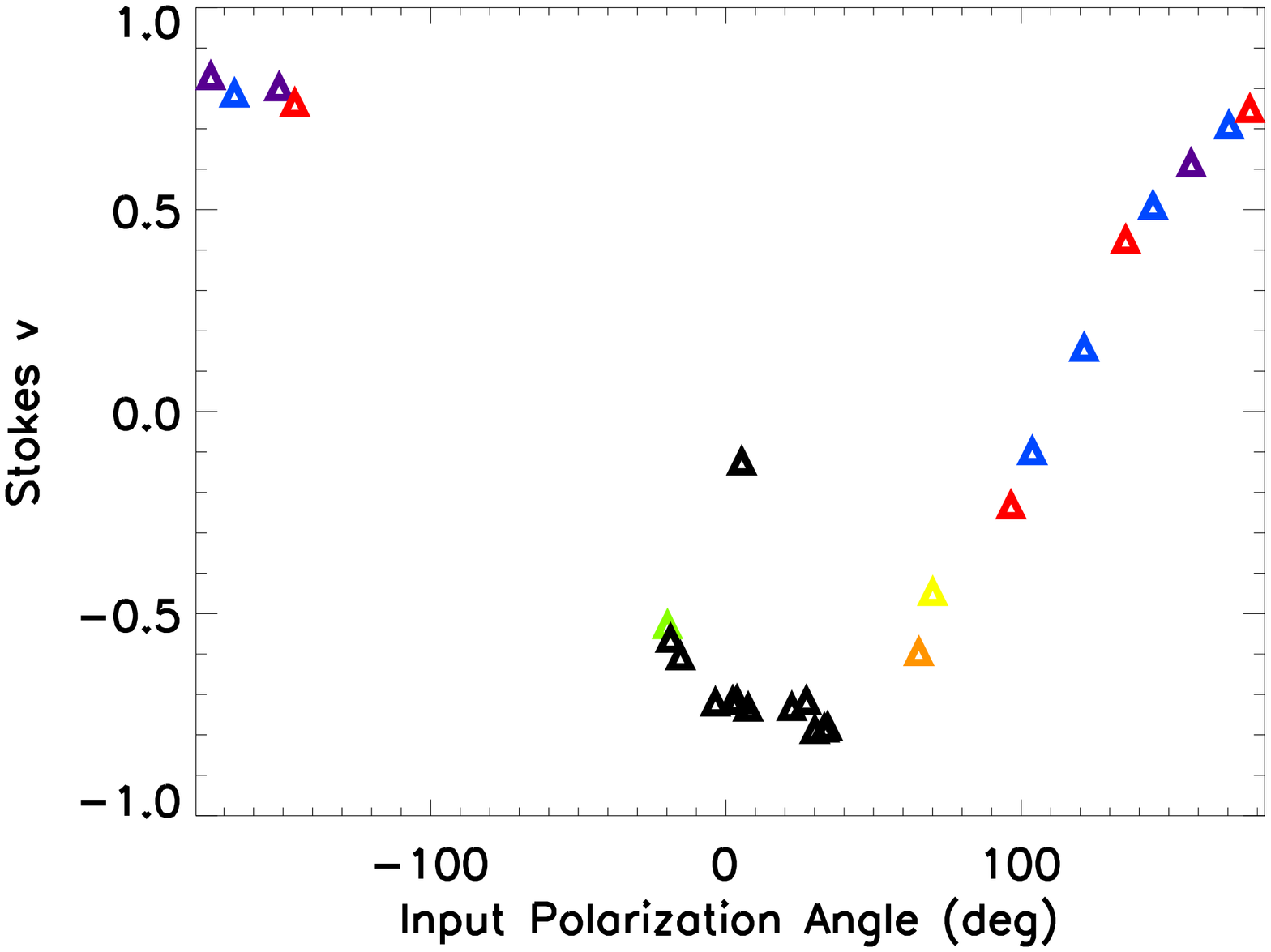}
\includegraphics[width=0.36\linewidth, angle=90]{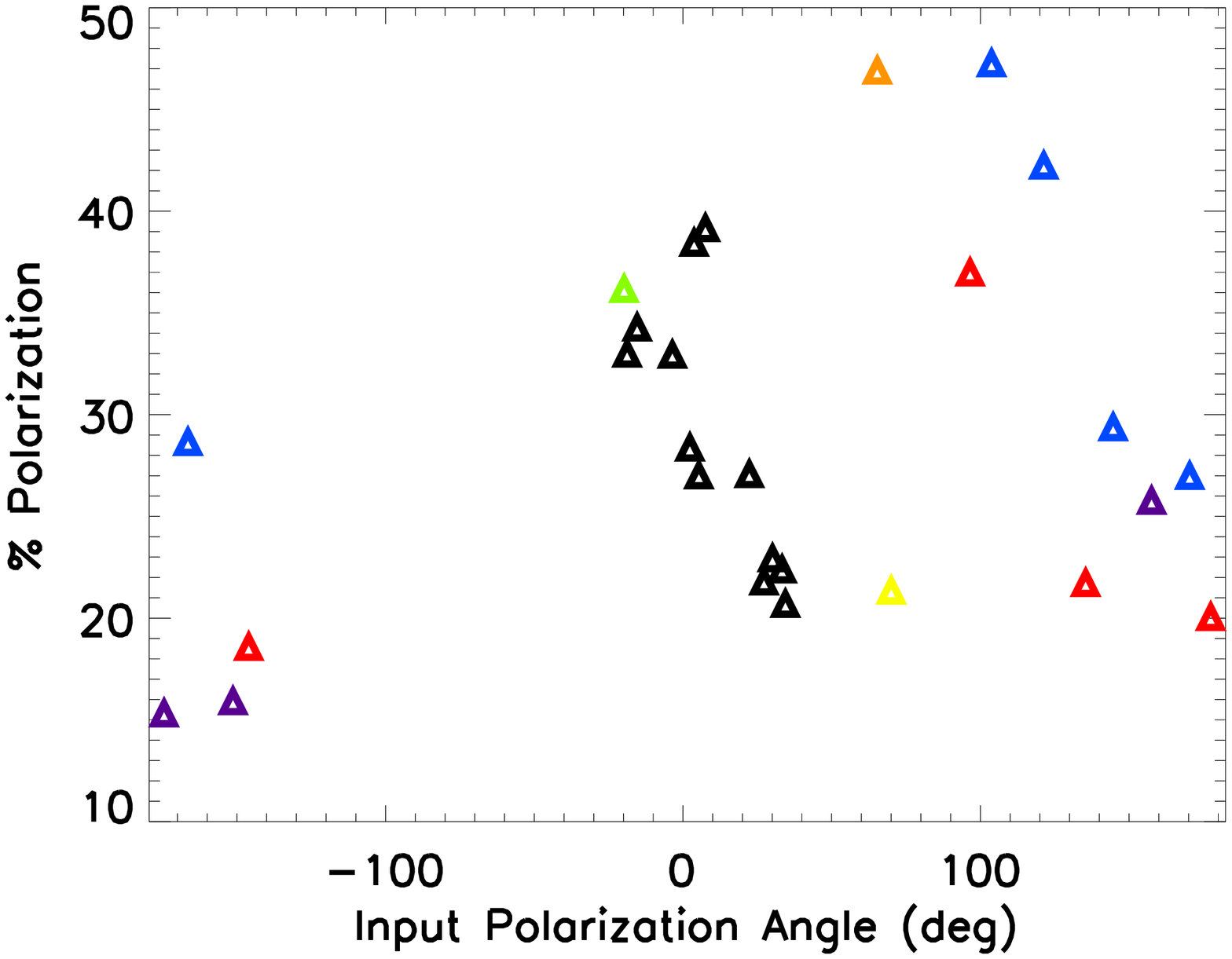}
\caption{ \label{stokes_vs_scattering_angle_060_25} The measured data over the entire campaign for a telescope pointing of azimuth 60$^\circ$, elevation 25$^\circ$. Stokes $quv$ parameters have been normalized by the measured DoP to create 100\% polarized $quv$ points.  The measured DoP used to do the normalization is shown in the lower right hand panel. All points are shown as functions of the input polarization angle. May data is shown in black.  October data is orange.  November data is red.  December data is blue.    Over the October to May timeframe, the sun moves across the sky and generates a wide range of input linear polarization angles (plotted as the x axis).  There is a direct functional relationship between the input polarization angle and the output $quv$ data. The smooth functional relationship (with some outliers to be rejected) illustrates the mapping between input $qu$ angles and output $quv$ angles on the Poincar\'{e} sphere. The demodulation and 100\% DoP scaling has been applied. The $<$10\% measured DoP filter has been applied to reject low DoP points.}
\end{center}
\end{figure}

By comparing the agreement between individual calibrated measurements and the least-squares solution for the calibration itself, we can identify outliers and reject them from the data set. We call this filter {\it consistency} because it checks the agreement between each individual measurement and the resulting average over all measurements. Statistical outliers can be rejected given a threshold. One would expect smooth variation in the rotation of the polarization caused by the telescope induced cross-talk. The measured $quv$ parameters should be a smooth functions of input polarization angle. As an example, the measured (demodulated, 100\% scaled) Stokes parameters ($quv$) are shown in Figure \ref{stokes_vs_scattering_angle_060_25} for a telescope pointing of azimuth of 60$^\circ$, elevation 25$^\circ$. There is strong rotation of the measured cross-talk as the input polarization angle varies with solar location throughout the year.

The angular rotation in $quv$ space between the initial input Stokes vector and subsequent measured Stokes vectors is a useful test of the Mueller-matrix-as-a-rotation-matrix assumption. Assuming the cross-talk is confined to a rotation matrix uninfluenced by the induced polarization or depolarization as shown by our sensitivity analysis, the angular rotation between subsequent measurements should also match the angular rotation of the modeled Rayleigh sky $qu$ inputs. In most cases, the modeled angular rotation in $quv$ space matches the measured angular rotation with errors of less than 10$^\circ$.

%As an example, Figure \ref{input_rotation_with_time_residual} shows the differences between measured Stokes vectors and modeled Stokes vectors in angle on the Poincar\'{e} sphere for a telescope pointing of azimuth 060, elevation 25. 
%
%
%\begin{figure} 
%\begin{center}
%\hbox{
%\hspace{-2.0em}
%\includegraphics[width=0.80\linewidth, angle=90]{rotation_with_time_residual_azi060_el25.eps}
%}
%\caption{ \label{input_rotation_with_time_residual} The difference between rotation of the input Rayleigh sky model Stokes vectors and the rotation of the measured Stokes vectors (demodulated, 100\% scaled to the Poincar\'{e} sphere). The telescope should preserve angles between subsequent measurements in the approximation of the $quv$ cross-talk as a rotation matrix.  The first 12 data points cluster about +10$^\circ$ degrees rotation and correspond to October-December data sets.  There is one outlier showing 60$^\circ$ rotation which will be subsequently rejected by the various filters.  Data sets 13 to 27 cluster about -10$^\circ$ degrees rotation angle and corresponds to data collected in May.  The large angular difference between winter and summer data sets shows drift in the telescope calibrations.  }
%\end{center}
%\end{figure}

\subsection{Filter: Iterative Filtering and Rotation Matrix solution consistency}

As another independent assessment of the data, we can ensure that the individual data points used to create Mueller matrix estimates give consistent results and are not statistical outliers compared to the average.  We can apply a data filter by requiring that the derived rotation matrix calibrate each individual measurement to within some threshold tolerance of the average. In order to check the consistency of the telescope Mueller matrix solution, we create an iterative process. We apply the above processes of fitting the telescope Mueller matrix elements and calibrating all individual polarization measurements. Once we derive an initial set of calibrations, we apply the calibration to all individual measurements. With calibrated measurements, we can check the differences between each calibrated individual measurement and the associated Rayleigh sky model. If these differences are above a threshold, we can identify and reject the data as an outlier. An iterative process was written to follow these steps:

\begin{itemize}
\item 1 - Compute Mueller matrix estimates from all measurements. 
\item 2 - Fit for Euler angles ${\bf \xi}_{i}(\alpha, \beta, \gamma)$
\item 3 - Compute rotation matrix elements for telescope Mueller matrix approximation
\item 4 - Calibrate all data points used to compute Mueller matrix estimates
\item 5 - Compute angular difference ( \textbf{S} $\cdot$ \textbf{R} ) between all calibrated data and predicted Rayleigh sky.
\item 6 - Check if all data points are below a threshold value for angular differences  (Convergence?)
\item 7 - IF no points are rejected, STOP
\item 8 - IF points are above the threshold, reject the data point with highest angular difference (nominally 25$^\circ$)
\item REPEAT all steps with newly filtered data set until convergence criteria is obtained.
\end{itemize}

Outliers are rejected and the process is repeated until convergence criteria are met.  Note that in this formalism, the Rayleigh sky model and the measured demodulated, scaled Stokes parameters both have 100\% DoP and are by definition vectors with a length of 1.  We can calculate the angle between the calibrated measurements and the model sky polarization at each measurement ($i$) with a simple dot product as: ${\bf \Theta}_{i} = ACOS( {\bf R}_{i} \cdot {\bf S}_{i} )$.

The first steps in computing Mueller matrix elements were shown above in Figure \ref{mueller_matrix_estimates_raw}.  These Mueller matrix estimates are then fit to a rotation matrix via the least squares process.  These rotation matrices are used to calibrate all individual $quv$ measurements. The rotation error between these calibrated measurements and the Rayleigh sky is computed.  As shown in Figure 14 of \cite{Harrington:2015dl}, the fitting of the Mueller matrix elements to a rotation matrix results in some differences.  The six individual Mueller matrix element estimates here are typically within 0.1 of the final rotation matrix fit values.  Depending on the number of points, filters, input polarization angular diversity and other factors, the cumulative error distribution functions show that 80\% of the Mueller matrix estimates are between 0.03 and 0.1 of the final fit value.  
%
%\begin{equation}
%{\bf \Theta}_{i} = ACOS( {\bf R}_{i} \cdot {\bf S}_{i} )
%\label{eqn_rotmat_multout}
%\end{equation}

\subsection{Uncertainty in optical window}

\begin{wrapfigure}{r}{0.55\textwidth}
\centering
\vspace{-0mm}
\hbox{
\hspace{-0.5em}
\includegraphics[width=0.98\linewidth, angle=0]{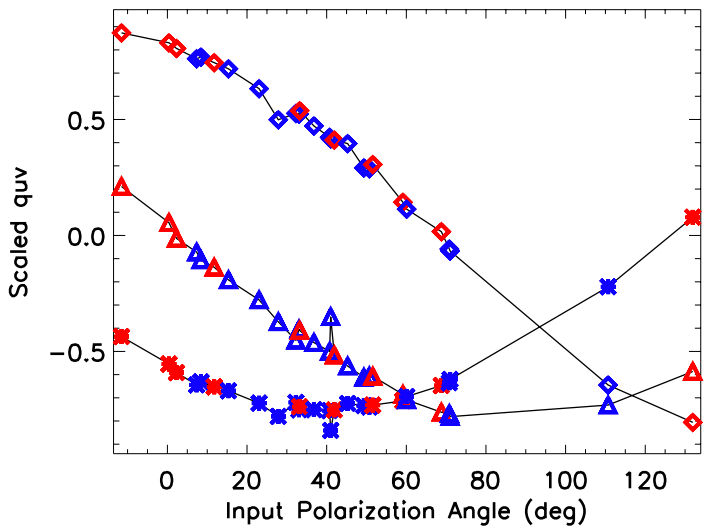}
}
\caption{ \label{quv_vs_solang_window} The measured sky $quv$ spectra as functions of input AoP for different coud\'{e} windows (BK7 and Infrasil). Blue symbols show the Infrasil window, red symbols are for the BK7 window substrate. The telescope was at a pointing of azimuth 330$^\circ$ elevation 20$^\circ$. Similar results are seen at other telescope pointings. The diamonds show measured Stokes $q$, the triangles show measured Stokes $u$ and the asterisks show measured Stokes $v$.  All points have been scaled to 100\% DoP and have been demodulated using the $+$ state. Observations have been minimally filtered using a minimum 15\% measured DoP threshold and a 30\% minimum $\delta_{max}$ filter to guarantee a reasonably polarized sky signal (rejecting clouds, obvious outliers or highly contaminated data sets). The measured $quv$ points are consistent between different window substrates, different times and different input polarization angles. We consider the two windows to not limit the calibration precision for this study. }
\vspace{-5mm}
 \end{wrapfigure}

	We have two different window options (BK7 and Infrasil).  As outlined in the schematic of \cite{Harrington:2015dl}, the telescope has a wheel for different window substrates that separate the coud\'{e} lab from the telescope optical feed. The windows are mounted in between the 6th and 7th mirrors in the optical train.  The windows are permanently mounted in the wheel and are in the vertical orientation as the incoming f/200 beam travels downward.  These two windows were used on alternating days. The AEOS operators had a mechanical issue discovered only after our October observing run.  This issue with the window wheel caused uncertainty on which window was used on which day in October. A new motor and computer controlled wheel rotation was installed between December and May runs.   We intentionally tested both windows.  With operator confusion and uncertainty for the observing run and we will treat the window parameters as unknown in our October measurements.  Certainly the window was fixed in one location on a particular day but it could have changed configurations between observing days. 

Several observing runs were done to assess the impact of changing windows from the BK7 window to the Infrasil window. Data was collected on several days often interchanging windows.  Figure \ref{quv_vs_solang_window} shows the measured $quv$ spectra taken in May where blue represents Infrasil and red represents BK7. Some data points show deviations from the smooth curves, but no systematic difference between window substrates is seen.  Outliers from the smooth curves are likely caused by a variety of observing-related issues and can easily removed via data filters.

\section{Sensitivity to assuming polarization preservation: a perturbation analysis}

Another assumption of this method is neglecting consideration of the first row and first column of the Mueller matrix. These intensity to / from $QUV$ terms can introduce errors. In this section, we show a simple sensitivity analysis and demonstrate that neglecting these terms is a good assumption. The measured induced polarization and depolarization is low, giving us some confidence in this assumption. For high reflectivity mirrors with low diattenuation, as our mirrors are, the predicted Mueller matrix first row and column are typically low. Cross-talk values are 100\% while induced polarization is less than 1\% to 3\%. Our method of fitting the 3x3 cross-talk elements of a Mueller matrix with a Rotation matrix is relatively robust against errors in the first row and column of the Mueller matrix. The analysis below will show that the 3x3 cross-talk elements of the Mueller matrix are minimally impacted to {\it second order} when including non-zero elements in the first row and column of the Mueller matrix.

A Mueller matrix is not a random combination of 16 numbers arranged in a square. A Mueller matrix is a group of numbers that must transform a Stokes vector into another Stokes vector through multiplication according to a set of rules that preserves certain properties of the vector. This means that any Mueller matrix is a transformation matrix that must belong to a group of transformations and must behave according to a set of rules. For instance, a familiar constraint would be that the Stokes vector cannot have greater than 100\% polarization expressed as $I^2-Q^2-U^2-V^2 > 0$.  This set of rules means Mueller matrixes are part of a group, formally called SO(1,3)$^+$ as part of the orthochronous Lorentz group. This group describes a set of functions that transform Stokes vectors in a given 4 dimensional space. Conveniently, it also provides for many conditions that a Mueller matrix must obey in order to be physically realizable. It also provides specific frameworks (technically, a Lie algebra) for the convenient mathematical analysis of the Mueller matrix, since you can generate any matrix as the exponential of a few simple constants multiplied by what are typically called {\it generators}.  These generators are simple matrices  that can be multiplied by a constant (for instance, a rotation angle) to define a physically realizable Mueller matrix.  With these generators, we can do error propagation and we can test for the sensitivity of any approximation to small perturbations. In our method, we write the Mueller matrix in terms of the Euler angles ($\alpha, \beta, \gamma$) and the {\it generators} used to create the rotation matrix. In our ZXZ convention for Euler angles, we rotate first about the Z axis, then about the X axis then about the Z axis in the rotated coordinates. To compute the Mueller matrix via exponentials, we use the generators for rotations about the Z axis and about the X axis. 
%
%\begin{equation}
%\label{mm_exponentiated}
%M = exp(\alpha S_1 + \beta S_2 + \gamma S_3) = exp(\Sigma d_i S_i)
%\end{equation}

We use the standard exponential notation to describe a Mueller matrix as an exponential of the three Euler angles ($d_i$) for i=1,2,3: $M = exp(\alpha S_1 + \beta S_2 + \gamma S_3) = exp(\Sigma d_i S_i)$. The S matrices describe the standard infinitesimal generators for rotation for 1=Z, 2=X, 3=Z with the ZXZ convention.  The Z rotation for $S_1$ is often denoted $J_3$ and is: 

\begin{equation}
\label{mm_form}
{\bf S}_{i} = J_3 = 
 \left ( \begin{array}{rrrr}
 0  		& 0	& 0		& 0		\\
 0	 	& 0	& -1		& 0		\\
 0 		& 1 	& 0		& 0		\\
 0 		& 0	& 0		& 0		\\ 
\end{array} \right )
\end{equation}

If we write a matrix using this notation for a rotation about the Z axis by and Euler angle $\alpha$ we recover a rotation matrix which is also the standard form for a circular retarder with a retardance $\alpha$ via $exp(\alpha J_3)$ :

\begin{equation}
\label{mm_exponentiated}
M = exp( \alpha 
 \left [ \begin{array}{rrrr}
 0  		& 0	& 0		& 0		\\
 0	 	& 0	& -1		& 0		\\
 0 		& 1 	& 0		& 0		\\
 0 		& 0	& 0		& 0		\\ 
\end{array} \right ] ) = 
 \left ( \begin{array}{rrrr}
 1  		& 0			& 0			& 0		\\
 0	 	& c_\alpha	& -s_\alpha	& 0		\\
 0 		& s_\alpha 	& c_\alpha	& 0		\\
 0 		& 0			& 0			& 1		\\ 
\end{array} \right ) 
\end{equation}

The standard infinitesimal generator for a X rotation is denoted called $J_1$ and if we rotate by an angle $\beta$, we get a Mueller matrix for a linear retarder via $exp(\beta J_1)$: 

\begin{equation}
\label{mm_form}
{\bf M}_{ij} = exp(\beta  
 \left [ \begin{array}{rrrr}
 0  		& 0	& 0		& 0		\\
 0	 	& 0	& 0		& 0		\\
 0 		& 0 	& 0		& -1		\\
 0 		& 0	& 1		& 0		\\ 
\end{array} \right ]  ) = 
\left ( \begin{array}{rrrr}
 1  		& 0			& 0			& 0			\\
 0 		& 1			& 0			& 0			\\ 
 0	 	& 0			& c_\beta		& -s_\beta	\\
 0 		& 0 			& s_\beta	 	& c_\beta	\\
\end{array} \right ) 
\end{equation}

We can use the same exponential notation for generating $IQ$ and $QI$ terms of the Mueller matrix using the generator $K_1$ and a small diattenuation term $\epsilon$. The matrix would be computed as $exp(\epsilon K_1)$  For this Mueller matrix, we take the approximation that $sinh(\epsilon)\sim\epsilon$ and $cosh(\epsilon)\sim1$ so that the Mueller matrix simplifies. 
 
\begin{equation}
\label{mm_form}
%\begin{split}
{\bf M}_{ij} = exp(\epsilon
 \left [ \begin{array}{rrrr}
 0  	& 1	& 0	& 0		\\
 1	& 0	& 0	& 0		\\
 0 	& 0 	& 0	& 0		\\
 0 	& 0	& 0	& 0		\\ 
\end{array} \right ]) = \\
\left ( \begin{array}{rrrr}
cosh(\epsilon)  		& sinh(\epsilon)		& 0	& 0		\\
sinh( \epsilon)		& cosh(\epsilon)	& 0	& 0		\\
 0 				& 0 				& 1	& 0		\\
 0 				& 0				& 0	& 1		\\ 
\end{array} \right )  = 
\left ( \begin{array}{rrrr}
 1  		& \epsilon	& 0	& 0		\\
 \epsilon	& 1		& 0	& 0		\\
 0 		& 0 		& 1	& 0		\\
 0 		& 0		& 0	& 1		\\ 
\end{array} \right ) 
%\end{split}
\end{equation}

The arbitrary form for the generators can be represented by the terms applicable to the first row and column of the Mueller matrix, $K_i$ and the terms corresponding to rotations, $J_i$.  The full set of generators can be written with a set of infinitessimal operators (often called {\it boosts}) in $QUV$ denoted as $\zeta_i$ and a set of {\it rotations} in $QUV$ with rotations denoted as $\theta_i$:

\begin{equation}
\label{lorentz_generators}
%\begin{split}
{\bf M}_{ij} = exp(\zeta_x K_1 + \zeta_y K_2 + \zeta_z K_3 + \theta_x J_1 + \theta_y J_2 + \theta_z J_3) = \\
exp(
 \left [ \begin{array}{rrrr}
 0  		& \zeta_x		& \zeta_y		& \zeta_z		\\
 \zeta_x	& 0			& -\theta_z	& \theta_y		\\
 \zeta_y 	& \theta_z 	& 0			& -\theta_x	\\
 \zeta_z 	& -\theta_y	& \theta_x		& 0			\\ 
\end{array} \right ])
%\end{split}
\end{equation}

%
%\begin{equation}
%\label{mm_exponentiated}
%{\bf M}_{ij} = exp(\epsilon K_1 + \alpha S_1 + \beta S_2 + \gamma S_3) = exp(\epsilon K_1 + \Sigma d_i S_i)
%\end{equation}

We can show how errors in the first row and column of the Mueller matrix propagate in to cross-talk elements for $quv$ to $quv$.  Consider a Mueller matrix of the form: ${\bf M}_{ij} = exp(\epsilon K_1 + \alpha S_1 + \beta S_2 + \gamma S_3) = exp(\epsilon K_1 + \Sigma d_i S_i)$. Using the infinitessimal generators outlined above, and the rotation matrix {\bf $\mathbb{R}_{ij}$} as defined in Equation 3 in the main text, we get a Mueller matrix: 

\begin{equation}
\label{mm_form}
{\bf M}_{ij} =
 \left ( \begin{array}{rrrr}
 1   		& \epsilon	& 0		& 0		\\
 \epsilon 	& R_{11}	& R_{21}	& R_{31}	\\
 0 		& R_{12}	& R_{22}	& R_{32}	\\
 0 		& R_{13}	& R_{23}	& R_{33}	\\ 
\end{array} \right ) 
\end{equation}

Each element ${\bf \mathbb{R}}_{ij}$ is part of the rotation matrix and $\epsilon$ is a small error caused by dichroism in the system. We will neglect the other dichroism terms ($IU$ and $IV$) for simplicity. In the limit of small $\epsilon$, we can derive a sensitivity of the ${\bf \mathbb{R}}_{ij}$ elements to $\epsilon$. Note that for AEOS, the ${\bf \mathbb{R}}_{ij}$ terms are as large as 1 while the induced polarization and depolarization terms $\epsilon$ are $<$0.05.  

Note that the Euler angles do not commute and we have an equation for the Mueller matrix as an exponential of a sum of terms representing the rotation matrices. We can use the Zassenhaus formula which expands non-commuting exponential functions of sums to an infinite series of terms, similar to other familiar expansions. The formula below includes the first of the additional terms and represents the expansion: $exp(X + Y) =  exp(X) \  exp( Y) \  exp(\frac{1}{2}[X,Y])$. Additional terms grow complex quickly but have small amplitudes. The next correction to this Equation involves nested commuting:  $\frac{1}{12}$ [X,[X,Y]] + [Y,[Y,X]].   For our case, the non-commuting terms would be $[X,Y] =  [K_1, S_i]$. We write our Mueller matrix where we denote the sum of non-commuting rotations $exp(\Sigma d_i S_i )$ with $S_i$ as the ZXZ rotation group as a single matrix ${\bf \mathbb{R}}_{ij}$ of the three Euler angles $d_i = (\alpha,\beta,\gamma)$ for i=1,2,3.

%\begin{equation}
%\label{mm_exponentiated_approx}
%exp(X + Y) =  exp(X) \  exp( Y) \  exp(\frac{1}{2}[X,Y]) 
%\end{equation}

\begin{equation}
{\bf M}_{ij} = exp(\epsilon K_1) \   {\bf \mathbb{R}(\alpha,\beta,\gamma})_{ij} \  exp(\frac{1}{2}[\epsilon K_1, \Sigma d_i S_i])
\label{eqn_three_terms}
\end{equation}

This gives us three terms multiplied to create the Mueller matrix. Since $\epsilon$ is small, we approximate the first term $exp(\epsilon K_1)$ as the identity matrix ($\mathbb{II}$) plus an additional term representing the diattenuation ($\epsilon K_1$) to first order.  

\begin{equation}
{\bf M}_{ij} = exp(\epsilon K_1) = \mathbb{II} + \epsilon K_1 = 
\left [ \begin{array}{rrrr}
 1  		& \epsilon	& 0	& 0		\\
 \epsilon	& 1		& 0	& 0		\\
 0 		& 0 		& 1	& 0		\\
 0 		& 0		& 0	& 1		\\ 
\end{array} \right ] 
\end{equation}

For the third term representing $[X,Y]$, we need to use the commutation relation $[X_i, Y_j] = \varepsilon_{ijk}X_k$ where $\varepsilon_{ijk}$ is the three dimensional Levi-Civita symbol. The commutation relation for the three Euler angles (i=1,2,3) gives $[K_1, S_i] = \epsilon_{1ik} K_k$. The Levi-Civita symbol will be +1 for the cyclic (1,2,3) ordering, adding the term $\beta K_3$. The Levi-Civita symbol will be -1 for the anti-cyclic (1,3,2) ordering, adding the term $\gamma K_2$. If we collect terms and include the proper Euler angles in the sum over generators $\Sigma d_i S_i$, we find that the term $[K_1, \Sigma d_i S_i]$ includes only $(\beta K_3 - \gamma K_2)$.  Therefore the third term in Equation \ref{eqn_three_terms} for the Mueller matrix becomes:
	
\begin{equation}
{\bf M}_{ij} = exp(\frac{1}{2}[\epsilon K_1, \Sigma d_i S_i] = exp(\frac{\epsilon}{2} (\beta K_3 - \gamma K_2))
\end{equation}

Using a similar approximation for small $\epsilon$ we can write the $[X,Y]$ term as.

\begin{equation}
{\bf M}_{ij} = exp(\frac{1}{2}\epsilon (\beta K_3 - \gamma K_2))  \sim \mathbb{II} + \frac{\beta \epsilon}{2} K_3 - \frac{\gamma \epsilon}{2} K_2
\end{equation}

Combining the three terms for Equation \ref{eqn_three_terms}, we get:

\begin{equation}
{\bf M}_{ij} = (\mathbb{II} + \epsilon K_1) {\bf \mathbb{R}}  (\mathbb{II} + \frac{\beta \epsilon}{2} K_3 - \frac{\gamma \epsilon}{2} K_2 )
\end{equation}

Multiplying out this equation and neglecting terms that are $\epsilon^2$, we collect 4 terms for the Mueller matrix with the original $IQ$ perturbation $\epsilon$, the Euler angles $\beta$, $\gamma$ and the generators $K_1$, $K_2$ and $K_3$:

\begin{equation}
\label{mm_first_order_epsilon}
{\bf M}_{ij} = {\bf \mathbb{R}} (\mathbb{II} +  \epsilon K_1 + \frac{\beta \epsilon}{2}  K_3  - \frac{\gamma \epsilon}{2} K_2 )
\end{equation}

This expression gives the approximation for the Mueller matrix under our assumptions for small errors $\epsilon$ in the polarization to and from intensity cross-talk level terms of the Mueller matrix. The Mueller matrix can be represented as a rotation matrix (${\bf \mathbb{R}}$) multiplied by a group of four correction terms.  For the case we examined here, a small error in the $QI$ and $IQ$ terms gives us a first order correction to the Mueller matrix.  When expanding out the terms in Equation \ref{mm_first_order_epsilon}, we get first order corrections that have no impact inside the cross-talk terms ${\bf \mathbb{R}}$ from errors in $\epsilon$. 

We can write the first Mueller matrix correction term $K_1 {\bf \mathbb{R}}$ as:

\begin{equation}
\begin{small}
\left ( \begin{array}{rrrr}
 0  		& 1		& 0	& 0		\\
 1 		& 0		& 0	& 0		\\
 0 		& 0 		& 0	& 0		\\
 0 		& 0		& 0	& 0		\\ 
\end{array} \right ) 
\left ( \begin{array}{rrrr}
 1   		& 0		& 0		& 0		\\
 0	 	& R_{11}	& R_{21}	& R_{31}	\\
 0 		& R_{12}	& R_{22}	& R_{32}	\\
 0 		& R_{13}	& R_{23}	& R_{33}	\\ 
\end{array} \right )  =
 \left ( \begin{array}{rrrr}
 0   		& R_{11}	& R_{12}	& R_{13}	\\
 1	 	& 0		& 0		& 0		\\
 0 		& 0		& 0		& 0		\\
 0 		& 0		& 0		& 0		\\ 
\end{array} \right ) 
\end{small}
\label{eqn_mm_term_1}
\end{equation}

This term is scaled by $\epsilon$ and does not include any terms in the $QUV$ to $QUV$ portion of the Mueller matrix. Similar corrections to the Mueller matrix are seen for the term $K_2 {\bf \mathbb{R}}$:

\begin{equation}
\begin{small}
\left ( \begin{array}{rrrr}
 0  		& 0		& 1	& 0		\\
 0 		& 0		& 0	& 0		\\
 1 		& 0 		& 0	& 0		\\
 0 		& 0		& 0	& 0		\\ 
\end{array} \right ) 
\left ( \begin{array}{rrrr}
 1   		& 0		& 0		& 0		\\
 0	 	& R_{11}	& R_{21}	& R_{31}	\\
 0 		& R_{12}	& R_{22}	& R_{32}	\\
 0 		& R_{13}	& R_{23}	& R_{33}	\\ 
\end{array} \right )  =
 \left ( \begin{array}{rrrr}
 0   		& R_{12}	& R_{22}	& R_{32}	\\
 0	 	& 0		& 0		& 0		\\
 1 		& 0		& 0		& 0		\\
 0 		& 0		& 0		& 0		\\ 
\end{array} \right ) 
\end{small}
\label{eqn_mm_term_2}
\end{equation}

A similar Mueller matrix correction would be generated for the $K_3$ term. From these corrections to the Mueller matrices in Equations \ref{eqn_mm_term_1} and \ref{eqn_mm_term_2}, we see that there is no correction in the rotation matrix terms ${\bf \mathbb{R}}$ due to this first-order approximation. In the limit of small $\epsilon$, we see that neglecting the $IQ$ and $QI$ terms only impacts the first row of the Mueller matrix.  From this sensitivity analysis we can conclude that our method of approximating the 3x3 cross-talk elements of the Mueller matrix as a rotation matrix is relatively insensitive to errors in the first row and column of the Mueller matrix. Corrections to M to first order in $\epsilon$ as above only effect the intensity to polarization terms (the dichroism in M). The rotation matrix fit terms are second order in $\epsilon$ and as such, have errors much smaller than other typical limiting noise sources.  For AEOS and HiVIS, the cross-talk terms are of order 1. The induced polarization and depolarization terms are of order 5\%.  The second order corrections from neglecting the first row and column in the rotation matrix fitting are thus of order $\epsilon \sim$0.05$^2$  which is 2.5 x 10$^{-3}$.  The method of fitting rotation matrices to the cross-talk elements of the Mueller matrix is robust against dichroism type errors in induced polarization and depolarization.

\bibliography{ms_ver03}
\bibliographystyle{spiebib}		% makes bibtex use spiebib.bst

\end{document}